\documentstyle[12pt,equations]{article}

\def\NLC{future $e^+ e^-$ linear collider }

\def\9{\phantom 0}      
\renewcommand\linebreak{\unskip\break} 
\def\lesssim{\mathrel{\mathpalette\vereq<}}

\makeatletter
\def\vereq#1#2{\lower3pt\vbox{\baselineskip1.5pt \lineskip1.5pt
\ialign{$\m@th#1\hfill##\hfil$\crcr#2\crcr\sim\crcr}}}
\makeatother

\def\alt{\stackrel{<}{\sim}}
\def\agt{\stackrel{>}{\sim}}
\def\eslt{E\llap/_T}
\def\etmiss{E\llap/_T}
\def\tg{\tilde g}
\def\tst{\tilde t}
\def\tq{\tilde q}
\def\tf{\tilde f}
\def\tl{\tilde \ell}
\def\tell{\tilde \ell}
\def\tnu{\tilde \nu}
\def\tz{\widetilde\chi^0}
\def\tw{\widetilde\chi^\pm}

\begin{document}
\input psfig.sty
\newlength{\captsize} \let\captsize=\small 
\newlength{\captwidth}                     
\hfill{FSU-HEP-950401}

\hfill{LBL-37016}

\hfill{UH-511-822-95}

\renewcommand{\thefootnote}{\fnsymbol{footnote}}
\thispagestyle{empty}
{\bf LOW ENERGY SUPERSYMMETRY PHENOMENOLOGY\footnote{Manuscript
prepared by H. Baer, H. Murayama and X. Tata.}\footnote{This work was
supported by the Director, Office of Energy Research, Office of High
Energy and Nuclear Physics, Division of High Energy Physics of the U.S.
Department of Energy under Contract DE-FG05-87ER40319,
DE-AC03-76SF00098 and DE-FG-03-94ER40833.}}

\vskip .5cm
\begin{center}
H. Baer$^{*a}$, A. Bartl$^b$, C.H. Chen$^a$, H. Eberl$^h$,
J. Feng$^c$, K. Fujii$^d$, J. Gunion$^e$,\\
T. Kamon$^f$, C. Kao$^g$, J.~L.~Lopez$^f$, W. Majerotto$^h$, P. McIntyre$^f$,
R. Munroe$^a$,\\
H. Murayama$^{*i}$, F. Paige$^j$,W. Porod$^b$, J. Sender$^k$,A. Sopczak$^l$,
X. Tata$^{*k}$,\\ T. Tsukamoto$^d$ and  J. White$^f$
\end{center}
\begin{center}
$^*$ Sub-group leaders

{\it $^a$Department of Physics, Florida State University,
Tallahassee, FL 32306 USA }\break
{\it $^b$Institut f\"{u}r Theoretische Physik, Universit\"{a}t Wien,
Austria}\break
{\it $^c$SLAC, Stanford University, Stanford, CA 94309 USA}\break
{\it $^d$National Laboratory for High Energy Physics (KEK),
Tsukuba, 305 Japan}\break
{\it $^e$Department of Physics, University of California, Davis, CA USA}\break
{\it $^f$Department of Physics, Texas A\&M University,
College Station, TX 77843 USA}\break
{\it $^g$Dep't of Physics and Astronomy, Univ. of Rochester,
Rochester, NY 14627 USA }\break
{\it $^h$Institut f\"{u}r Hochenergiephysik, \\d. \"{O}sterreichischen
Akademie der Wissenschaften, Wien, Austria}\break
{\it $^i$Theoretical Physics Group, Lawrence Berkeley Laboratory\\
Berkeley, CA 94720 USA }\break
{\it $^j$Brookhaven National Laboratory, Upton, NY 11973 USA}\break
{\it $^k$Dep't of Physics and Astronomy, University of Hawaii\\
Honolulu, HI 96822 USA }\break
{\it $^l$PPE Division, CERN, CH-1211 Geneva 23, Switzerland}\break

\end{center}

\setlength{\baselineskip}{2.6ex}

\begin{center}
\parbox{13.0cm}
{\begin{center} ABSTRACT \end{center}
{\small \hspace*{0.3cm}
We summarize the current status and future prospects for low energy
(weak scale) supersymmetry.
In particular, we evaluate the capabilities of various
$e^+e^-$, $p\bar p$ and $pp$ colliders to discover evidence for
supersymmetric particles.
Furthermore, assuming supersymmetry is discovered,
we discuss capabilities of future facilities to dis-entangle the
anticipated spectrum of super-particles, and, via precision measurements,
to test mass and coupling parameters for comparison with various
theoretical expectations.
We comment upon the complementarity of
proposed hadron and $e^+e^-$ machines for a comprehensive study
of low energy supersymmetry.
}}\end{center}
\eject

\renewcommand{\thepage}{\roman{page}}
\setcounter{page}{2}
\mbox{ }

\vskip 1in

\begin{center}
{\bf Disclaimer}
\end{center}

\vskip .2in

\begin{scriptsize}
\begin{quotation}
This document was prepared as an account of work sponsored by the United
States Government. While this document is believed to contain correct
 information, neither the United States Government nor any agency
thereof, nor The Regents of the University of California, nor any of their
employees, makes any warranty, express or implied, or assumes any legal
liability or responsibility for the accuracy, completeness, or usefulness
of any information, apparatus, product, or process disclosed, or represents
that its use would not infringe privately owned rights.  Reference herein
to any specific commercial products process, or service by its trade name,
trademark, manufacturer, or otherwise, does not necessarily constitute or
imply its endorsement, recommendation, or favoring by the United States
Government or any agency thereof, or The Regents of the University of
California.  The views and opinions of authors expressed herein do not
necessarily state or reflect those of the United States Government or any
agency thereof of The Regents of the University of California and shall
not be used for advertising or product endorsement purposes.
\end{quotation}
\end{scriptsize}

\vskip 2in

\begin{center}
\begin{small}
{\it Lawrence Berkeley Laboratory is an equal opportunity employer.}
\end{small}
\end{center}

\newpage
\renewcommand{\thepage}{\arabic{page}}
\setcounter{page}{1}

\renewcommand{\thefootnote}{\arabic{footnote}}
\setcounter{footnote}{0}

\section{Goal of this study and outline}

The recent demise of the Superconducting Supercollider project in the
United States has led to the need for a re-evaluation of directions
for not only the U.S., but indeed for the world
High Energy physics community.
The goal of this report is to evaluate
the capabilities of current and future experimental facilities
with respect to the search for weak scale supersymmetric particles.
To this
end, we only review analyses that attempt to make a more or less detailed study
of experimental signatures and backgrounds in $e^+e^-$, $p\bar p$, and $pp$
interactions. Other aspects of supersymmetry model building and
phenomenology are discussed in the
accompanying review by Drees and Martin\cite{MODELS}.
It is hoped
that the information reviewed in this report will serve as an
aid to the decision making
process of how to most wisely allocate limited resources such that progress
in supersymmetry phenomenology (in particular) and high energy
physics (in general) can be maximized.

This report is organized into the following sections.
\begin{enumerate}
\item Goal and outline
\item Introduction (theoretical framework and experimental facilities)
\item Production, decay and simulation of super-particles
\item Current status of sparticle searches
\item The reach of LEP II
\item Search for SUSY at the Tevatron and upgrades
\item Search for SUSY at CERN LHC
\item Supersymmetry at future linear $e^+e^-$ colliders
\item Overview and complementarity of facilities
\item Conclusions
\end{enumerate}

\section{Introduction}

\subsection{Theoretical Framework}

Supersymmetry (SUSY)\cite{SUSY} is a novel type of symmetry that relates
properties of bosons to those of fermions. It is the largest known
symmetry\cite{HLS}
of the $S$-matrix. Locally supersymmetric theories necessarily incorporate
gravity\cite{GRAV}.
SUSY is also an essential ingredient of superstring theories\cite{STRING}
which,
today,
offer the best hope for a consistent quantum theory of gravitation.
Although no compelling supersymmetric model has yet emerged,
and despite the fact that there is no direct experimental evidence for SUSY,
the remarkable theoretical properties of SUSY theories have provided ample
motivation for their study. Of importance to us is the fact that SUSY leads
to an amelioration of divergences in quantum field theory. This, in turn,
protects the electroweak scale from large quantum corrections, and stabilizes
the ratio $\frac{M_W}{M_X}$,
when the Standard Model (SM)
is embedded
into a larger theory, involving an ultra-high energy scale $M_X$
({\it e.g.} $M_{GUT}$ or $M_{Planck}$). In other words, SUSY models do
not require\cite{HIER} the incredible fine-tuning endemic to the Higgs sector
of the SM,
provided only that the super-partners exist at or below the TeV energy scale.
On the experimental side, while the measurements
of the three SM
gauge couplings at LEP\cite{COUP} are incompatible with unification in
the minimal SU(5) model, they unify\cite{UNIF}
remarkably well in the simplest supersymmetric SU(5) GUT, with SUSY
broken at the desired scale $\sim 1$ TeV.
The last two arguments are especially important in that they
bound the SUSY breaking scale and strongly
suggest that supersymmetric partners of ordinary particles should
be accessible at colliders designed to probe the TeV energy scale\cite{REV}.

To evaluate the experimental consequences of low energy supersymmetry, one
must set up a Lagrangian including the various particles and
partner sparticles,
and their interactions. Such a theory should reduce to the well-tested
SM when the supersymmetric degrees of freedom are integrated over.
The simplest possibility\cite{MODELS}, the Minimal Supersymmetric Standard
Model (MSSM),
is a direct supersymmetrization of
the SM\cite{REV}. It is a Yang-Mills type gauge theory based on the SM gauge
group,
with electroweak symmetry spontaneously broken via vacuum expectation values
(VEVs) of two different Higgs superfields that respectively couple
to $T_3 = \frac{1}{2}$ and $T_3 = -\frac{1}{2}$ fermions. The (renormalizable)
superpotential
that determines the Yukawa interactions of quarks and leptons is required to
conserve baryon and lepton numbers;
it is then possible to define a multiplicatively conserved
$R$-parity quantum number which is +1 for ordinary particles and -1 for
supersymmetric partners.
The MSSM is thus minimal in that it
contains the smallest number
of new particles and new interactions to be compatible with
phenomenology. An important consequence of $R$-parity conservation is that
the lightest supersymmetric particle (LSP) is stable.
The LSP, which would have been abundantly produced in the early
universe, is unlikely\cite{WOLF} to be colored or electrically charged since
it would then be able to bind to nuclei or atoms to make heavy isotopes,
for which searches\cite{EXOT} have yielded negative results. The LSP, which is
the end product of every sparticle decay, thus escapes experimental
detection, resulting in apparent non-conservation of
energy/momentum in SUSY events.

Supersymmetry must, of course, be a broken symmetry. In the absence
of fundamental understanding of the origin of supersymmetry breaking,
supersymmetry breaking is parametrized by incorporating
all soft supersymmetry breaking terms (defined to be those that do not
destabilize the ratio $\frac{M_W}{M_X}$ introduced above) consistent with
the SM symmetries. These terms
have been classified by Girardello and Grisaru\cite{GG}. For the MSSM, they
consist
of
\begin{itemize}
\item gaugino masses ($M_1$, $M_2$ and $M_3$ for each of the $U(1)$,
$SU(2)$ and $SU(3)$ gauge groups),
\item mass terms for
various left- and right- spin-0 (squark, slepton, Higgs) fields,
\item trilinear ($A$-term) interactions amongst the scalars, and
\item analogous bilinear ($B$-term) interactions.
\end{itemize}
In addition to these soft-breaking terms, the ratio $\tan\beta$ of the two
Higgs
field VEVs and a supersymmetric Higgsino mixing parameter $\mu$ must be
specified. Aside from the particles of the Standard Model, the physical
spectrum of the MSSM consists of the following additional states.
\begin{itemize}
\item squarks (spin-0): $\tilde d_L$,$\tilde u_L$,$\tilde s_L$,
$\tilde c_L$,$\tilde b_1$,$\tilde t_1$, $\tilde d_R$,$\tilde u_R$,$\tilde s_R$,
$\tilde c_R$,$\tilde b_2$,$\tilde t_2$;
\item sleptons (spin-0): $\tilde e_L$,$\tilde \nu_{eL}$,
$\tilde\mu_L$,$\tilde\nu_{\mu L}$, $\tilde\tau_1$,$\tilde\nu_{\tau L}$,
$\tilde e_R$,$\tilde\mu_R$,$\tilde\tau_2$;
\item charginos (spin-$1\over 2$): $\tw_1$, $\tw_2$;
\item neutralinos (spin-$1\over 2$): $\tz_1$, $\tz_2$, $\tz_3$, $\tz_4$
\item gluino (spin-$1\over 2$): $\tg$;
\item Higgs bosons: (spin-0): $h$, $H$, $A$, $H^\pm$.
\end{itemize}

Here, $\tilde{t_i},\tilde{b_i}$, and $\tilde{\tau_i}$ ($i=1,2$) are
mixtures of the corresponding left- and right- chiral scalar fields,
charginos are mixtures of charged higgsino and wino, and neutralinos are
mixtures of two neutral higgsinos, bino and the neutral wino.
In our analysis, we neglect any inter-generational sfermion mixing. The
intra-generational mixing (being proportional to the corresponding {\it
fermion} mass) is negligible for the first two generations of sfermions.

An independent value for each one of the above masses and
couplings leads to a proliferation of new parameters, making phenomenological
analyses intractable.
It is customary to assume that higher symmetries, which are broken
at some ultra-high scale, relate these parameters.
An especially appealing and economic class of models is based on minimal
supergravity (SUGRA) GUTs, where it is assumed the three gauge couplings
unify at some ultra-high energy scale $M_X$.
It is also assumed that SUSY breaking in the effective theory
defined at $M_X$ arises due to gravitational
interactions which, being universal, allow\cite{REV,MODELS} only a {\it few}
independent soft SUSY
breaking parameters, renormalized at $M_X$:
these are
\begin{itemize}
\item a common gaugino mass ($m_{1/2}$),
\item  a common scalar mass ($m_0$),
\item a common trilinear interaction ($A_0$), and
\item the bilinear coupling ($B_0$).
\end{itemize}

The various MSSM masses and couplings have to be evolved\cite{RGE} from the
common value at $M_X$
to the electroweak scale to sum the large logarithms
arising from the disparity between the two scales.
This generally involves solving 26 coupled differential
equations, with the values of the four GUT-scale parameters as boundary
conditions.
A bonus of this
framework is that the same radiative corrections that give rise to these
large logs also yield a mechanism for the breakdown of
electroweak gauge symmetry, leaving
colour and electromagnetic gauge symmetries unbroken\cite{RGE,RAD}.
The electroweak symmetry breaking constraint allows one to
eliminate $B_0$
in favour of $\tan\beta$, and also to adjust the magnitude (but not the sign)
of the $\mu$ parameter to get the measured $Z$ boson mass.
Thus, the Renormalization Group (RG) evolution
of these four parameters, renormalized at the  GUT scale where
the physics is simple, results
in a rich pattern of sparticle masses and couplings at the weak scale relevant
for phenomenology. The various SUSY parameters, masses and mixings
are then determined in terms of the four plus a sign parameter set
\begin{eqnarray}
m_0,\ m_{1/2},\ \tan\beta ,\ A_0\ {\rm and}\ sign(\mu ).
\end{eqnarray}
In addition, as with the SM, the top mass $m_t$ must be specified.
The simplest supergravity model (with minimal kinetic energy terms\cite{WSFN})
leads to
(approximate) degeneracy of the first two generations of squarks, and so,
is automatically consistent with constraints\cite{FCNC} from flavour changing
neutral
currents in the $K$-meson sector\cite{FCNCFN}.
The masses of third generation squarks can
be significantly different\cite{ER}:
this can lead to interesting phenomenology\cite{TPHEN,BST} as
discussed below. When these supergravity constraints are incorporated,
one finds (approximately)
$m_{\tq}^2 \simeq m_{\tell}^2 + (0.7-0.8)m_{\tg}^2$; thus
sleptons may be significantly lighter than the first two generations of
squarks. Furthermore, the value of $|\mu|$ is generally large compared to the
electroweak gaugino masses, so that the lighter neutralinos
($\tz_{1,2}$)
and the lighter chargino ($\tw_1$)
are gaugino-like, while the heavy
chargino and the heavier neutralinos are dominantly Higgsinos. If
$m_{\tq}\simeq m_{\tg}$ so that sleptons are significantly lighter than
squarks, the leptonic decays of $\tz_2$, and sometimes also of $\tw_1$
can be significantly enhanced relative to those of $Z$ and $W$ bosons,
respectively; this has important implications\cite{BT}
for detection of sparticles
at hadron colliders.
Within the SUGRA framework,
the lightest SUSY particle is a viable candidate for dark matter\cite{DM},
provided that the sfermions are not too heavy (the LSP, being mostly a
hypercharge gaugino, mainly annihilates via sfermion exchange, so that the
annihilation rate is proportional to $\frac{1}{m_{\tf}^4}$).

While minimal supergravity models indeed provide an economic and
elegant framework, it should be recognized that the assumptions
(about the physics at an ultra-high energy scale) on which they are based
may ultimately prove to be incorrect.
The point, however, is that these models
lead to rather definite correlations between various sparticle
masses\cite{SPECTRA}
as well as between the cross-sections\cite{TSUK,SUGCS,BCMPT}
for numerous signals. We will see that
these predictions, which serve as tests of the underlying assumptions, can be
directly  tested at future accelerator facilities.
Thus the discovery of
sparticles and a determination of their properties\cite{TSUK}
may provide a window
to the nature of physics at an ultra-high energy scale.

It is also worth considering various extensions of the minimal framework
that we have been describing. On the one hand, in some string-inspired
models\cite{IBAN},
the four SUGRA input parameters are
not all independent so that these models are even more tightly constrained.
On the other hand, there has been some recent
interest in non-universal SUSY-breaking terms\cite{POKORSKI},
so that the correlations
would be modified from their expectations within the minimal framework.
In addition, there is no reason why the grand unification scale should
exactly coincide
with the scale at which the boundary conditions
for the RGEs are specified. Furthermore, some models
prefer unification at the string scale ($\sim 0.5\times 10^{18}$ GeV),
so that the
apparent unification at $\sim 10^{16}$ GeV is only coincidence\cite{NOSCALE}.
It may also be that the particle content of the low energy
theory goes beyond that of the MSSM\cite{EXT} ({\it e.g.} there might be
additional
Higgs singlets or gauge bosons),
or $R$-parity might not be conserved by superpotential
interactions\cite{RVIOL}. In this latter case, there would be additional
unknown parameters
and the phenomenology might differ substantially from what is expected in
minimal supergravity. However, in the absence of large $R$-violating
interactions, many of the gross features of minimal supergravity seem likely
to be manifest if low energy supersymmetry exists.
But practically speaking, minimal supergravity is a simple and
phenomenologically viable framework that offers distinctly testable
consequences from a small parameter set. It serves as the paradigm for
most phenomenological investigations, although one must bear in mind
the consequences of possible modifications to this scheme.

In the early literature\cite{EARLY}, it is common to see analyses based upon a
more general
supersymmetry parameter set, but frequently with various SUGRA GUT inspired
assumptions built in. For instance, assuming all gaugino masses
evolve as in a SUSY GUT allows the correlation of the various gaugino masses,
which are frequently parametrized in terms of $m_{\tg}$
($m_{\tg}\simeq |M_3|$, modulo the distinction\cite{MT} between the
$\overline{DR}$
and pole mass\footnote{See discussions on Fig.~\ref{fig64}}),
or $M_2$. In addition, the various squarks or sleptons are assumed
approximately degenerate, as predicted by minimal SUGRA with its common
GUT scale $m_0$ mass. A common but more general parameter set (which we will
loosely refer to as the MSSM parameter set to distinguish it from the SUGRA
set (2.1)) is frequently
specified by
\begin{eqnarray}
m_{\tg},\ m_{\tq},\ m_{\tell},\ A_t,\ A_b,\ \mu ,\ m_A ,\ \tan\beta ,
\end{eqnarray}
where the
supersymmetric Higgsino mass $\mu$ and the pseudoscalar Higgs boson mass
determine (at tree level) all the other parameters of the  Higgs sector,
and the weak scale soft SUSY-breaking trilinear
couplings $A_t$ and $A_b$ mainly only influence the phenomenology of third
generation sfermions. As in the SM, $m_t$ must be input as well.
We caution the reader that the SUGRA model specified
by the parameter set (2.1) is referred to as the MSSM by some
authors\cite{MODELS}. Here, we
will reserve the term MSSM for the broader framework specified by the set
(2.2) above, and refer to the more constrained framework as minimal SUGRA.
The parameter set (2.2) yields much of what is expected
in minimal SUGRA, but (with some modification) can also accommodate other
models with non-universal soft-breaking terms, which can yield substantially
different
Higgs masses ($m_A$) and $\mu$ values (leading to Higgsino-like LSP's)
than the usual SUGRA prediction (for an example, see Ref. \cite{POKORSKI}).

\subsection{Experimental facilities}

The formalism of supersymmetry was developed in the 1970's,
and viable phenomenological
(low energy) models ({\it e.g.} the MSSM) were formulated in the early
1980's.
A reasonable picture of how
supersymmetry might manifest itself has emerged, and now the urgent
need is to either discover supersymmetry, or experimentally rule out
the existence of weak scale sparticles.
It is possible that the first evidence for supersymmetry could arise in
a number of experiments: {\it e.g.} dark matter detectors, study of
rare $B$ or $K$ decays, proton decay. However, it is usually expected that
unambiguous evidence for supersymmetry must be obtained at colliding
beam experiments, where super-particles can be directly produced, and
their decay products detected and analyzed. Hence, in this report, we
focus on future colliding beam facilities, and their ability to cover regions
of the supersymmetric parameter space. For $e^+e^-$ colliders, we consider,
along with current constraints from LEP and SLC, future searches at LEP II
and at hypothetical linear colliders \cite{JLC} operating at
$\sqrt{s}=500-1000$ GeV.
For $p\bar p$ colliders, we consider current constraints from the Tevatron
collider ($\sqrt{s}=1.8$ TeV), and its possible high luminosity (TeV$^*$) and
high energy (DiTevatron ($\sqrt{s}=4$ TeV)) upgrades\cite{AMED}. Finally, we
also
examine the capability of the CERN LHC $pp$ collider, at
$\sqrt{s}=14$ TeV, approved at CERN Council Meeting in December 1994.

\section{Production, decay and simulation of sparticles}

If $R$-parity is conserved,
then sparticles ought to be pair-produced at colliders with sufficiently high
energy, and furthermore, sparticles ought always to decay into other
sparticles, until the decay cascade terminates in the stable LSP. Within
the minimal framework, the only viable candidates are the sneutrino
and the lightest neutralino $\tz_1$. If we further assume that the LSP
is the dominant component of galactic dark matter, the sneutrino is
heavily disfavoured\cite{BDT} when the negative results of experiments
searching
for double beta decay are combined with LEP constraints discussed below.
In what follows, we will assume that $\tz_1$ is the LSP.

\subsection{Hadron colliders}

At hadron colliders, sparticles can be produced via the following lowest order
reactions (particles/anti-particles not distinguished):
\begin{itemize}
\item $qq,\ gg,\ qg\to\tg\tg ,\ \tg\tq ,\ \tq\tq ,$ (strong production)
\item $qq,\ qg\to \tg\tz_i ,\ \tg\tw_i ,\ \tq\tz_i ,\ \tq\tw_i$
(associated production)
\item $qq\to \tw_i\tilde{\chi}_j^\mp ,\ \tw_i\tz_j ,\ \tz_i\tz_j$ ($\tilde\chi$
pair production)
\item $qq\to \tell\tnu ,\ \tell\tell ,\ \tnu\tnu$ (slepton pair production)
\end{itemize}
In addition, the Higgs bosons of the MSSM can be produced via
direct $s$-channel subprocess,
\begin{itemize}
\item $qq,\ gg\to h,\ H,\ A,\ H^{\pm}H^{\mp}$,
\end{itemize}
in addition to production in association with other heavy quarks and
vector bosons, and in some cases, production via vector boson fusion.

Once produced, sparticles rapidly decay to other sparticles through a cascade
ending in the LSP\cite{CAS}. The decay modes and branching fractions
of sparticles are too numerous to be listed here. However, a number of
groups have generated computer programs to calculate some or all of
the sparticle decays. Perhaps the most complete listing is available
as a public access program called ISASUSY, and can be extracted from
the ISAJET program\cite{ISAJET}
described below. Given a point in MSSM parameter
space, ISASUSY lists all sparticle and Higgs
masses, decay modes, decay widths and branching fractions.

The crucial link between the theoretical framework of supersymmetry
(discussed above), and the detection of long lived particles such as
$\pi$'s, $K$'s, $\gamma$'s, $e$'s, $\mu$'s, {\it etc.} at colliding beam
experiments, lies with event generator programs. Many groups have
combined sparticle production and decay programs, to create parton level
Monte Carlo programs. Some have added in as well parton showers and
hadronization. ISAJET 7.13 is currently the most comprehensive of the
supersymmetry event generators available for hadron colliders.

To simulate sparticle production and decay at a hadron collider, the following
steps are taken using ISAJET:
\begin{itemize}
\item input the parameter set ($m_0, m_{1/2},A_0,\tan\beta ,sign(\mu )$),
(or the less constrained MSSM set (2.2)),
\item all sparticle and Higgs masses and couplings are computed,
\item all sparticle, top and Higgs decay modes and branching fractions
are calculated,
\item all lowest order $2\rightarrow 2$ sparticle production processes
are calculated (if desired, subsets of the reactions can be selected),
\item the hard scattering is convoluted with CTEQ2L PDF's\cite{CTEQ},
\item initial and final state QCD radiation is calculated with the parton
shower
model,
\item particles and sparticles decay through their various cascades,
\item quarks and gluons are hadronized, and heavy hadrons are decayed,
\item the underlying soft scattering of beam remnants is modelled,
\item the resulting event and event history is generated for interface
with detector simulations, or for direct analysis.
\end{itemize}

There are a variety of shortcomings to (the current) ISAJET supersymmetry
simulation. Some of these include:
\begin{itemize}
\item direct Higgs production mechanisms for hadron colliders are
currently absent,
\item large $\tan\beta$ solution, (currently, ISAJET is only valid
for $\tan\beta\alt 10$, primarily because the mixing between $b$-squarks
and third generation sleptons, due to the corresponding Yukawa interactions
has not yet been incorporated---this can be very important if {\it
e.g.} the resulting mass splitting opens up new gluino, chargino or
neutralino decay channels).
\item first two generations of squarks are assumed mass degenerate,
\item lack of spin correlation between initial and final state sparticles,
\item although decay branching fractions are calculated with full matrix
elements, decays in event generation are modelled using only phase space.
\end{itemize}
Some of these deficiencies should be corrected in the near future.

Separate programs can be extracted
from ISAJET 7.13 which generate just the SUGRA mass solution plus decay table
(ISASUGRA),
and also the sparticle decay table with MSSM input (ISASUSY).
ISAJET is maintained in PATCHY format.
The complete card image PAM file for ISAJET 7.13 can be copied across
HEPNET, the high energy physics DECNET, from \hfil\break
\verb|bnlcl6::$2$dua14:[isajet.isalibrary]isajet.car|.
A Unix makefile \hfil\break \verb|makefile.unix|
and a VMS \verb|isamake.com| are available in the
same directory. The same files can be obtained by anonymous ftp from
\verb|bnlux1.bnl.gov:pub/isajet|.

\subsection{$e^+e^-$ colliders}

At $e^+e^-$ colliders, the following production mechanisms can be important:
\begin{itemize}
\item $e^+e^-\to\tq\bar{\tq}$ (squark pair production)
\item $e^+e^-\to\tell\bar{\tell}$ (slepton pair production)
\item $e^+e^-\to \tw_i\tilde{\chi}_j^\mp,\ \tz_i\tz_j$ ($\tilde\chi$ pair
production)
\item $e^+e^-\to hZ,\ HZ, hA,\ HA,\ H^+H^-$ (Higgs production),
\end{itemize}
in addition to vector boson annihilation to Higgs particles.
After production, the sparticles and Higgs bosons decay through the usual
cascades.

ISAJET 7.13 includes all lowest order $e^+e^-\rightarrow SUSY$
and Higgs particle production processes\cite{BBKMT}, so that cascade
decays and minimal
SUGRA can be simulated for electron machines as well.
However, ISAJET does not include initial state photon radiation, spin
correlations, decay matrix elements or polarizable beams. All these effects
are expected to be more important in the environment of $e^+e^-$
colliders.

An event generator for $e^+e^-\to SUSY$ particles, containing
the above first three sets of reactions, has been released by
Katsanevas, under the name SUSYGEN\cite{katsan}. SUSYGEN also includes
cascade decays as calculated by Bartl {\it et. al.}\cite{alfred}.
SUSYGEN includes initial state photon radiation, and is interfaced to
the LUND JETSET string hadronization program. Like ISAJET, SUSYGEN
currently lacks spin correlations, final state decay matrix elements, and
beam polarizability.

Various groups have created generators that correct some or all of the
deficiencies in the SUSYGEN or ISAJET $e^+e^-$ generators, but these are
limited to specific production and decay configurations\cite{hitfeng}.
The problem then becomes that it is difficult to simulate all reactions
and decays simultaneously, so a separate program is needed for each
possible configuration.  However, it is desirable to incorporate the
angular correlation in the decay for the studies at an $e^+ e^-$
collider since it affects the detection efficiencies and background
contaminations which are important especially for precision studies. The
helicity amplitude technique is better suited for this purpose. A set of
FORTRAN subroutines HELAS \cite{HELAS} can be used to calculate helicity
amplitudes numerically, where each of the subroutine calls correspond to
each of the vertices in the Feynman diagrams. A repeated use of HELAS
calls computes helicity amplitudes rather easily. It can be
obtained via anonymous ftp at {\tt tuhep.phys.tohoku.ac.jp}.

\section{Current status of supersymmetry searches}

\begin{figure}
\let\normalsize=\captsize
\centerline{\psfig{file=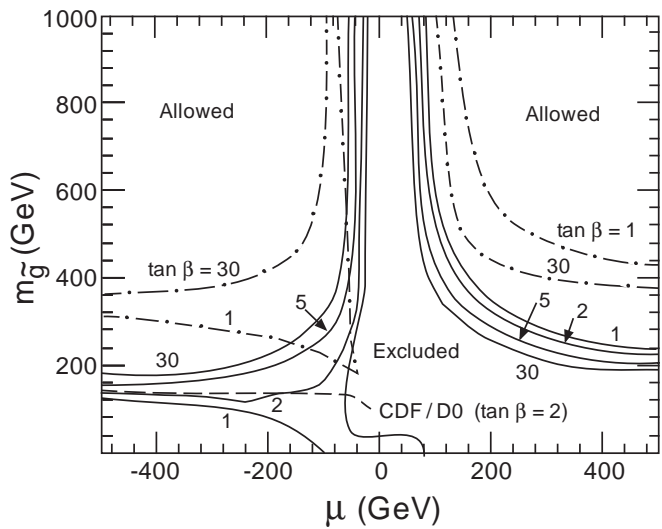,height=7cm}}
\caption[]{Regions in the $\mu\ vs.\ m_{\tg}$ plane excluded by
various constraints from LEP experiments described in the text, for
$\tan\beta =1$, 2, 5 and 30. We also show the approximate reach of LEP II
via the dot-dashed curves ($m_{\tw_1}=90$ GeV), and the region excluded by
CDF and D0 gluino and squark searches at the Tevatron.}
\label{fig41}
\end{figure}

The most direct limits on sparticle masses come from the non-observation of
any SUSY signals at high energy colliders, and from the precision
measurements of the properties of $Z$ bosons in experiments at LEP (for
a compilation of constraints, see \cite{PDG}).
The agreement\cite{COUP} of the measured value of $\Gamma_Z$ with its
expectation in
the SM gives model-independent constraints on decays of the $Z$-boson into any
new particles with known $SU(2)\times U(1)$ quantum numbers\cite{BDT,OTHWID}.
This translates to a lower
limit $\sim 30-45$ GeV on the masses of the sneutrinos, squarks
and charginos of the MSSM, independent of the decay patterns of these
sparticles. Likewise, the measurement of the invisible width of the $Z$-boson
which leads to the well-known bound on the number of light neutrino species,
leads the lower bound of $m_{\tnu}>43$ GeV,
even when the sneutrinos
decay invisibly via $\tnu \rightarrow \nu \tz_1$\cite{BDT,OTHWID}. These bounds
are relatively insensitive to the details of the model. In contrast,
even within the MSSM framework, the
corresponding bounds on neutralino masses are sensitive to model parameters.
This is because in the limit $\mu \gg M_1,M_2$, the lighter neutralinos are
dominantly gaugino-like, so that their couplings to the $Z$-boson are
strongly suppressed by electroweak gauge invariance\cite{BDT,OTHWID}. The LEP
experiments\cite{DIRECT}
have also directly searched for charginos, sleptons and squarks in $Z$ decays.
These searches assume that the charginos and sfermions decay directly
to the LSP which is assumed to be the lightest neutralino. The typical
signature of SUSY events is a pair of acollinear leptons, acollinear jets
or a lepton-jet pair recoiling against $\eslt$ (missing transverse
energy). The non-observation of such
spectacular event topologies have led to lower bounds very close to
$\frac{M_Z}{2}$ on the masses of these sparticles. Finally, LEP experiments
have also searched\cite{NEUT} for neutralino production via $Z \rightarrow
\tz_1\tz_2$
and $Z \rightarrow \tz_2\tz_2$ decays, assuming that $\tz_2 \rightarrow
\tz_1 f \bar{f}$ ($f = q$ or $\ell$). The non-observation of acollinear
jet or lepton pairs from this process excludes certain regions of the
parameter space, but does not (for reasons already explained) lead to an
unambiguous lower bound on $m_{\tz_2}$. The region of the $\mu -m_{\tg}$
plane excluded by LEP searches for charginos and neutralinos is illustrated
in Fig.~\ref{fig41} for various values of $\tan\beta$. Also shown in this
figure
are corresponding contours of $m_{\tw_1}$ = 90~GeV, which roughly denote
the reach of LEP II, discussed in the next Section.

\begin{figure}
\centerline{\psfig{file=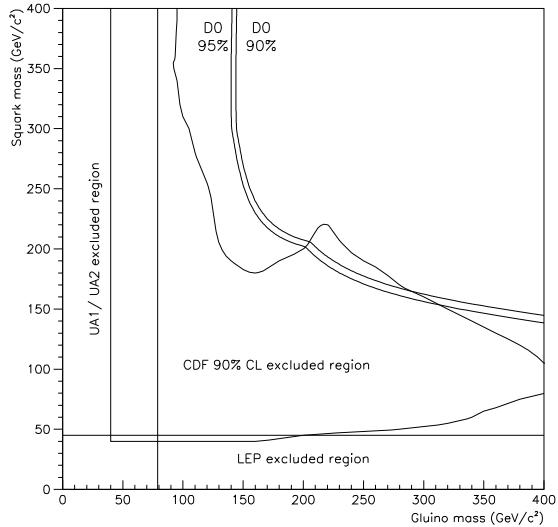,height=8cm}}
\caption[]{Regions in the $m_{\tg}\ vs. m_{\tq}$ plane excluded by
searches for $\etmiss +jets$ events at
various colliders, for $\tan\beta =2$ and $\mu =-250$ GeV. The curves
labeled D0 exclude the regions below the curves at 90 and 95~\%
confidence levels, respectively. The figure is
taken from Ref. \cite{D0SQ}.}
\label{fig42}
\end{figure}

Although LEP experiments have yielded a limit of $m_{\tq}\agt\frac{M_Z}{2}$,
the search for the strongly interacting squarks and gluinos
is best carried out at high energy hadron colliders such as the Tevatron
via $\tq\tq, \tg\tq$ and $\tg\tg$ production as discussed in Sec. 3. The
final state from the subsequent cascade decays\cite{CAS} of squarks and gluinos
consists
of several jets plus (possibly) isolated leptons (from
$\tw_1$ and $\tz_2$ production via their primary decays) and $\eslt$ from the
two LSPs in each final state. For an integrated luminosity of $\sim 20 \
pb^{-1}$ that has been accumulated by the CDF and D0 experiments at
Tevatron run IA, the classic $\eslt$ channel offers the best hope for
detecting SUSY.
The non-observation of an excess of $\eslt$ events above SM
background expectation has enabled the D0 collaboration\cite{D0SQ} to infer
a lower limit of $\sim 150$~GeV on their masses, improving on the
published CDF limit\cite{CDFSQ} of $\sim 100$~GeV.
The region of the $m_{\tg}-m_{\tq}$
plane excluded by these anlyses depends weakly on other
SUSY parameters, and is shown in Fig.~\ref{fig42} for $\mu = -250$~GeV and
$\tan\beta = 2$. We see that the lower bound on the mass improves to about
205~GeV if $m_{\tq}=m_{\tg}$. As the experiments at the Tevatron continue
to accumulate more data, they will also become sensitive to leptonic signals
from cascade decays of squarks and gluinos. Although the single lepton signals
are overwhelmed by the background from $(W \rightarrow \ell\nu) + jet$ events,
we will see in Sec. 6 that
the multilepton signals offer new ways of searching for supersymmetry
at various Tevatron upgrades\cite{TEVLEP,RPV}.

Before closing this Section, we briefly remark upon some constraints from
``low energy'' experiments and from cosmology. Some judgement must be
exercised in evaluating these constraints which, unlike the
direct constraints from collider experiments, can frequently be evaded by
relatively minor modifications of the model framework. For instance,
an overabundance of LSPs produced in the early universe
leads to significant restrictions on SUGRA parameters\cite{DM}. This bound
can, however, be simply evaded by allowing a small amount of $R$-parity
violation that causes the LSP to decay, although at a rate that has no other
impact on particle physics phenomenology. Likewise, constraints from
proton decay\cite{PDECAY} are sensitive to assumptions about physics at the GUT
scale.
Supersymmetry also allows for new sources of CP violation\cite{CPVIOL}
in the form of new
phases in gaugino masses or SUSY breaking trilinear scalar interactions.
Indeed, for sparticle masses $\sim 100$~GeV,
these phases (which are usually set to zero in the MSSM) are
limited\cite{CPVIOL,EDM}
to be $\sim 10^{-3}$ in order that
the induced electric dipole moment of the neutron or electron not exceed
its experimental upper limit. If, however, these phases are zero at some
ultra-high unification scale, it has been checked that their values at
the weak scale induced via renormalization group evolution do not lead to
phenomenological problems. This only pushes the problem to the unification
scale where the physics is as yet speculative\cite{DIMOP}. There
are also constraints from the universality of the charged-current and
neutral-current weak interactions.  The Cabbibo universality between the
$\mu$-decay and $\beta$-decays put constraints only on rather light
sparticles $\lesssim 20$~GeV \cite{BBGD}.  $Z$-decay partial widths into
different species of light fermion are more sensitive than the
low-energy experiments.  However most of the decoupling effects do not
put constraints better than that from the direct search \cite{HM}.
Non-decoupling effects in $\rho$-parameter \cite{rho} or $\Gamma(Z
\rightarrow b\bar{b})$ \cite{BF} are relatively sensitive to the virtual
exchange of sparticles.  Indeed, it has been claimed that the
experimental value of $R_b = \Gamma(Z \rightarrow
b\bar{b})/\Gamma(Z\rightarrow \mbox{hadrons})$ prefers a light top
squark and chargino\cite{zbbkane}. Even so, it is hard to obtain a
large enough effect to explain the ``anomaly'' \cite{Altarelli}. These
measurements do not currently lead to any significant restrictions on sparticle
masses.

Finally, we turn to the flavour violating
inclusive decay $b \rightarrow s\gamma$ recently measured by the CLEO
collaboration\cite{CLEO}. Even within the minimal SUSY
framework, there are several additional contributions to this amplitude.
Of course, the agreement of the SM computations with the experimental data
lead to an interesting limit (within theoretical and experimental errors)
on the {\it sum} of various new physics contributions. Since it is
possible\cite{BSGAM}
for these new contributions to (partially) cancel over
a significant range of model parameters, these measurements do not lead to
unambiguous bounds on the masses of various sparticles. Like the neutralino
search at LEP, they do, however, exclude significant
regions of parameter space. It should, however, be mentioned that
complete calculations of QCD
corrections
(which are known to be significant within the SM\cite{BSQCD})
to these amplitudes\cite{ANLAUF} are not yet available, so that there is still
considerable uncertainty\cite{BURAS} in the theoretical estimates of the
$b\rightarrow s\gamma$ decay rate.

To summarize: a wide variety of empirical constraints have served to
restrict the parameter ranges of supersymmetric
models. It is, however, interesting that even the simplest, highly
constrained supergravity GUT model, is
consistent with {\it all} experimental data
including those from cosmology.

\section{Search for SUSY at LEP II}

For LEP-II, the second phase of the LEP program, the center-of-mass energy
will be raised to about 175~GeV in 1996 and could reach about 200~GeV
at a later stage.
The reach of LEP II experiments on the search for sparticles has been
discussed extensively in the literature. The basic result is that all
sparticles close to the kinematic limit can be discovered except
special cases of small visible energies or very low cross sections.
The measurements of various
SUSY parameters have also been discussed, and are shown to be possible for
particular combinations of the parameters by studying the first chargino
alone. These are the topics which will be described in this
section.

\subsection{Characteristics of LEP-II experiments}

First of all, $e^+ e^-$ experiments gradually raise the center of
mass energy,\footnote{Even if LEP-II immediately goes from $\sqrt{s} =
m_Z$ to $\simeq 175$~GeV, this step is in a much smaller ratio compared to
that of the Tevatron to the LHC.} so that one does not expect the production of
many different
sparticles at the same time. We expect the discovery of the lightest
visible sparticle (LVSP) first, which does not decay in a (long) cascade
as for squarks or gluinos at hadron colliders.
This fact makes the search for LVSP at
an $e^+ e^-$ collider very simple. We look for the production of just one
new particle which leaves clear signatures.

In the MSSM framework, there are three candidates for LVSP: (1) slepton
(mostly right-handed $\tilde{e}_R$, $\tilde{\mu}_R$, $\tilde{\tau}_R$),
(2) chargino $\tilde{\chi}_1^\pm$, or (3) stop $\tilde{t}$ (or sbottom
in some cases).  When we
relax the theoretical assumptions built in to the
MSSM, one may also expect other
sparticles to be LVSP. We will discuss each case separately below.

The following point is worth emphasizing. A search for a sparticle
below $m_W$ is relatively easy because one can set the center-of-mass
energy below the $W$-pair threshold, so that most of the signatures are
nearly background-free. This is a continuation of searches done at lower
energy colliders PEP, PETRA and TRISTAN. Cuts on missing $p_T$
and on the acoplanarity angle removes almost all the QED
backgrounds. Above the $W$-pair threshold, $W$-pair production becomes the
most severe background. Most of the time, we assume $\sqrt{s} =
190$~GeV, and an integrated luminosity of 100 to 500~pb$^{-1}$.

\subsection{$\tilde{l}$-pair production}

We assume that one of the sleptons is the LVSP in this subsection. We
also assume that $R$-parity is conserved, and the LSP is the lightest
neutralino.
Then the only possible decay mode is $\tilde{l} \rightarrow l
\tilde{\chi}_1^0$ further assuming lepton family number conservation.

\begin{figure}
\centerline{\psfig{file=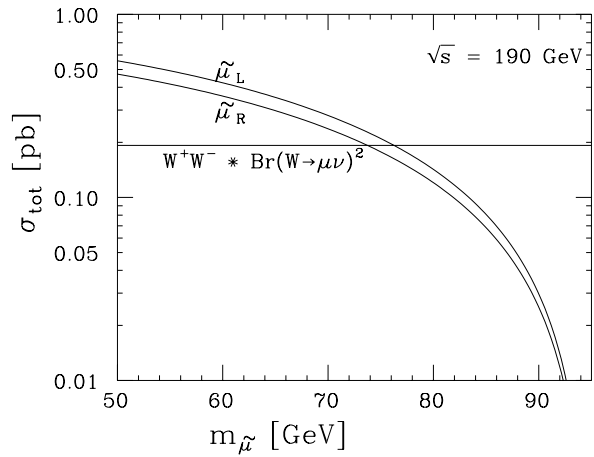,height=6cm,angle=90}}
\caption[smu-LEP2]{Total cross sections of $\tilde{\mu}_R$ and
$\tilde{\mu}_L$ pair production at $\sqrt{s} = 190$~GeV.}
\label{smu-LEP2}
\end{figure}

Fig.~\ref{smu-LEP2} shows the total cross sections for right-handed and
left-handed $\tilde{\mu}$ pair production. The threshold behavior is
$\beta^3$ characteristic of scalar pair production. The experimental
signature is a lepton pair with the same flavors with large missing
energy or acoplanarity. Here we summarize the analysis of
Ref.~\cite{Dionisi,Chen} based on a rather conservative LEP detector
simulation using the resolutions of
$18\%/\sqrt{E}$ for ECAL at $5^\circ \leq \theta \leq 175^\circ$,
$120\%/\sqrt{E}$ for HCAL at $10^\circ \leq \theta \leq 170^\circ$, and
the angular resolutions $\Delta \phi = \Delta \theta = 0.01$ (ECAL) and
0.02 (HCAL). We made a slight modification: the level of the $W$-pair
background is recalculated with $m_W = 80$~GeV and we discuss
right-handed sleptons only without assuming the degeneracy with their
left-handed counter parts.\footnote{More detailed studies with the real
detector simulations are ongoing in LEP-II workshop.} The reach for
the left-handed sleptons is the
same because their production cross sections are almost the same with
the right-handed ones. One standard set of cuts is
\begin{enumerate}
\item Two isolated muons well inside the detector $|\cos \theta_{\mu^\pm}| <
0.9$, and $E_{\mu^\pm} > 5~\mbox{GeV}$,
\item large acoplanarity angle\footnote{We define the acoplanarity angle
as $\theta_{acop} = \pi - \Delta \phi$
where $\Delta \phi$ is a difference between the azimuths of the two
final-state momenta.
Some people use $180^\circ - \theta_{acop}$ as the definition.
\label{acopdef}} $\cos \theta_{acop} < 0.9$,
\item $\pm \cos \theta_{\mu^\pm} > 0$, where the polar angle $\theta$ is
defined as an angle from the {\it electron}\/ beam axis.
\end{enumerate}
A typical efficiency for the signal is about 35\% \cite{Dionisi,Chen}. Then
the discovery reach (5~$\sigma$) extends to 65~GeV (80~GeV) with
100~pb$^{-1}$ (500~pb$^{-1}$).

\begin{table}
\begin{center}
\begin{minipage}{9cm}
\let\normalsize=\captsize
\caption[Dionisi]{Cross sections for $\tilde{\mu}$-pair signals and
standard model backgrounds at $\sqrt{s} = 190$~GeV, with $m_{LSP} = 20$~GeV.}
\label{smu-LEP2-table}
\centerline{
\begin{tabular}{llll}
process & $\sigma_{tot}$ (pb) & $\sigma_{acc}$ (pb)\\
\hline
$\tilde{\mu}_R$(75)-pair & 0.18 & 0.058\\
$\tilde{\mu}_R$(80)-pair & 0.11 & 0.0375\\
$\tilde{\mu}_R$(85)-pair & 0.064 & 0.020\\
$\mu\mu(\gamma)$ & 7.8 & 0.0 \\
$WW \rightarrow \mu\mu+\nu$'s & 0.26 & 0.034\\
\hline
\end{tabular}
}
\end{minipage}
\end{center}
\end{table}

The production of $\tilde{e}$-pairs has a larger cross section than
$\tilde{\mu}$-pairs due to the $t$-channel neutralino exchange diagram.
The additional background specific to the $\tilde{e}$-pairs is $e\nu_e
W$ final state, but it can be safely neglected at LEP-II energy.
Therefore, the reach is in general higher than $\tilde{\mu}$-pair
depending on the neutralino mass spectrum, and ranges between
85--90~GeV with 500~pb$^{-1}$ \cite{Dionisi,Chen}.

The study of $\tilde{\tau}$ requires a special treatment due to the decay
of $\tau$ in the final state. There are purely hadronic final state,
mixed hadronic-lepton final state, and purely leptonic final state
depending on the decay of each of the $\tau$ leptons. The purely
leptonic mode suffers the most from the $W$-pair background, because the signal
is reduced by the $\tau$ leptonic branching ratios $0.36^2 = 0.13$,
while the $W$-pair background is four times as big as the
$\tilde{\mu}$-pair case because both $e$ and $\mu$ final states
contribute (see Table~\ref{smu-LEP2-table}).
The hadronic decay of the $\tau$
leptons can be distinguished from that from $W \rightarrow q\bar{q}$ by
requiring small invariant mass of the hadronic system. However, the
mixed mode still suffers from $W$-pair where one $W$ decays into $\tau
\nu_\tau$ and the other into $e\nu_e$ or $\mu\nu_\mu$, whose level is
twice as high as in the $\tilde{\mu}$-pair case, while
the signal level is reduced by a factor of $0.36\times 0.74 = 0.23$.
The hadronic decay of both $\tau$'s is the most efficient mode to avoid
the background from $W$-pair. The background level is the same as in
$\tilde{\mu}$-pair case, while the signal is reduced only by a factor of
$0.74^2 = 0.55$. If one uses the following selection criteria \cite{Chen}:
\begin{enumerate}
\item topology of 1-1 or 1-3 charged particles (with an efficiency of
30\%),
\item mass of each clusters to be less than $m_\tau$,
\item $|\cos \theta|<0.9$ for both clusters,
\item each clusters with $E > 2$~GeV,
\item total $E_{vis} > 10$~GeV,
\item no isolated $\gamma$,
\item acoplanarity $-0.8 < \cos \theta_{acop} < 0.87$,
\end{enumerate}
a typical efficiency is about 10\%. The discovery reach at 5~$\sigma$
is estimated to be about 75~GeV with 500~pb$^{-1}$.

\subsection{$\tilde{\chi}_1^\pm$-pair production}

At the mass range explorable at LEP-II, $\tilde{\chi}_1^\pm$ decays into
three-body states $\tilde{\chi}_1^0 f \bar{f}'$ where $f$ and $f'$ belong
to the same weak isodoublet.
We assume in this subsection that $\tilde{\chi}_1^\pm$ is the LVSP, and
hence the decay $\tilde{\chi}_1^+ \rightarrow \tilde{l}^+ \nu_l$ or
$\tilde{\nu}_l l^+$ is not allowed.\footnote{Even when
$\tilde{\chi}_1^\pm$ is the LVSP, there is a possibility that
$\tilde{\nu}$ decays invisibly $\tilde{\nu} \rightarrow \nu
\tilde{\chi}_1^0$ and has been missed experimentally, and
$\tilde{\chi}_1^\pm$ decays mainly into $\tilde{\nu}_l
l^+$. Then the search for $\tilde{\chi}_1^\pm$-pairs is similar to the
search for slepton pairs, and the reach is expected to be higher than
that for sleptons because of larger cross sections and sharper threshold
behavior ($\beta$ vs. $\beta^3$ for sleptons). We do not discuss this
case below.}

\begin{figure}
\centerline{\psfig{file=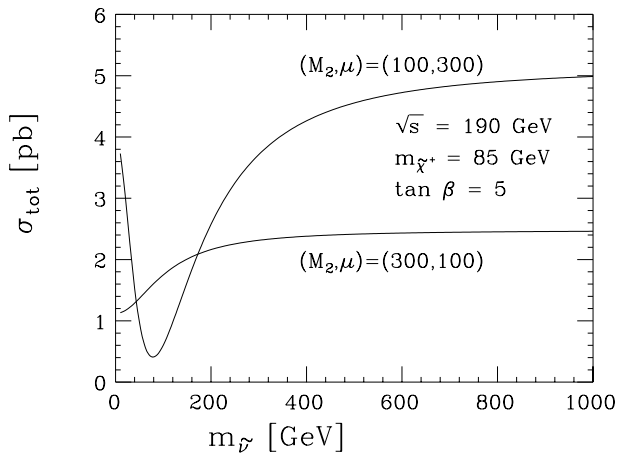,height=6cm,angle=90}}
\caption[chargino-LEP2]{Total cross sections of $\tilde{\chi}_1^\pm$ pair
production at $\sqrt{s} = 190$~GeV, as a function of
$m_{\tilde{\nu}_e}$, for two representative cases of gaugino-rich and
higgsino-rich $\tilde{\chi}_1^\pm$. }
\label{chargino-LEP2}
\end{figure}

The production cross section is sensitive to the mass of
$\tilde{\nu}_e$, which is exchanged in the $t$-channel.
Fig.~\ref{chargino-LEP2} shows
the cross sections for varying $m_{\tilde{\nu}_e}$. The destructive
interference between the $s$-channel $\gamma$, $Z$-exchange and the
$t$-channel $\tilde{\nu}_e$ exchange can suppress the cross section by
about an order of magnitude if $m_{\tilde{\nu}_e} \sim
\sqrt{s}/2$.

\begin{figure}
\centerline{\psfig{file=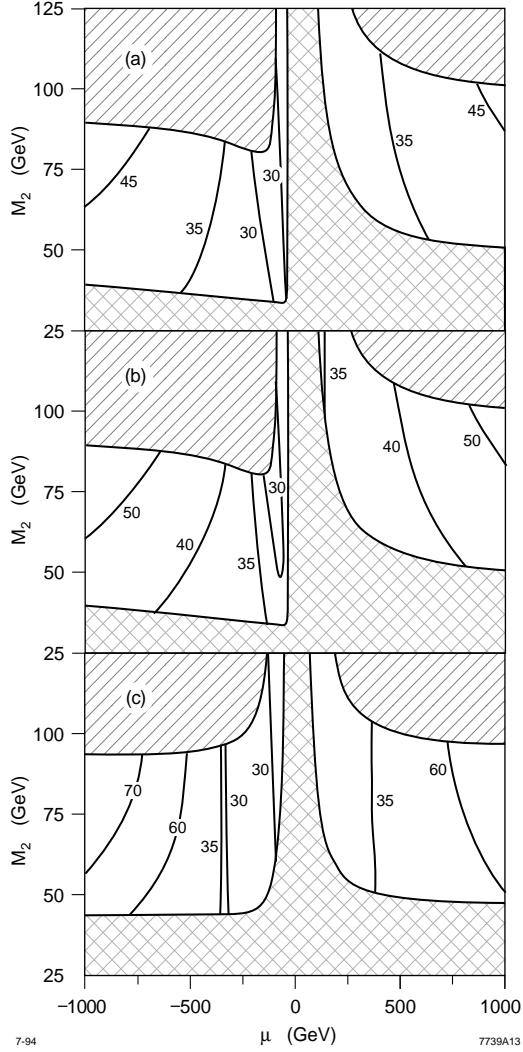,height=14cm}}
\caption[FS13]{Contours of constant value of the leptonic branching
fraction $B_l$ (in \%) in the $(\mu, M_2)$ plane for $M_1/M_2=0.5$ and
three sets of parameters $(\tan\beta,
m_{\tilde{l}}, m_{\tilde{q}})$:
(a) (2, 200, 200), (b) (2, 200, 800), and (c) (10, 200,
200). Note that we used different scales for $\mu$ and $M_2$
to make the numbers visible and to expand the gaugino region. For all
figures, the value of $B_l$ is $\frac{1}{3}$ in the Higgsino region and
grows as one approaches the gaugino region.  The growth is faster for
large $\tan\beta$ (c) than for low $\tan\beta$ (a). In (a) and (b) the $B_l$
contours differ by approximately 5\% in the far gaugino region.
Note also the ``pocket'' in
the $\mu<0$ near gaugino region, where $B_l < \frac{1}{3}$. The exchange
of the charged
Higgs is neglected in the decay process. Taken from Ref.~\cite{FS}.}
\label{FS13}
\end{figure}

\begin{figure}
\centerline{\psfig{file=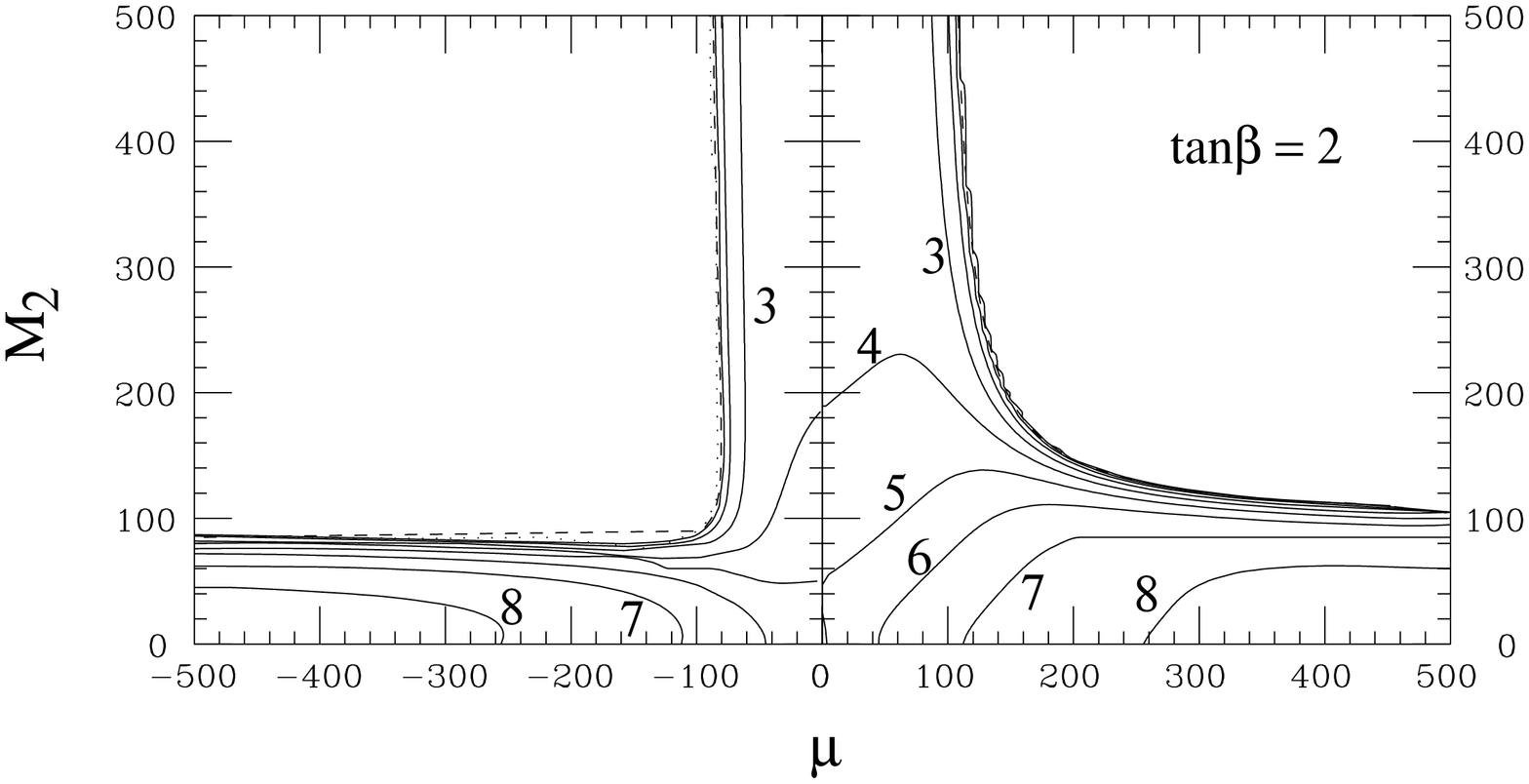,height=6cm,angle=90}}
\caption[chi1000-LEP2]{Contour of the total cross sections in pb of
$\tilde{\chi}_1^\pm$ pair production at $\sqrt{s} = 190$~GeV, for
$m_{\tilde{\nu}_e} = 1$~TeV. The kinematic limit is shown in dashed
line, while the discovery reach at 500~pb$^{-1}$ is shown in dotted
line. Units are in GeV.}
\label{chi1000-LEP2}
\end{figure}

\begin{figure}
\centerline{\psfig{file=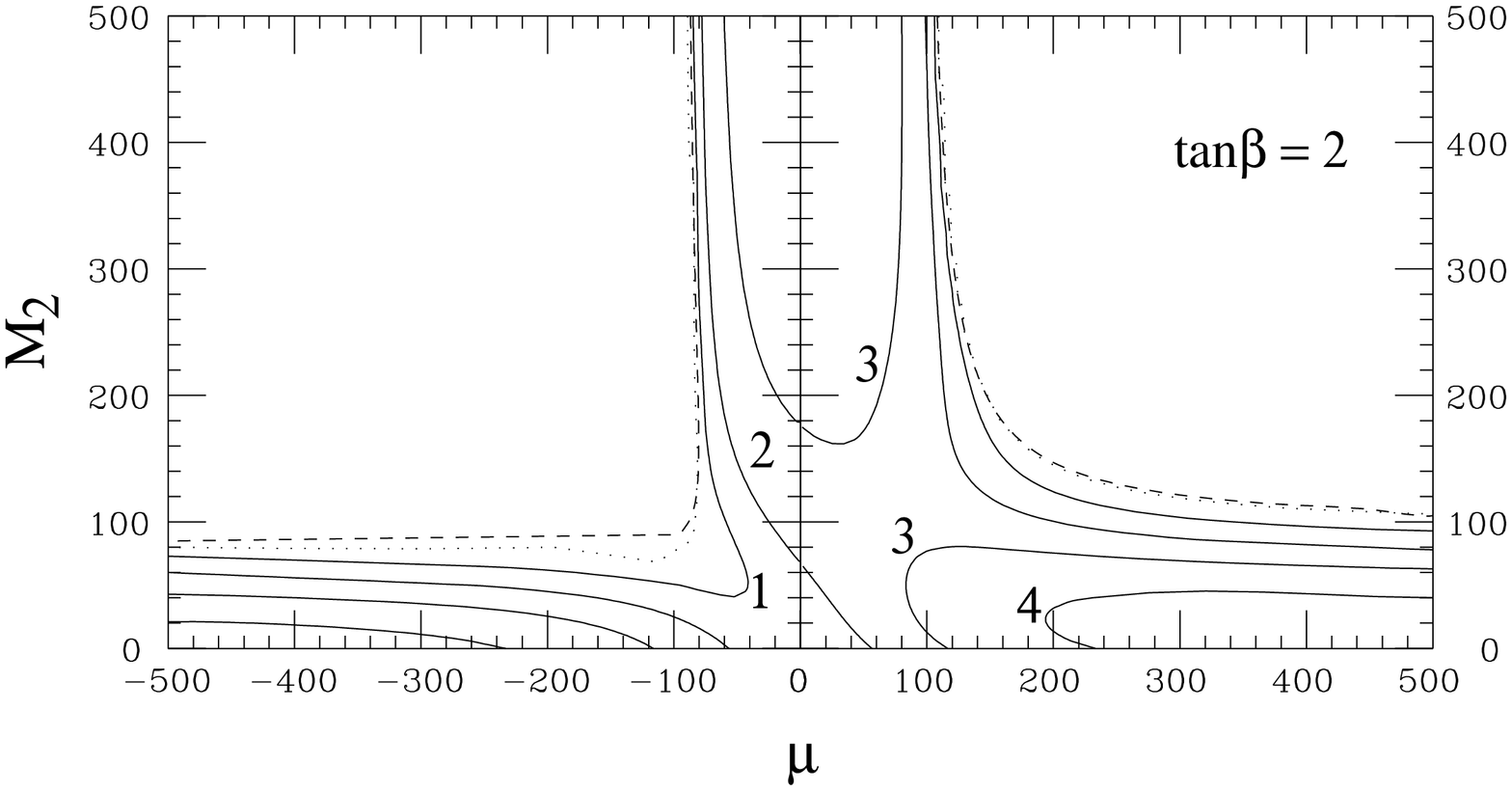,height=6cm,angle=90}}
\caption[chi100-LEP2]{Contours of the total cross sections in pb of
$\tilde{\chi}_1^\pm$ pair production at $\sqrt{s} = 190$~GeV, for
$m_{\tilde{\nu}_e} = 100$~GeV. The kinematic limit is shown in dashed
line, while the discovery reach at 500~pb$^{-1}$ is shown in dotted
line. Units are in GeV.}
\label{chi100-LEP2}
\end{figure}

The decay proceeds via virtual $W$-, $\tilde{l}$- and
$\tilde{q}$-exchange. Then the leptonic and hadronic branching ratios
vary as a function of $\tilde{l}$ mass. For a wide
range of parameter space, $W$-exchange dominates \cite{FS} and the
leptonic branching ratio is roughly 30\%.  The leptonic branching ratio
is larger when the chargino is gaugino-dominant and the sleptons are
light.  The dependence of the leptonic branching ratio on the underlying
parameters is shown in Fig.~\ref{FS13}.  Below, the dominance of
$W$-exchange is assumed with a leptonic (hadronic) branching fraction of
$\sim 30$\% (70\%). Even when one has different branching ratios, the
efficiency of mixed leptonic-hadronic mode used below obviously does not
change much (and could be even better).

The mixed leptonic-hadronic mode is the most efficient one for the
search \cite{Grivaz-Eloisatron}. Even though the pure hadronic mode has
the largest branching ratio, it suffers from the background of $W$-pair
where one $W$ decays hadronically and the other $W$ into $\tau$, with
its subsequent hadronic decay.\footnote{This is not a problem if
charginos are well within the threshold. Then the typical total
cross section is
of order $>3$~pb, with roughly half of this in the hadronic final state,
while
the $jj\tau\nu_\tau$ final state from $W$-pair has a cross section of
3.5~pb, and it is not difficult to find the excess.  The advantage of
the mixed mode is that one can go very close to the kinematic reach.
Indeed, we also use pure hadronic mode in studying the branching ratios
and cross sections later in this section.}
The pure leptonic mode suffers the most from the $W$-pair production
where both of them decay into leptonic final states.\footnote{This is
indeed much more difficult than the other modes. Signal cross section is
of order 0.06~pb, while the background from $W$-pair is 0.8~pb, assuming
the dominance of the $W$-exchange in the chargino decay.
However, the leptonic branching ratio of the chargino may be much larger
if the slepton-exchange dominates in the chargino decay (see
Fig.~\ref{FS13}); in this case, the purely leptonic mode is also useful.}

Here we summarize the analysis done in Ref.~\cite{Grivaz-Eloisatron}.
The detector simulation is based on the parameters of ALEPH detector at
LEP. One possible set of cuts is
\begin{enumerate}
\item \# charged particles $>$ 5.
\item $\not\!p_T > 10$~GeV.
\item $e$ or $\mu$ with $p_l > 5$~GeV and with a good isolation (no
energy deposited larger than 1~GeV within a cone of 30$^\circ$
half-angle).
\item squared missing mass $> 4000$~GeV$^2$.
\item mass of the hadronic system $< 45$~GeV.
\item $m_{l\nu} < 70$~GeV with $W$-pair hypothesis.
\end{enumerate}
The efficiency of the signal is $\agt 12$\% including the branching
ratio, while the $W$-pair background after the cuts is 7~pb; other
standard model backgrounds are 2~pb. Given 100~ (500)~pb$^{-1}$, a
5~$\sigma$ signal can be found if the $\tilde{\chi}_1^\pm$-pair cross
section exceeds 0.40~(0.17)~pb assuming 12\% efficiency.

The coverage on the parameter space $(\mu, M_2)$ is shown in
Figs.~\ref{chi1000-LEP2},\ref{chi100-LEP2}.
For a typical choice of MSSM parameters, $\mu=-100$~GeV and $\tan \beta =
2$, charginos up to 94~GeV can be discovered at 5~$\sigma$ level even
with 100~pb$^{-1}$ if $m_{\tilde{\nu}_e}$ is large enough. If
$m_{\tilde{\nu}_e} \sim 100$~GeV, the discovery reach reduces to 81~GeV,
but can be pushed up to 90~GeV if 500~pb$^{-1}$ is accumulated.

\begin{figure}
\centerline{\psfig{file=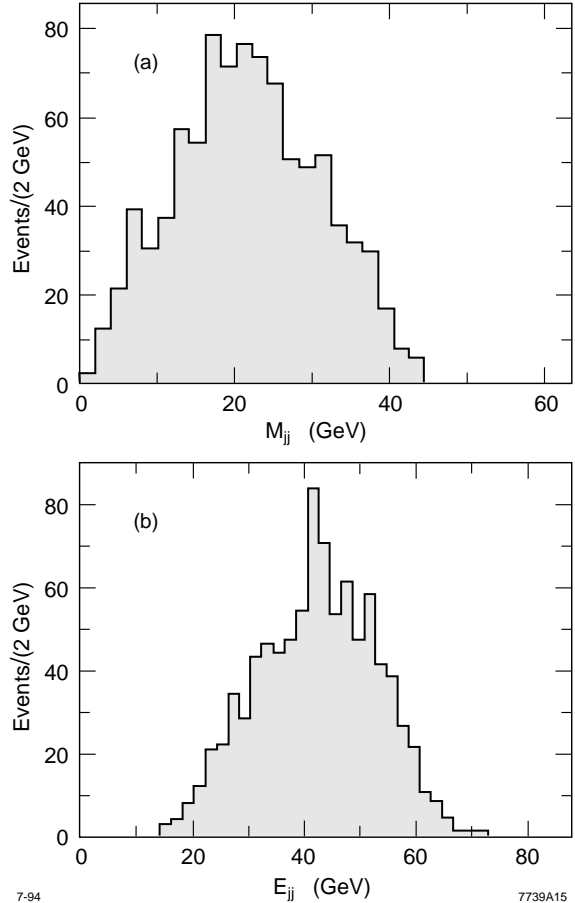,height=12cm}}
\caption[FS15]{
The dijet (a) mass spectrum and (b) energy spectrum, after cuts, for
the gaugino case $(\mu, M_2, \tan\beta, M_1/M_2, m_{\tilde{l}},
m_{\tilde{q}})$ =
$(-400,75,4,0.5,200,300)$ with integrated luminosity 1~fb$^{-1}$. In these
distributions, hadrons from $\tau$ lepton decays have not been
included. The finite detector resolution effects cause the
spectra to have tails that exceed the theoretical limits, but despite
this, the endpoints are fairly sharp. We estimate that the 1$\sigma$
uncertainty of $m_{jj}$ is 2 GeV, and that for $E_{jj}$ is 3 GeV.
Note that very few events have dijets with low invariant mass, and it
is therefore possible to distinguish hadrons that result from $\tau$
decays and those that result from hadronic chargino decays.}
\label{FS15}
\end{figure}

Once the chargino-pair is found, one can determine the masses of
$\tilde{\chi}_1^\pm$ and $\tilde{\chi}_1^0$, and decay branching ratios.
Here we use both mixed and purely hadronic modes to determine the
branching ratios.  This is possible if the chargino does not lie too
close to the kinematic limit.
For a sample parameter set\footnote{Here, we take the value $M_1/M_2$ as
predicted from GUT, but try to reproduce the number by taking it as a free
parameter in the analysis.}
\begin{equation}
(\mu, M_2, \tan \beta, M_1/M_2, m_{\tilde{l}}, m_{\tilde{q}})
	= (-400, 75, 4, 0.5, 200, 300)
\end{equation}
in units of GeV for dimensionful parameters, a parton-level analysis
was performed with the same selection criteria as above \cite{FS}. The quark
parton and lepton energies are smeared with the resolutions of ALEPH
detector. It is assumed that the jet energy measurement can be
improved by 25\% by matching the tracks with the hits in HCAL and use
momentum measurements by the tracking detector for charged particles
\cite{Miyamoto}.
The di-jet invariant mass and energy distributions are shown in
Fig.~\ref{FS15}. For an integrated luminosity of 1~fb$^{-1}$,
the resolutions for the mass determination from the end
points of the di-jet energy distribution are estimated to be,
\begin{equation}
\Delta m_{\tilde{\chi}_1^\pm} = 2.5~\mbox{GeV} \hspace{1cm}
\mbox{and} \hspace{1cm} \Delta m_{\tilde{\chi}_1^0} = 2.2~\mbox{GeV}.
\end{equation}
 Similarly, the total
production cross section and the leptonic branching ratio can be
determined as
\begin{equation}
\Delta \sigma_{tot}/\sigma_{tot} = 5.0\% \hspace{1cm}
\mbox{and} \hspace{1cm} \Delta B_l/B_l = 4.8\%.
\end{equation}

\begin{figure}
\centerline{\psfig{file=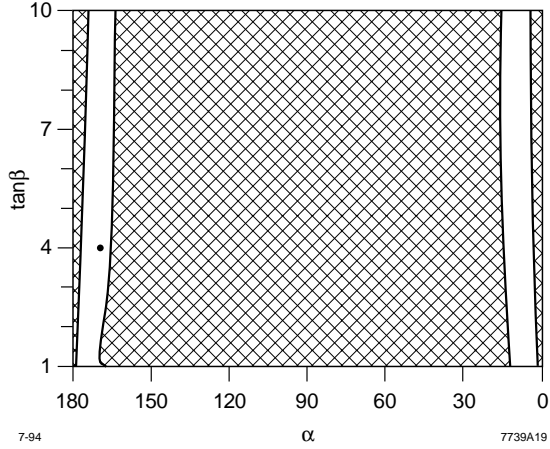,height=6cm}}
\caption[FS19]{
The allowed region in the $(\alpha, \tan\beta )$ plane for the gaugino case
study with both signs of $M_1$ allowed, where $\alpha = \arctan
\mu/M_2$. The hatched regions are
excluded by the measurements of $m_{\tilde{\chi}_1^\pm}$,
$m_{\tilde{\chi}_1^0}$, $\sigma_{tot}$, and
$B_l$, which confine the allowed region to narrow strips in the gaugino
region. The dot indicates the underlying value of $(\alpha, \tan\beta )$
for the gaugino case study.}
\label{FS19}
\end{figure}

The important point is that we can often determine some of the original SUSY
parameters even though we have six parameters (the rough degeneracy of
all sleptons and all squarks is assumed) while there are only four
observables. This is because one cannot reproduce observed masses, cross
section and branching ratios by choosing the parameters arbitrarily. For
instance, see the strong dependence of the leptonic branching ratio on
the underlying slepton mass and chargino parameters in Fig.~\ref{FS13}.
Also the parameter space is cut off by boundary
conditions ({\it e.g.}\/, lower bound on sparticle masses). For the
above sample parameter set, the ratio $|\mu|/M_2$ can be determined rather
well. The result is shown in Fig.~\ref{FS19}, on the two-dimensional
parameter space $(\alpha, \tan \beta)$ where $\alpha \equiv \arctan
\mu/M_2$. Other parameters are determined as
\begin{equation}
\begin{array}{rcccl}
0.97    &<& \rho_{\tilde{\chi}^{\pm}_1}        &<& 1.00 \\
0.97    &<& \rho_{\tilde{\chi}^0_1}        &<& 1.00 \\
180~\mbox{GeV} &<& m_{\tilde{l}}          &<& 225~\mbox{GeV} \\
0.43    &<& \frac{M_1}{M_2} &<& 0.58 \\
-1~\mbox{TeV}   <  \mu              < -290~\mbox{GeV}
&\quad& {\rm or} &\quad&
300~\mbox{GeV} <  \mu              <  1~\mbox{TeV} \\
63~\mbox{GeV}  &<& M_2             &<& 93~\mbox{GeV}\ ,
\end{array}
\end{equation}
assuming the positive relative sign for $M_1$ and $M_2$.
Here, $\rho_{\tilde{\chi}^{\pm}_1}$, $\rho_{\tilde{\chi}^0_1}$ are
``gaugino-ness'' of the chargino and LSP, respectively; they are defined
as the squared sum of the coefficients of the
gaugino fields in the linear expansion of the mass eigenstates
in terms of the interaction eigenstates.
Note that one could obtain an upper bound on the slepton mass.
It is also interesting to note that one can verify the GUT relation
$M_1/M_2 = 0.5$, illustrating the ability of precision measurements
to shed light on physics at the very high energy scale.

\subsection{Other sparticles and Higher Order Processes}

The associated $\tilde{\chi}_1^0 \tilde{\chi}_2^0$ production is
kinematically allowed for a larger range than
$\tilde{\chi}_1^\pm$-pair production in the gaugino-region $M_2 <
|\mu|$\cite{Chen}.
The total cross section for this process is typically quite
small since the gaugino components of the neutralinos don't couple to
the $Z$. However, if $m_{\tell}$ is small enough, a detectable rate is
possible. An analysis has been performed in the framework of the
minimal SUGRA model\cite{BBMT}.
They find that
the dilepton channel from
$\tz_2\to l\bar{l}\tz_1$ is plagued by background from $WW$ production,
but suitable cuts can allow a region of observability. The dijet
channel, which occurs when $\tz_2\to q\bar{q}\tz_1$, is also observable,
but only in a small region of SUGRA space, when chargino pair production
is not allowed.

If squarks are within the reach of LEP-II, but the chargino
$\tilde{\chi}_1^\pm$ is not, they decay directly into $\tilde{q}
\rightarrow q \tilde{\chi}_1^0$. This case was studied in
Ref.~\cite{Dionisi}, with the same detector parameters shown in the case
for $\tilde{\mu}$-pair study. The selection criteria are:
\begin{enumerate}
\item 85~GeV $< \not\!E < $ 160~GeV,
\item 40~GeV $< \not\!p_T < $ 100~GeV,
\item no isolated leptons, where an isolated lepton is defined as one
with $E_l > 10$~GeV and with hadronic
energy smaller than 2~GeV inside the 20$^\circ$ cone around the lepton
momentum.
\end{enumerate}
Requirements
(2) and (3) completely eliminates $Z$-pair and most of the $W$-pair and
$q\bar{q}\gamma$ backgrounds. The efficiency of the signal is found to
be 51\% for $m_{\tilde{q}} = 85$~GeV and $m_{\tilde{\chi}_1^0} =
20$~GeV. The background levels are
0.05~pb from $q\bar{q}\gamma$, 0.15~pb from $W$-pair and none from
$Z$-pair. A 5~$\sigma$ signal can be obtained up
to $m_{\tilde{q}_L} = 85$~GeV with 100~pb$^{-1}$ at $\sqrt{s} = 190$~GeV
\cite{Dionisi}, with 10 degenerate flavors assumed, which is
a common assumption made at a hadron collider.

If one does not assume the degeneracy between squark flavors, the
reaches are very different for each of the quantum numbers.  The
production cross sections are very different for $\tilde{u}_L$,
$\tilde{u}_R$, $\tilde{d}_L$, $\tilde{d}_R$, with a ratio of roughly
7.3:4.0:6.0:1.  Assuming the same efficiencies as above, the
5~$\sigma$ discovery reach for individual squarks is about 72~GeV, 59~GeV,
70~GeV, and none at 100~pb$^{-1}$, and 83~GeV, 76~GeV, 80~GeV and 31~GeV
at 500~pb$^{-1}$.  It is noteworthy that one can have better discovery
reach for $\tilde{d}_R$ below the $W$-pair threshold: 35~GeV with
100~pb$^{-1}$ and 54~GeV with 500~pb$^{-1}$ at $\sqrt{s} = 150$~GeV.

An interesting possibility is that the scalar top $\tilde{t}_1$ is
light.  When $\tilde{t}_1$ is lighter than $\tilde{\chi}_1^\pm$, it
mostly decays\cite{HK} into $c \tilde{\chi}_1^0$. The signature is the same as
$\tilde{q}$ discussed above, acoplanar $c\bar{c}$. Therefore a discovery
 reach of about 76--83~GeV is expected with 500~pb$^{-1}$
depending on the mixing angle $\tilde{t}_1 = \tilde{t}_L
\cos \theta_t + \tilde{t}_R \sin \theta_t$.

If the ``visible'' sparticles do not lie within the kinematic reach of
their pair productions, we can still look for three-body final states
such as
\begin{eqnarray}
e^+ e^- &\rightarrow& \tilde{\chi}_i^0 \tilde{\chi}_j^0 \gamma ,\\
	&\rightarrow& \tilde{\nu} \tilde{\nu}^* \gamma,\\
	&\rightarrow& e^\pm \tilde{e}^\mp \tilde{\chi}_i^0,\\
	&\rightarrow& e^\pm \tilde{\chi}_i^\mp \tilde{\nu}_e^{(*)},\\
	&\rightarrow& \mu^\pm \tilde{\mu}^\mp \tilde{\chi}_i^0,\\
	&\rightarrow& q \tilde{q} \tilde{\chi}_i^0 .
\end{eqnarray}
However, their cross sections are usually suppressed due to the small
phase space and because they are higher order in coupling constants than
the pair production.  The first four processes above were studied in
detail in Ref.~\cite{Chen}.  However, the cross sections are $\lesssim
0.1$~pb for $\tilde{\nu} \tilde{\nu}^* \gamma$, and even smaller for the
other three processes given the current LEP constraint on $\tilde{\nu}$
mass and the assumption that either $\tilde{e}$ or $\tilde{\chi}_1^\pm$
are beyond the threshold of their pair productions.  The standard model
background from the $\nu \bar{\nu} \gamma$ final state was studied in
\cite{DDR} with a similar cut and found to be $\simeq 0.36$~pb.
Therefore, the discovery potential of these signals via the
observation of an excess of single
photon events over SM expectations is marginal.
Other signals have not yet been fully studied.

\section{Search for SUSY at the Tevatron and its upgrades}

\begin{figure}[t]
\centerline{\psfig{file=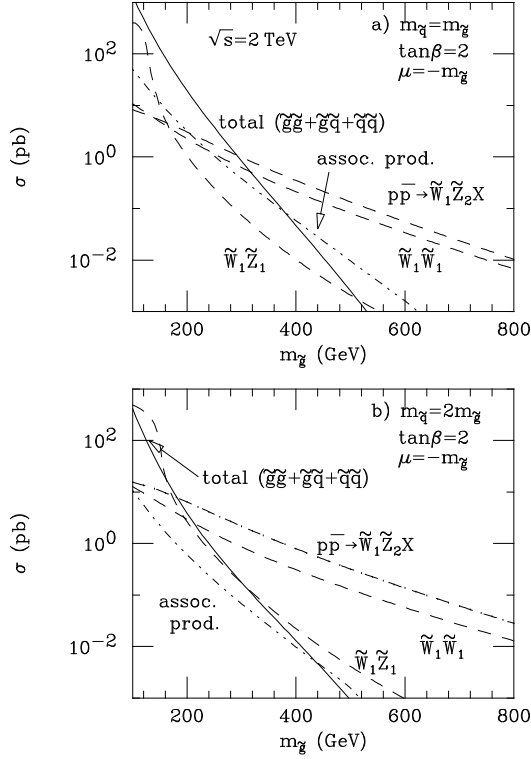,height=11cm,angle=90}}
\caption[]{Total cross sections for various sparticle pair reactions for
$p\bar p$ collisions at $\sqrt{s}=2$ TeV.}
\label{fig61}
\end{figure}

Various sparticle pair production cross sections are shown in Fig.~\ref{fig61}
for $\sqrt{s}=2$ TeV $p\bar p$ collisions at the Tevatron collider.
In this plot, we take ({\it a}) $m_{\tq}=m_{\tg}$ and
({\it b}) $m_{\tq}=2m_{\tg}$, with $\tan\beta=2$ and $\mu
=-m_{\tg}$.\footnote{This is motivated by supergravity models.}
We convolute with CTEQ2L parton distribution functions. From
({\it a}), we see that strong production of $\tg\tg$, $\tg\tq$ and
$\tq\tq$ is the dominant production cross section for $m_{\tg}\alt 325$ GeV.
As one goes to higher gluino masses, the gluino and squark cross sections
become kinematically suppressed.
However, the charginos and neutralinos are still
relatively light ($m_{\tw_1}\sim m_{\tz_2}\sim {1\over 4}-{1\over 3} m_{\tg}$);
then, when $\mu$ is large compared to $M_1$ and $M_2$, $\tw_1\tz_2$ and
$\tw_1\tilde{\chi}_1^\mp$ become the dominant cross sections.
It is instructive to see that if $m_{\tq}=2m_{\tg}$, $\tw_1\tz_2$ production
begins to be the dominant source of SUSY events for gluinos heavier than
just 200~GeV. The sum of all
associated production mechanisms remains below these other cross sections,
and never dominates.
This gives a good idea of what to search for, depending
on $m_{\tg}$: as long as one is probing $m_{\tg}\alt 200$~GeV, one should
focus on $\tg$ and $\tq$ production, with their subsequent cascade decays;
for gluinos with masses between 200-325~GeV, this is still the best channel
to search for SUSY if $m_{\tq}=m_{\tg}$, but for heavier squarks,
$\tw_1\tz_2$ production dominates even if gluinos are as light as $\sim
200$~GeV. Note, however, that given sufficient integrated luminosity,
the highest
reach in $m_{\tg}$ will ultimately be reached by exploring the
$p\bar p\to\tw_1\tz_2$ and $\tw_1\tilde\chi_1^\mp$ reactions.

\subsection{Sparticle cascade decay signatures at the Tevatron}

In general, sparticle pair production at the Tevatron collider is followed
by sparticle decays through a cascade until the LSP ($\tz_1$) state is
reached\cite{CAS}. Hence, sparticle production is signalled by events with
$n$-jets plus $m$-isolated leptons plus $\etmiss$ ($n,m$ non-negative
integers).
These
events can be broken down into various distinct classes:
\begin{itemize}
\item $multi-jet+\etmiss$ events (veto isolated leptons);
\item $1\ell +jets+\etmiss$ events (these have huge backgrounds from direct
$W$ production);
\item $2\ell +jets+\etmiss$ events (these can further be broken
down into opposite-sign (OS)
isolated dileptons, which have substantial backgrounds from $t\bar t$, $WW$
and $\tau^+\tau^-$ production, and same-sign (SS) isolated dileptons, for
which the SM backgrounds are expected to be much smaller);
\item $3\ell +jets+\etmiss$ events (these further
sub-divide into those containing
jets, which usually come from gluino and squark cascade decays, or events
with three isolated leptons, plus little or no jet activity, which usually
come from $\tw_1\tz_2$ production, followed by their leptonic decays. Assuming
that leptonic decays of the $Z$ can be identified with high efficiency,
SM backgrounds to these events are expected to be small);
\item $4\ell +jets+\etmiss$ events (these events usually come from the presence
of two $\tz_2$'s in an event; they have low cross section, but tiny
backgrounds as well).
\item $\geq 5\ell +jets+\etmiss$ events (these can only be produced by
multi-step cascades of very heavy sparticles, or, in $R$-parity violating
models where the LSP decays into leptons via lepton number violating
interactions).
\end{itemize}

In Ref. \cite{RPV}, ISAJET 7.13 was used to generate {\it all}
lowest order $2\to 2$ subprocesses, with a complete event simulation.
Experimental conditions were simulated using a toy
calorimeter with segmentation $\Delta\eta \times \Delta\phi = 0.1 \times
0.09$ and extending to $|\eta| = 4$. An energy resolution of
$\frac{0.7}{\sqrt{E}}$ ($\frac{0.15}{\sqrt{E}}$) for the hadronic
(electromagnetic) calorimeter was assumed. Jets were
defined to be hadron clusters
with $E_T > 15$~GeV in a cone with
$\Delta R=\sqrt{\Delta\eta^2+\Delta\phi^2}=0.7$. Leptons with $p_T > 8$~GeV
and within $|\eta_{\ell}| < 3$ were considered to be isolated if the hadronic
scalar $E_T$ in a cone with $\Delta R = 0.4$ about the lepton was smaller
than $\frac{E_T(\ell)}{4}$.
Finally, $\eslt > 20$~GeV was required in all events.
For each of the event topologies introduced above, we impose
the following additional requirements:
\begin{enumerate}
\item $\eslt$ events, are required to have
$n_{jet} \geq 4$ with at least one of the jets
in the central region, $|\eta| < 1$, and following the recent analysis by
the D0 collaboration\cite{D0SQ}, $\eslt \geq 75$~GeV.
We veto events
with either isolated leptons with $E_T \geq 15$~GeV (to reduce $W$
backgrounds), or a jet within $30^o$ of $\vec{\rlap/p_T}$.
\item Single lepton events were defined to have exactly one isolated
lepton with $E_T \geq 15$~GeV.
We reject events with 60~GeV $\leq m_T(\ell,\eslt) \leq 100$~GeV
which have large backgrounds from $W$ production.
\item The OS dilepton sample was defined to have two opposite
sign isolated leptons with $p_T \geq 15$~GeV and $30^o \leq
\Delta\phi_{\ell^+\ell'^-} \leq 150^o$ and no other isolated leptons.
To eliminate backgrounds from $Z$
production, events with 80~GeV $\leq m(\ell^+\ell^-) \leq$ 100~GeV were
rejected.
\item The SS dilepton sample was required to have exactly two
isolated leptons, each with $p_T \geq 15$~GeV, and no other isolated leptons.
At least two jets are also required.
\item The $n_{\ell} \geq 3$ event sample was defined to have exactly $n_{\ell}$
isolated leptons, with $p_T(\ell_1) \geq 15$~GeV and $p_T(\ell_2) \geq
10$~GeV.
\end{enumerate}

\begin{figure}[tbh]
\centerline{\psfig{file=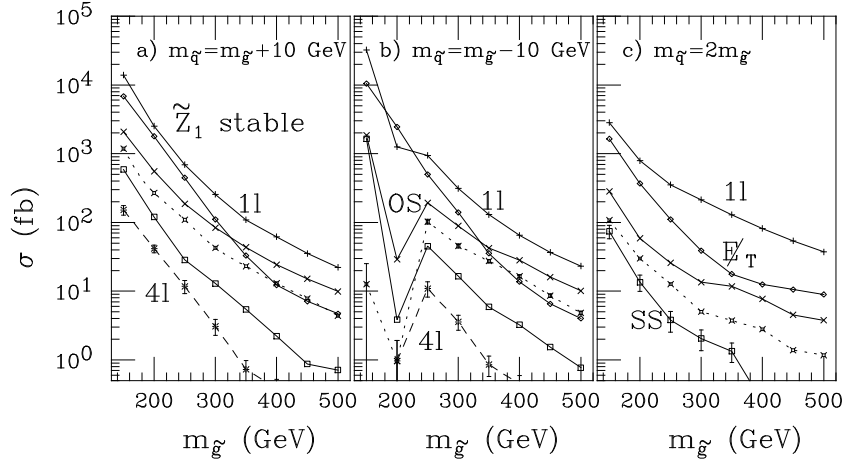,height=7cm,angle=90}}
\caption[]{Cross sections at the Tevatron ($\sqrt{s}=1.8$ TeV)
in {\it fb} for various event topologies after
cuts given in the text for the MSSM,
for three choices of
squark mass. We take $\mu =-m_{\tg}$, $\tan\beta =2$, $A_t=A_b=-m_{\tq}$
and $m_A =500$ GeV. The $\eslt$ events are labelled with diamonds, the
1-$\ell$ events with crosses, the $\ell^+\ell^-$ events with x's and
the $SS$ with squares. The dotted curves are for 3-$\ell$ signals, while
dashes label the 4-$\ell$ signals. For clarity, error bars are
shown only on the lowest lying curve; on the other curves the error bars are
considerably smaller. We note that the $m_{\tg}= 150$~GeV case in {\it b}
is already excluded by LEP constraints on the $Z$ width, since this implies
$m_{\tnu}=26$~GeV.}
\label{fig62}
\end{figure}
The cross sections for the various SUSY signals calculated within the
SUGRA-inspired MSSM framework are shown in Fig.~\ref{fig62},
for ({\it a}) $m_{\tq}=m_{\tg}+10$~GeV, ({\it
b}) $m_{\tq}=m_{\tg}-10$~GeV, and ({\it c}) $m_{\tq}=2m_{\tg}$. Here,
$\tan\beta=2$, $\mu = -m_{\tg}$,
$m_t = 170$~GeV, and the pseudoscalar Higgs boson mass is taken to be
500~GeV. The slepton masses are determined in terms of $m_{\tg}$
and $m_{\tq}$ using renormalization group equations to evolve from a common
sfermion mass at the GUT scale.

The {\it physics} backgrounds to these event topologies within the SM
framework are shown in Table 2 for a top quark mass of 150 GeV and 175~GeV.
Detector-dependent backgrounds to multilepton signals from
misidentification of jets as isolated leptons\cite{KAMON} or to the $\eslt >
75$ GeV
signal from mismeasurement of QCD jets should be small.

\begin{table}
\caption[]{Standard Model background cross sections in $fb$ for
various event topologies after cuts
described in the text, for $p\bar p$ collisions at $\sqrt{s}=1.8$~TeV.
The $W+jet$ and $Z+jet$ results include decays to $\tau$ leptons.}

\bigskip
\centerline{
\begin{tabular}{ccccccc}
case & $E\llap/_T$ & $1\ \ell$ & $OS$ & $SS$ & $3\ \ell$ & $\ge 4\ \ell$ \\
\hline
$t\bar t(150)$ & 270 & 1200 & 190 & 0.8 & 0.7 & -- \\
$t\bar t(175)$ & 145 & 590 & 90 & 0.3 & 0.3 & -- \\
$W+jet$ & 710 & $1.2\times 10^6$ & -- & -- & -- & -- \\
$Z+jet$ & 320 & 2200 & 69 & -- & -- & -- \\
$WW$ & 0.4 & 110 & 130 & -- & -- & -- \\
$WZ$ & 0.04 & 4.3 & 1.2 & 2.1 & 0.4 & -- \\
$total\ BG(150)$ & 1300 & $1.2\times 10^6$ & 390 & 2.9 & 1.1 & -- \\
$total\ BG(175)$ & 1175 & $1.2\times 10^6$ & 290 & 2.4 & 0.7 & -- \\
\end{tabular}
}
\end{table}

We see that while SUSY signals and SM backgrounds
are of comparable magnitude in the $\eslt$ and OS dilepton channels,
the signal cross sections substantially exceed backgrounds in the SS and
$n_{\ell} =3$, and in some cases, $n_{\ell} \geq 4$ isolated lepton channels.
The reach of the Tevatron is estimated by requiring that the SUSY
signal (in any channel) exceed the background by 5$\sigma$; {\it i.e.}
$N_{sig} > 5\sqrt{N_{back}}$, where $N_{sig}$ ($N_{back}$) are
the expected number of signal (background) events in a collider run,
and where the $m_t =150$ GeV background numbers have been used.
In addition, to incorporate systematic uncertainties inherent to these
calculations, it is further required\cite{RPV} (somewhat arbitrarily) that
$N_{sig}>0.25N_{back}$. For further discussion of this point, see Sec. 9.
The reach of the Tevatron is illustrated in Table 3,
both for an integrated luminosity
of 0.1 $fb^{-1}$ that is expected to be accumulated by the end of
Tevatron run IB, and, in parenthesis, for an integrated
luminosity of 1 $fb^{-1}$ that
should be accumulated after one year of Main Injector (MI) operation.
The multi-lepton signals have only been evaluated for negative values of
$\mu$. For $\mu > 0$ and somewhat heavy squarks, the leptonic decay of
$\tz_2$ is strongly suppressed by complicated interference effects. This
could lead to a substantial reduction of the 3$\ell$ signal.
In Table 3, a minimum of five signal events (ten for the MI reach)
are required in each channel.
For the SS and $3\ell$ samples where the expected
background is very small (so that the $5\sigma$ criterion is not meaningful),
we have checked that the Poisson probability for the background to fluctuate
to this minimum event level is $\leq 2\times 10^{-4}$ and $<10^{-5}$,
respectively.

\begin{table}
\caption[]{Reach in $m_{\tg}$ via various event topologies
for the SUGRA-inspired MSSM, assuming an integrated
luminosity of 0.1 fb$^{-1}$ (1 fb$^{-1}$), at the Tevatron collider.
The reach has been conservatively calculated using $m_t =150$ GeV, and will be
slightly larger since the top quark mass has recently been measured to be
$m_t\simeq 175$ GeV.}

\bigskip

\begin{tabular}{ccccccc}
case & $E\llap/_T$ & $1\ \ell$ & $OS$ & $SS$ & $3\ \ell$ & $\ge 4\ \ell$ \\
\hline
$m_{\tq}=m_{\tg}+10$ GeV & 240 (260) & --- (---) & 225 (290) &
230 (320) & 290 (425) & 190 (260) \\
$m_{\tq}=m_{\tg}-10$ GeV & 245 (265) & --- (---) & 160 (235) &
180 (325) & 240 (440) & --- (---) \\
$m_{\tq}=2m_{\tg}$ & 185 (200) & --- (---) & --- (180) &
160 (210) & 180 (260) & --- (---) \\
\end{tabular}
\end{table}

Several conclusions can be drawn from Table 3.
We have checked that in run IA of the Tevatron collider
(integrated luminosity $\sim 0.02\ fb^{-1}$),
the greatest reach in $m_{\tg}$ is achieved in the $\etmiss +jets$
channel, with essentially no reach via multi-lepton signals.
However, for run IB
(integrated luminosity $\sim 0.1\ fb^{-1}$), there is now a comparable
reach also in each of the $OS$ and $SS$ dilepton channels, and for $\mu < 0$,
especially in the
$3\ell$ channel. In the main injector era
(integrated luminosity $\sim 1\ fb^{-1}$), the reach in the $\etmiss +jets$
channel will be
background limited. However, in the $SS$ dilepton and $3\ell$ channels,
a much larger range of masses can be explored. In the $3\ell$ channel, for
$m_{\tq}\sim 2m_{\tg}$, $m_{\tg}\sim 260$ GeV can be explored, while for
$m_{\tq}\sim m_{\tg}$, squarks and gluinos as heavy as
425-440~GeV might be detectable!

The reach in $m_{\tg}$ has also been evaluated in Ref. \cite{KAMON} for
the TeV$^*$ and the DiTevatron.
At the DiTevatron, the strong production of SUSY particles
is greatly enhanced relative to the $\etmiss$ background,
which is mainly electroweak. In the $\etmiss +jets$ channel, it is estimated
the DiTevatron has a reach of $m_{\tg}\sim 500-600$ GeV.

\subsection{$\tw_1\tz_2\to 3\ell$ signal}

The usefulness of the reaction $p\bar p\to W\to\tw_1\tz_2\to 3\ell +\etmiss$
was suggested long ago in Ref. \cite{dnt}, while complete calculations
for {\it on-shell} $W$'s were carried out in Ref. \cite{bht}.
Arnowitt and Nath pointed out that
with an integrated luminosity of 100 $pb^{-1}$, signals from
the $p\bar p\to \tw_1\tz_2$ reaction
remains substantial even when the intermediate $W$ is {\it off-shell}\cite{an}.
Even so, subsequent full calculations of the $3\ell$ signature reached rather
pessimistic conclusions
for Tevatron energies\cite{barb}, although these assumed $\tz_2$ and
$\tw_1$ decays only via virtual $W$ and $Z$ bosons. Ultimately, it was
pointed out that, in SUGRA models, where $m_{\tell}$
is frequently much lighter than $m_{\tq}$, the $\tz_2$ and sometimes also
$\tw_1$
leptonic branching ratios enjoy a large enhancement (for $\mu < 0$), if all
decay diagrams
are included\cite{BT}. This allows Tevatron collider experiments to probe
much deeper into the SUGRA parameter space than previously expected.
Further calculations, some using the SUGRA framework with particular choices of
string-motivated soft-breaking boundary conditions\cite{lopez,KAMON}, others
performed within the minimal SUGRA framework\cite{kane}, confirmed these
expectations.

\begin{figure}
\centerline{\psfig{file=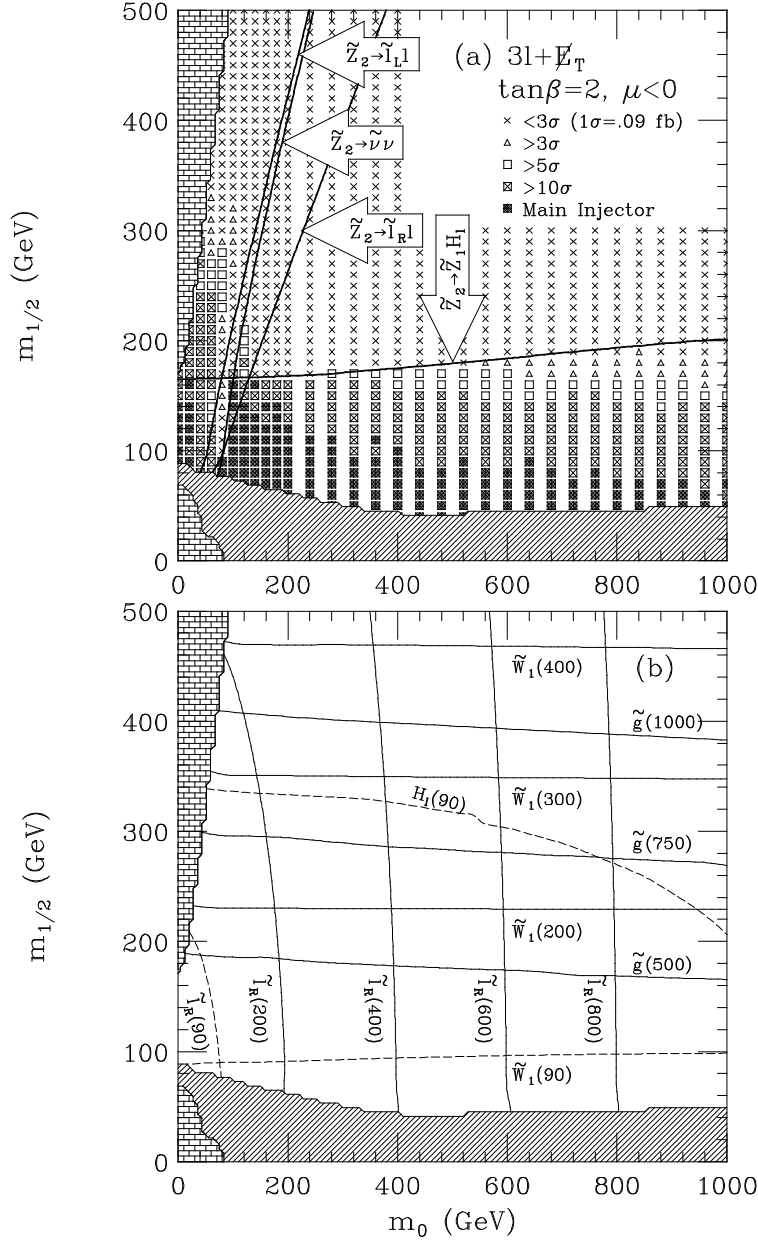,height=18cm,angle=90}}
\caption[]{
Regions of $m_0\ vs.\ m_{1/2}$ plane where clean trilepton signal is
observable over background, for Tevatron MI project (1 $fb^{-1}$) and
TeV$^*$ (25 $fb^{-1}$). We take $A_0=0$ and $\mu <0$.
The decay
$\tz_2\to\tz_1 h$ is allowed above the indicated contour, while the
decays $\tz_2\to\tnu\nu$ and $\tz_2\to\tell_R\ell$ are allowed to the left
of the labelled contours.
}
\label{fig63}
\end{figure}

\begin{figure}
\centerline{\psfig{file=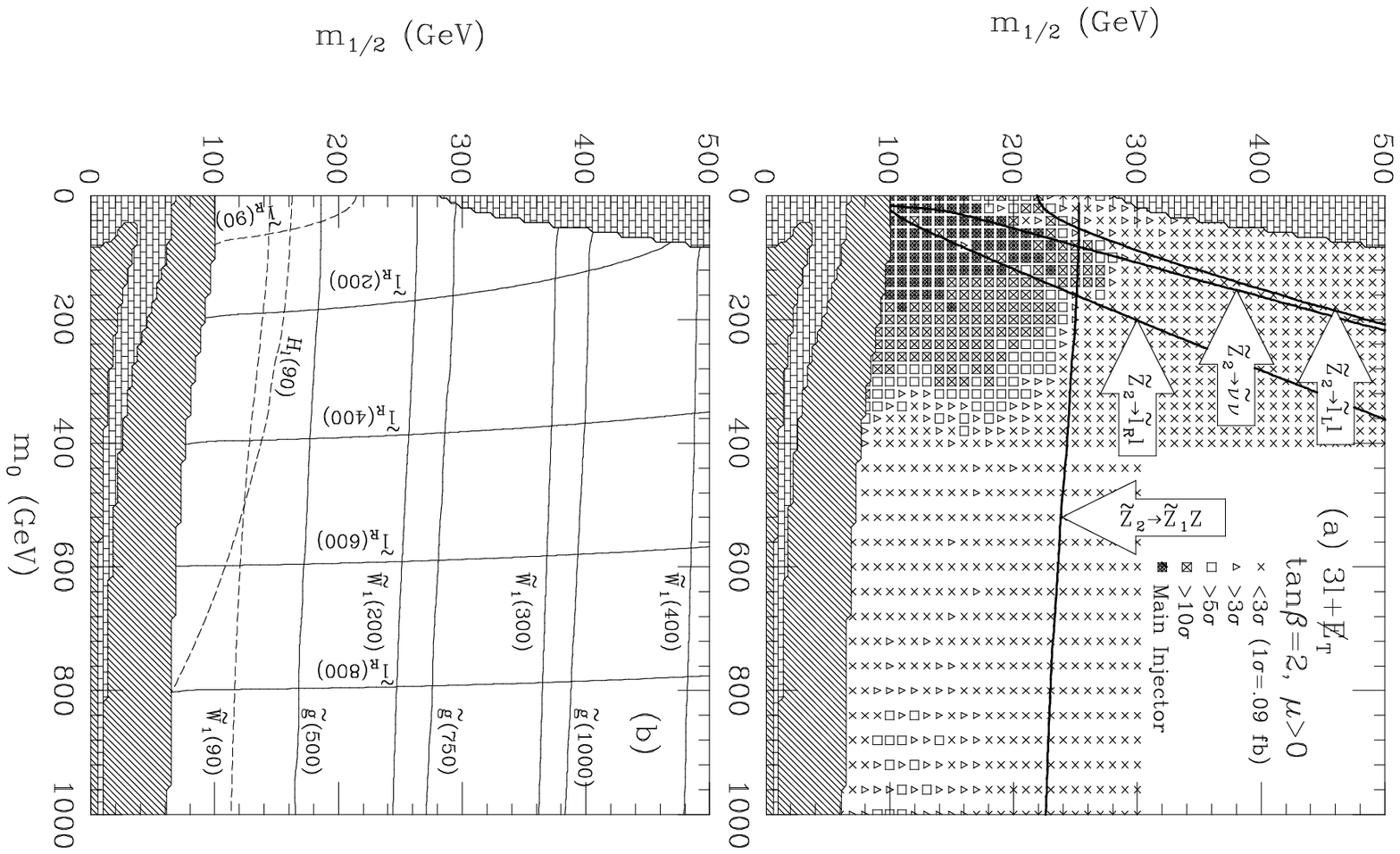,height=18cm,angle=90}}
\caption[]{
Regions of $m_0\ vs.\ m_{1/2}$ plane where clean trilepton signal is
observable over background, for Tevatron MI project (1 $fb^{-1}$) and
TeV$^*$ (25 $fb^{-1}$). We take $A_0=0$, and $\mu >0$.
The decay
$\tz_2\to\tz_1 h$ is allowed above the indicated contour, while the
decays $\tz_2\to\tnu\nu$ and $\tz_2\to\tell_R\ell$ are allowed to the left
of the labelled contours.
}
\label{fig63p}
\end{figure}

In Ref. \cite{BCKT}, simulations of clean trilepton signal
and background were performed using the minimal SUGRA framework.
We show in Fig.~\ref{fig63}{\it a} and Fig.~\ref{fig63p}{\it a}
(taken from Ref. \cite{BCKT})
the regions in the $m_0\ vs.\ m_{1/2}$ plane where the clean trilepton
signal ought to be observable (after cuts)
at the Tevatron Main Injector with an integrated luminosity
of 1~$fb^{-1}$ (black squares), and
at a luminosity upgraded (25 $fb^{-1}$) Tevatron, TeV$^*$, at the $10\sigma$
(squares with x's) and $5\sigma$ level (open squares). In
Fig.~\ref{fig63}{\it b} and Fig.~\ref{fig63p}{\it b}, we show corresponding
mass contours
for $\tg$, $\tw_1$ and $\tell_R$, for comparison.

We see that the reach of the MI and TeV$^*$ are variable in parameter
space, but that the largest reach is attained when $m_0$ is small, so
that $\tz_2\to\tell\ell$ two body decay dominates the branching fractions.
In this case, for both signs of $\mu$, TeV$^*$ can probe to
$m_{\tg}\sim 600-700$ GeV. As $m_0$ increases, $\tz_2\to\tell_L\ell$ closes,
and
decays are dominated
by $\tz_2\to\tnu\nu$, and there is a gap in trilepton
reach, even though $\tz_2\to\tell_R\ell$ is accessible.
For even higher values of $m_0$, $\tz_2$ decays via 3-body decays,
unless the $\tz_2\to\tz_1 h$ or $\tz_2\to\tz_1 Z$ decay modes
are kinematically allowed. For $\mu <0$ and large $m_0$,
the MI (TeV$^*$) can see to $m_{\tg}\sim 300$ GeV (500 GeV). However, for
$\mu >0$, we see that for large $m_0$ there is {\it no reach}\footnote{This
is due to a strong suppression of the leptonic
branching fraction of $\tz_2$ which mainly occurs due to
negative interference between $Z$ and
$\tell_L$ exchange diagrams in the $\tz_2$ leptonic width.
The result presented here is at variance with
results presented in Ref. \cite{lopez,KAMON,kane},
where the authors suggest that
trilepton searches will significantly extend the reach in $\tw_1$
independently of the sign of $\mu$. Of course, for very large values
of $m_0$, the sfermion mediated amplitudes become suppressed, and the
neutralino branching ratios are those of the Z boson.}
via trileptons
for either MI or TeV$^*$.
For large values of $m_{1/2}$, the $\tz_2\to\tz_1 h$ or $\tz_2\to\tz_1 Z$
decay modes turn on, dominating the branching fractions,
and spoiling the signal. Thus, the onset of these ``spoiler modes''
provides a natural limit beyond which the $3\ell$ signal is no longer viable,
(for large $m_0$).

In Ref. \cite{KAMON}, the reach in the $3\ell$ channel for the
proposed DiTevatron upgrade is calculated, with $\sqrt{s}=4$ TeV.
Their results show that the reach via trileptons of the DiTevatron
is not enhanced much
beyond what a $2$ TeV Tevatron can do. This is easy to understand.
$\tw_1\tz_2$ production takes place mainly via valence quark annihilation,
and hence doubling the Tevatron energy only increases the cross section in this
channel by a factor of $\sim 3$. Meanwhile, a very significant SM background
comes from $t\bar t$ production, which takes place via $gg$ as
well as $q\bar q$ fusion. The $t\bar t$ background increases by a factor
of $\sim 20$ when going from $2$ to $4$ TeV collisions\footnote{It may,
however, be
possible to suppress the $t\bar{t}$ background with about a 50\% loss of
signal using a cut described in Sec. 7.2 where we discuss the extraction of
this signal at the LHC.}.

It has also been pointed out that observation of a sufficient number of
$3\ell$ events can allow for a relatively precise $m_{\tz_2}-m_{\tz_1}$
mass measurement, by measuring the $m(\ell\bar{\ell})$ distribution in
{\it e.g.} $e\bar{e}\mu$ events\cite{BT}.
Such a measurement requires a significant
trilepton sample that is devoid of contamination from SM backgrounds or from
other SUSY sources. While this measurement may turn out to be
possible at the Tevatron (for some ranges of parameters),
it will require the highest attainable luminosity,
and will be contingent upon how well the detectors function in the high
luminosity environment.
In the next section, we will also see that this measurement should be
relatively easy at the LHC, unless the leptonic decay of the neutralino is
strongly suppressed.
A measurement of $m_{\tz_2}-m_{\tz_1}$ may thus
be a starting point for disentangling the various sparticle masses.

\subsection{Top squark search}

In minimal SUGRA, the soft-breaking masses $m_{\tst_L}^2$ and $m_{\tst_R}^2$
are driven to lower values than for the other squarks, due to the large
top quark Yukawa coupling. The $\tst_L-\tst_R$ mixing induced by
Yukawa interactions reduces
the light top squark mass $m_{\tst_1}$ even further. Hence, the light
top squark
is frequently much lighter than the other squark species; in this
case, Tevatron limits on $m_{\tq}$, which are derived assuming ten degenerate
squark types, are not applicable to the top squark.
Top squarks require an independent search effort at Tevatron experiments.
For identical top and stop masses, the stop pair total cross section is
typically $\sim {1\over 5}-{1\over 10}$ of the top pair cross
section\cite{TPHEN}.

If $m_{\tst_1}> m_b+m_{\tw_1}$, then $\tst_1\to b\tw_1$ decays are expected
to dominate for the stop masses $m_{\tst_1}\sim 50-125$ GeV accessible to
Tevatron experiments. In the case where other squarks and sleptons
are all relatively heavy, the lighter chargino decays dominantly via
virtual $W$ exchange so that the branching fractions for the leptonic decay
$\tw_1 \rightarrow \ell\nu\tz_1$ is $\sim$11\% per lepton family.
Then, top squark pair signatures are
identical to top quark pair signatures, except that the decay products
are softer. In the
$p\bar p\to\tst_1\bar{\tst_1}\to b\bar{b}\ell\nu q\bar{q}'\tz_1\tz_1$
channel, it has been shown that\cite{BST} given a data set of $0.1\ fb^{-1}$,
a signal may be detectable for $m_{\tst_1}\lesssim 100$ GeV, but
only if adequate $B$ micro-vertex tagging is available.
In the
$p\bar p\to\tst_1\bar{\tst_1}\to b\bar{b}\ell\bar{\ell}'\nu
'\bar{\nu}\tz_1\tz_1$
channel, the $t$-squark signal can be separated from
the top quark background by searching for {\it soft} dilepton
events: one cuts, for instance, on the sum
$|p_T(\ell_1)|+|p_T(\ell_2)|+|\etmiss |<100$ GeV to remove harder dilepton
events from top quark pairs. This cut works well if $m_t \geq 150$~GeV, which
now appears to be the case. We also note that scalar top signals may be
significantly enhanced if the sleptons are significantly lighter than squarks
and $|\mu|$ is large, in which case the branching fractions for the leptonic
decays of chargino may be considerably larger than those for $W$.

If the decay $\tst_1\to b\tw_1$ is kinematically closed,
then the $\tst_1$ will
decay\cite{HK} dominantly via a flavor changing loop to $c\tz_1$. In this
case $\tst_1\bar{\tst_1}$ production is signalled by $\etmiss$ events,
exactly as for squark production (but without any cascade decays).
Hence, the relevant reaction is
$p\bar p\to\tst_1\bar{\tst_1}\to c\bar{c}\tz_1\tz_1$, and one looks for
two jets plus $\etmiss$. Again, suitable cuts will allow Tevatron
experiments to probe
$m_{\tst_1}\alt 100$ GeV with $0.1\ fb^{-1}$ of data, even if the LSP is
rather heavy\cite{BST}.

\subsection{Slepton search}

The search for sleptons at the Tevatron collider was addressed in
Ref.\cite{slep}. In that work, all channels of slepton production and decay
were simulated using ISAJET. The best bet for observing slepton signals
appeared to be in the OS $2\ell +\etmiss$ channel. Even here, there were
large irreducible backgrounds from $WW$ production (as well as
possible contamination from other SUSY
souces such as $\tw_1\tilde\chi_1^\mp$ production) which precludes clear
identification of the slepton signal. The conclusions were that,
even using the Tevatron main injector, a signal would be very difficult
to see for slepton production via off-shell $Z$ bosons-- {\it i.e.}, if
$m_{\tell}>45-50$ GeV.

\subsection{Multichannel search for minimal SUGRA}

\begin{figure}[tbh]
\centerline{\psfig{file=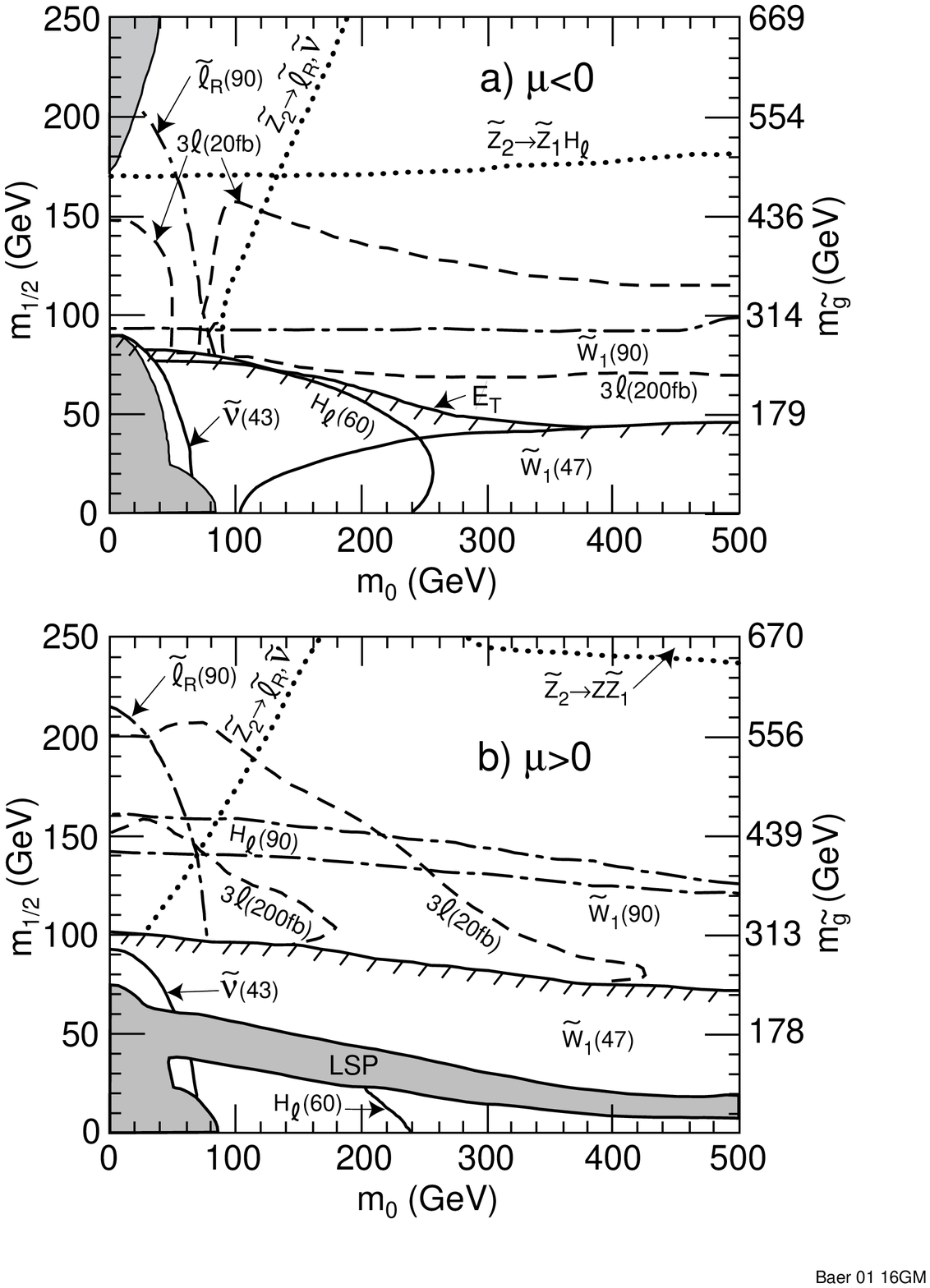,height=11cm}}
\caption[]{Regions in the $m_{0}\ vs. m_{1/2}$ plane explorable by
Tevatron and LEP II experiments.}
\label{fig64}
\end{figure}

How do the various search strategies outlined above correlate to each other?
Which SUSY searches are complementary, and which overlap? How do searches for
SUSY at the Tevatron compare to SUSY searches at LEP, or LEP-II? These
questions can most sensibly be addressed by working within the relatively
constrained framework of minimal supergravity with radiative electroweak
symmetry breaking, and universal soft-breaking terms at the unification
scale. In this framework, all sparticle masses and couplings are determined
by specifying the parameter set $m_0 ,m_{1/2}, A_0,\tan\beta ,sign(\mu )$,
introduced in Sec. 2.
The parameter $A_0$ mainly affects the 3rd generation sparticle masses,
and the
spectrum changes slowly with variation in $\tan\beta$ (of course, the
phenomenology may be significantly altered if new decays, {\it e.g.}
$\tw_1\rightarrow\tst_1 b$ become allowed).
Hence, the
$m_0\ vs.\ m_{1/2}$ plane seems to provide a
convenient panorama in which to plot results.
In Fig.~\ref{fig64}, we take $A_0=0$, $\tan\beta =2$ and $m_t=170$ GeV, and
show results
for both signs of $\mu$. We note the following features:
\begin{itemize}
\item the gray regions are excluded on theoretical grounds, since in the lower
left region, electroweak symmetry breaking does not occur, or
cannot attain the proper $Z$-boson mass. Other parts of the
gray regions are excluded because
sparticles other than the $\tz_1$ become the LSP.
\item In Fig.~\ref{fig64}{\it a}, the currently experimentally excluded region
is below
the hatch marks, and is due to four separate limits: the LEP limits that
$m_{\tw_1}>47$ GeV, $m_h \agt 60$ GeV, and $m_{\tnu}>43$ GeV, and
also the Tevatron $\etmiss +jets$ search. In Fig.~\ref{fig64}{\it b}, the
experimentally
excluded region is made up entirely of the LEP chargino mass bound.
\item Future searches at LEP II should probe $m_{\tw_1},\ m_{\tell}$ and
$m_h$ up to nearly 90 GeV. These regions are denoted by dot-dashed lines.
\item The corresponding value of $m_{\tg}$ is plotted on the right axis,
for comparison with reaches calculated in the summary Table in Sec. 10,
or Ref. \cite{KAMON}.
These values vary only slightly with $m_0$ due to inclusion of
differences between the $\overline{DR}$ and pole values of $m_{\tg}$\cite{MT}.
\item The total $\tw_1\tz_2\to 3\ell +\etmiss$ cross section is plotted
for $200\ fb$ (roughly attainable by Tevatron run IB), and for $20\ fb$
(roughly attainable by the Tevatron MI run with $1\ fb^{-1}$
of data). Since no simulation has been performed, these must be regarded
as maximal regions, because in some areas, lepton $p_T$ values may be too soft
for detection. We remind the reader that the results of the complete
simulation of the trilepton signal are shown in Fig.~\ref{fig63} and
Fig.~\ref{fig63p}.
\item We also show with the dotted lines the regions where the $\tz_2$
spoiler modes turn on. These regions show the limit beyond which
no Tevatron upgrade is likely to probe. We also show regions where
$\tz_2$ can decay via two-body modes into real sleptons or sneutrinos.
In these regions, there can be large fluctuations of signal due to
branching fraction and kinematical effects. Again, these regions may be
compared with the corresponding regions in Fig.~\ref{fig63} and
Fig.~\ref{fig63p}.
\end{itemize}

It is easy to see from Fig.~\ref{fig64} that the regions explorable by Tevatron
$\etmiss$ searches (across all $m_0$ values) are complementary to the regions
explorable via $3\ell +\etmiss$ searches (which favor small $m_0$).
Also, the complementarity of searches at LEP and Tevatron, and LEP II and
Tevatron Main Injector, is also easily seen. Similar plots for
a different $A_0$ value (where the $\tst_1$ search also enters),
and $\tan\beta =10$ are shown in Ref. \cite{BCMPT}. Since this
constrained SUGRA framework has been embedded into ISAJET, various
experimental search efforts should be able to combine results within the
GUT scale SUGRA parameter space.

Additional assumptions from theories at the GUT or Planck scale can further
reduce the parameter space of supersymmetric models. For example, in
superstring models with supersymmetry breaking in the dilaton sector,
the universal soft breaking parameters are related such that
\begin{eqnarray}
m_{1/2}=-A_0=\sqrt{3} m_0.
\end{eqnarray}
In models with supersymmetry breaking in the moduli sector, one expects
the ``extreme'' (sometimes called ``no-scale'') boundary conditions
\begin{eqnarray}
m_0=A_0=0,
\end{eqnarray}
although quantum corrections can distort these simplified values.
Implementation of such boundary conditions can reduce the parameter space
to just one or two dimensions (depending on whether one further specifies the
soft-breaking $B$ parameter), plus the usual sign of $\mu$ ambiguity.
Various analyses have been performed within these contexts, assuming
gauge unification at the GUT scale, or sometimes at the string scale,
where extra heavy particles ($m\sim 10^7 - 10^{12}$ GeV) must be added
to maintain unification\cite{anton}. In Ref. \cite{BGKP}, these various
classes of models have been examined, and rather complete parameter space
scans have been carried out for Tevatron energies. ISAJET was used to
simulate all sparticle subprocesses and decay mechanisms. In general, the
low values of $m_0$ in these models lead to slepton masses
much smaller than squark masses.
Consequently, one generates a large number of leptons in the final state.
In the best cases, one can probe to $m_{\tg}\sim 500$ GeV using the Tevatron
MI, by searching for trilepton and SS dilepton events.
In the worst case (Dilaton models with GUT scale unification and
negative $\mu$), the light chargino and neutralino dominantly decay to real
sneutrinos, so that only $m_{\tg}\stackrel{<}{\sim} 300$ GeV can be explored at
the MI.
Further details regarding the features of these models may be found in
Ref. \cite{lopez} and Ref. \cite{KAMON}.

We emphasize, finally, that the various GUT or string scale
assumptions made in this section are just that-- assumptions. Whenever
possible, search efforts should be performed using the most general,
and encompassing, framework available.

\subsection{What if $R$-parity is violated?}

Additional interactions which respect gauge symmetry, but violate
baryon ($B$) and/or lepton ($L$) number conservation, can be added to
the superpotential of the MSSM. In this case, $R$-parity is violated.
Simultaneous presence of both $B$ and $L$
violating interactions can lead to catastrophic proton decay, so
usually one or
the other (or both!) of the interactions is assumed absent. In general,
a large number of different $R$-violating interactions can be introduced,
which can lead to very complicated (and intractable)
phenomenology\cite{RVIOL}. If the $R$-violating
interactions are large, sparticle production and decay patterns are altered,
and in particular, single production of sparticles is possible. In addition,
these interactions can also affect the renormalization group flow of various
masses and couplings.
If the $R$-violating interactions are sufficiently small, then gauge
interactions still dominate sparticle production and decay rates but the LSP
$\tz_1$ becomes unstable so that the classic $\etmiss$ signature may
be destroyed.

This latter case of $R$-violation has been examined for Tevatron experiments
in Ref. \cite{RPV}, for a ``worst case'' and ``best case'' situation.
The worst case is where $\tz_1\to qqq$ ($B$-violating),
in which case the $\etmiss$ signature might be lost, and
additional hadronic activity
would cause leptons from cascade decays to fail the isolation requirements
causing the SUSY signal to be lost beneath SM backgrounds.
The best case might be where $\tz_1\to \ell\bar{\ell}\nu$, which may lead
to a plethora of isolated leptons in the final state\cite{BARGRPV}.

In Ref. \cite{RPV}, for the baryon number violating case it is assumed
that the only effect of
$R$-violation is to cause $\tz_1\to cds$ or $\bar{c}\bar{d}\bar{s}$.
Then sparticle production and decay are simulated exactly as in Sec. 6.1,
with the same detector characteristics. The resulting mass reach is
listed in Table 4. We see that there is almost no reach in the $\etmiss$
channel, even given the Main Injector integrated luminosity of
$\sim 1\ fb^{-1}$. However, the presence of cascade decays still allows
hard leptons and neutrinos to be produced in the final state, although fewer
leptons will be isolated due to the additional hadronic activity from LSP
decays. We see that in the SS dilepton channel, and especially in
the isolated $3\ell +jets+\etmiss$ channel, there remains enough signal
to search for substantial ranges of $m_{\tg}\sim 200-350$ GeV
with the Tevatron MI\footnote{Again, it should be kept in mind that the
reaches in Table 4 have been obtained with $\mu < 0$.}. The reach of the
Tevatron would be somewhat
smaller if $\tan\beta$ is large since the MSSM SS and $3\ell$ cross sections
are known to be smaller for $\tan\beta=10-20$.

\begin{table}
\caption[]{Reach in $m_{\tg}$ via various event topologies
for R-parity violating SUGRA-inspired MSSM, assuming an integrated
luminosity of 0.1 fb$^{-1}$ (1 fb$^{-1}$), at the Tevatron collider.
We use $m_t =150$ GeV for the background. In {\it a}), we show results
for $B$-violating interactions, while in {\it b}) we show results for
$L$-violating interactions.}

\bigskip

\begin{tabular}{ccccccc}
case & $E\llap/_T$ & $1\ \ell$ & $OS$ & $SS$ & $3\ \ell$ & $\ge 4\ \ell$ \\
\hline
$a) BNV$ & & & & & & \\
$m_{\tq}=m_{\tg}+10$ GeV & --- (---) & --- (---) & 165 (210) &
200 (280) & 220 (350) & --- (165) \\
$m_{\tq}=m_{\tg}-10$ GeV & 200 (210) & --- (---) & 150 (165) &
165 (235) & --- (360) & --- (---) \\
$m_{\tq}=2m_{\tg}$ & --- (---) & --- (---) & --- (---) &
--- (200) & --- (190) & --- (---) \\
\hline
$b) LNV$ & & & & & & \\
$m_{\tq}=m_{\tg}+10$ GeV & --- (150) & --- (---) & 240 (300) &
330 (450) & 480 (650) & 540 (740) \\
$m_{\tq}=m_{\tg}-10$ GeV & 160 (180) & --- (---) & 250 (300) &
330 (450) & 460 (640) & 520 (710) \\
$m_{\tq}=2m_{\tg}$ & --- (---) & --- (---) & 190 (260) &
340 (540) & 540 (730) & 600 (840) \\
\end{tabular}
\end{table}

For the ``best case'' scenario studied in Ref. \cite{RPV} it is assumed
that the LSP exclusively decays into (readily identifiable)
muons and electrons via
$\tz_1\to \mu\bar{e}\nu_e$ (and related processes) which take
place via $L$-violating
interactions, with all other production and decay mechanisms remaining
unaltered.
In this case, four additional hard, potentially isolated leptons can exist in
the final state. The reach is again listed in Table 4. We see
that in the isolated multilepton channels, very large reaches in $m_{\tg}$
are possible. In particular, in the $4\ell$ channel at the main injector,
equivalent gluino masses of $\sim 700-800$ GeV can be probed!

\section{Search for SUSY at the CERN LHC}

\begin{figure}[tb]
\centerline{\psfig{file=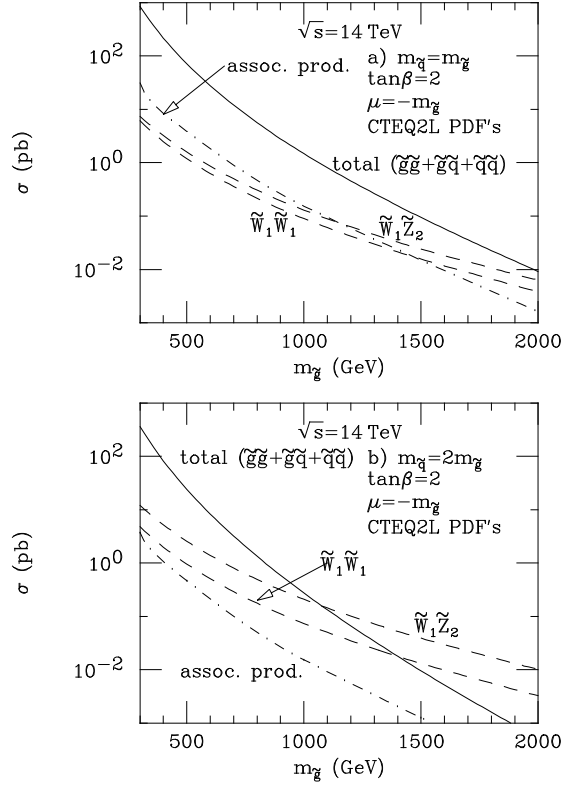,height=11cm,angle=90}}
\caption[]{Total production cross sections for various reactions
at the LHC.}
\label{fig71}
\end{figure}

The CERN Large Hadron Collider, a $pp$ collider to operate at $\sqrt{s}=14$
TeV,
is frequently regarded as a machine capable of a thorough search for
supersymmetric particles below the TeV scale. In Fig.~\ref{fig71}, we show
total cross
sections for various sparticle pair production reactions as a function of
$m_{\tg}$. We set $\tan\beta =2$, and $\mu =-m_{\tg}$; the plots are
insensitive to $\mu$ as long as $\mu$ is large so that the lighter -inos are
mainly gaugino rather than higgsino.
In {\it a}), we take $m_{\tq}=m_{\tg}$, while in {\it b})
we take $m_{\tq}=2m_{\tg}$. We see that in {\it a}), the summed strong
production of $\tg\tg +\tg\tq +\tq\tq$ is the dominant cross section
over the complete range of $m_{\tg}$ all the way up to 2 TeV. In case {\it b}),
however, we see that the chargino/neutralino pair production reactions
become the dominant SUSY particle production mechanism above
$m_{\tg}\sim 1100$ GeV.
In this mass range, $\tw_1\tz_2\to (W\tz_1)+(h\tz_1)\to \ell b\bar{b}+\etmiss$
should have major background problems from $Wb\bar{b}$, $Wh$ and $t\bar{t}$
production.
In these figures, there is assumed five
degenerate species of $L$- and $R$- squarks.
If the light top squark is
much lighter than other squarks, then its production mechanism may
dominate other squark production mechanisms.
Associated production is always a sub-dominant component of sparticle
pair production for both cases {\it a}) and {\it b}).

\subsection{Sparticle cascade decay signatures}

\begin{figure}[tb]
\centerline{\psfig{file=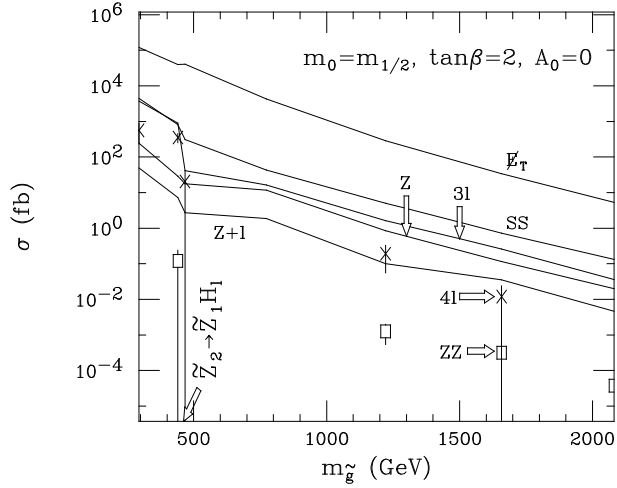,height=7cm,angle=90}}
\caption[]{Cross sections after cuts for various event topologies
at the CERN LHC $pp$ collider.}
\label{fig72}
\end{figure}

As in the case of the Tevatron collider, sparticle pair signatures at LHC
divide into the various $\etmiss$, $1\ell$, $2\ell$ (SS and OS), $3\ell$
and $\ge 4\ell$ classes. In addition, the rate for events containing
high $p_T$ leptonically
decaying $Z$ bosons together with $\etmiss$
can be substantial at LHC. In Ref. \cite{BTW}, these various signals
were plotted with a set of nominal cuts, and compared
with the corresponding SM backgrounds.
We update\cite{BCPTLEP} this plot in Fig.~\ref{fig72},
using
ISAJET 7.13, for the same set of cuts as in Ref.\cite{BTW},
except for now using
a calorimeter out to $|\eta |<5$, and requiring $|\eta_{jet}|<3$.
Error bars denote the statistical uncertainty in our simulation
of the smallest cross sections.
This sampling
of a particular slice of SUGRA parameter space ($m_0=m_{1/2}$,
$A_0=0$ and $\tan\beta =2$) illustrates several points:
\begin{itemize}
\item the $\etmiss +jets$ signal\cite{TDR,POLESELLO}, although dependent on the
specific cuts,
occurs at a significant rate for a large range of $m_{\tg}$ well beyond
1 TeV,
\item the SS dilepton\cite{bgh} and $3\ell$ signal rates are also substantial
over a large range of $m_{\tg}$. The kink just below
$m_{\tg}=500$~GeV marks where
the previously mentioned $\tz_2$ ``spoiler modes'' turn on.
We see that while the trilepton signal drops by an order of magnitude, almost
a hundred events are nevertheless expected annually at the LHC even if the
gluino is as heavy as 1~TeV.
This implies that a significant number of these events
comes from other than $\tz_2$ decays, {\it e.g.} $\tg\to t\tst_1$ decays.
The other cross sections are less sensitive to the ``spoiler mode''.

\item the signal from $4\ell$ events is likely to be visible in only
limited regions of parameter space, and
\item there also exist regions of parameter space where signals
from leptonically decaying high $p_T$ $Z$ bosons plus jets plus $\etmiss$
may be observable. Within the MSSM, this region is sensitive to the value of
$\mu$; within the SUGRA framework, $\mu$ is no longer a free parameter.
\end{itemize}

\subsubsection{$\etmiss +jets$ signature}

Previous studies on $\etmiss +jets$ signal from gluinos and squarks
by the GEM and SDC collaborations for the SSC concluded that
$m_{\tg}\sim 0.3-$ 1-2 TeV should have been detectable\cite{TDR}.
Recently, the Atlas\cite{POLESELLO}
collaboration has performed detailed studies on the $\etmiss +jets$
signal for the LHC. They found that gluinos as low as 300 GeV should easily be
seen above SM backgrounds. Then, requiring rather stiff cuts
\begin{eqnarray}
\etmiss >600\ {\rm GeV},\ p_T(jet_1,jet_2,jet_3)>200\ {\rm GeV},\
p_T(jet_4)>100\ {\rm GeV},
\end{eqnarray}
and transverse sphericity $S_T>0.2$, they found an upper reach for $m_{\tg}$
listed in Table 5, for various choices of integrated luminosity. We see
that if squarks and gluinos have mass below $\sim 1$ TeV, and if the MSSM
is a reasonable approximation of nature, then supersymmetry is unlikely to
escape detection at the LHC.
\begin{table}
\caption[]{5$\sigma$
discovery limits in $m_{\tg}$ (in TeV) via $\etmiss +jets$ events
at LHC Atlas detector for various
choices of squark and gluino mass ratios, and collider integrated
luminosities. Variation of MSSM
parameters can cause these limits to vary by $\sim 150$ GeV.}

\bigskip
\centerline{
\begin{tabular}{cccc}
case & $10^3\ pb^{-1}$ & $10^4\ pb^{-1}$ & $10^5\ pb^{-1}$ \\
\hline
$m_{\tq}=m_{\tg}$ & 1.8 & 2.0 & 2.3 \\
$m_{\tq}=2m_{\tg}$ & 1.0 & 1.3 & 1.6 \\
\end{tabular}
}
\end{table}

\begin{figure}
\centerline{\psfig{file=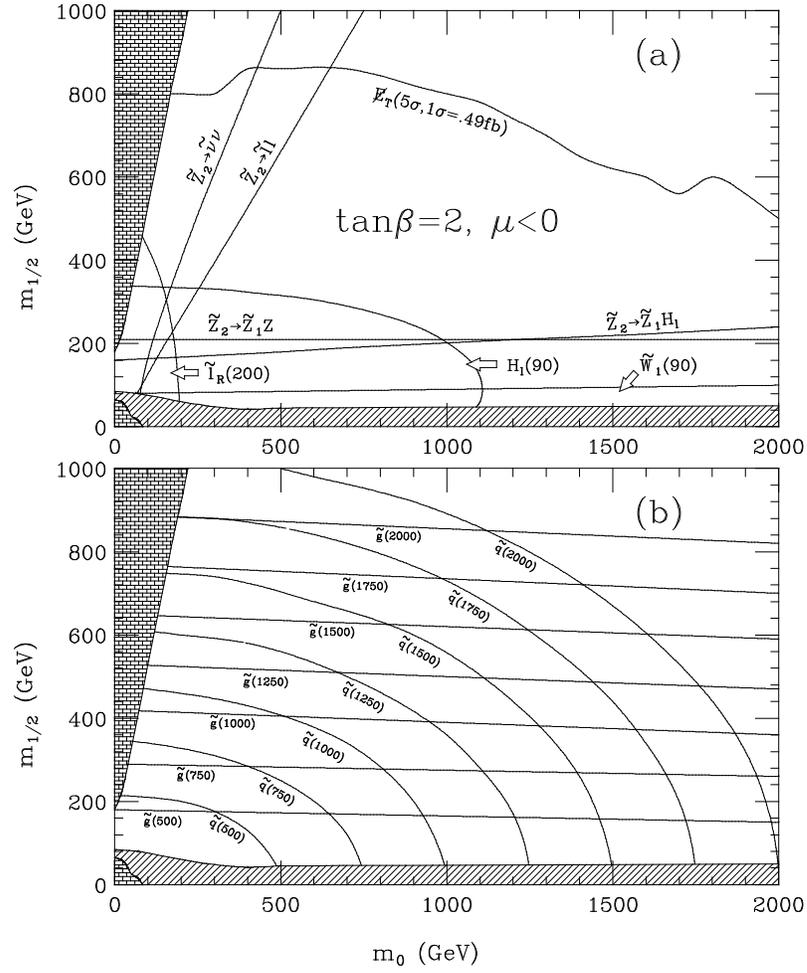,height=14cm,angle=90}}
\caption[]{Contour in the $m_0\ vs.\ m_{1/2}$ plane where the
multi-jet $+\eslt$ signal is observable above SM backgrounds.}
\label{fig73}
\end{figure}

A similar analysis has been performed in Ref. \cite{bcpt}, where
similar reach values were obtained. In addition, the regions of
the minimal SUGRA model explorable via multi-jets $+\eslt$ signature
were mapped out. Sample results for $\mu <0$, $A_0=0$ and $\tan\beta =2$
are shown in Fig.~\ref{fig73}{\it a}. Results differ only slightly for $\mu
>0$, or
$\tan\beta =10$. In Fig.~\ref{fig73}{\it b}, contours of $m_{\tg}$ and
$m_{\tq}$
are shown for comparison.

An interesting question to ask is: if a signal in the $\etmiss +jets$
channel is found at LHC, what information can be gleaned about
sparticle properties? In general, many different subprocesses can be
contributing to the SUSY signal, and the subsequent cascade decay channels
can be numerous and complicated, especially for relatively heavy sparticles.
Mapping out the size of the signal cross section, and the shapes of
different jet distributions and $\etmiss$ distributions, and matching
against Monte Carlo predictions, will significantly constrain the SUSY
parameter space. Also, in first approximation, one expects $\tg\tg$
events to have higher jet multiplicity than $\tq\tq$ events. In practice,
the cascade decays, along with substantial QCD radiation, distort this
picture. Jet multiplicity distributions have been evaluated in
Ref. \cite{bcpt},
where it was found that mixed $\tq\tq +\tg\tg +\tg\tq$ subprocesses
typically yield average jet multiplicities a half unit lower than pure
$\tg\tg$ production if $m_{\tq} \simeq m_{\tg}$.

Can one measure $m_{\tg}$ or $m_{\tq}$ in the $\etmiss +jets$ channel?
A rough mass determination can be made just based on the size of the total
cross section, and the hardness of distributions such as $p_T(jet)$ and
$\etmiss$. Direct measurement of $m_{\tg}$ is difficult. Even in the idealistic
case of constructing a two-jet mass from $\tg\to q\bar{q}\tz_1$, the
mass distribution is a smear of values, with an end-point at
$m_{\tg}-m_{\tz_1}$. Realistic situations can only do worse. In Ref.
\cite{bgh},
it was suggested to construct the mass value $M_{est}$ (in the context of
SS dilepton events) by a series of cuts designed to hemispherically
separate the decay products of each gluino. This method suffers from
the fact that it is difficult to obtain a reasonably pure sample of gluino
dilepton events. Nonetheless, in Ref.\cite{bcpt}
this technique has been pursued in the $\etmiss +jets$
channel, using ISAJET to simultaneously generate all sparticles.
The procedure here is to divide the transverse plane into
hemispheres using the transverse sphericity eigenvector, and then calculate
the invariant mass of all hard jets in each hemisphere.
The larger of the two values is taken to be $M_{est}$.
\begin{figure}
\centerline{\psfig{file=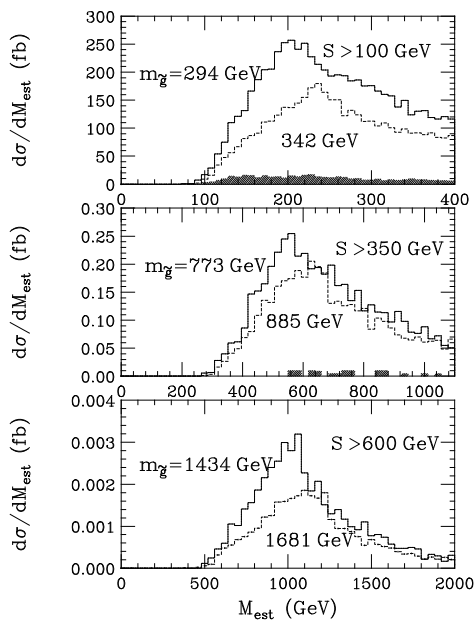,height=9cm,angle=90}}
\caption[]{Distribution in $M_{est}$ for various values of $m_{\tg}$
at the CERN LHC.}
\label{fig74}
\end{figure}

In Fig.~\ref{fig74}, we show histograms of $M_{est}$ for $\etmiss +jets$ events
with the following cuts:
\begin{enumerate}
\item veto isolated leptons,
\item $S_T$ (transverse sphericity) $>0.2$,
\item $p_T(jet)>100$ GeV,
\item $n(jets)\ge 2$,
\item $\etmiss >S$, and $p_T(jet_1,jet_2)>S$.
\end{enumerate}
Then, $M_{est}$ is calculated using only hard jets, with $p_T>S$. Events
are rejected if there is only one qualifying jet in each hemisphere.
In the figure, we take SUGRA parameters $m_0 =m_{1/2}$, $A_0 =0$,
$\tan\beta =2$, and $\mu <0$. For these parameters, $m_{\tq}\sim m_{\tg}$.
The corresponding values of $m_{\tg}$ are listed in the figure, as are
background events from SM sources. We see that the distributions
shown are able to distinguish $m_{\tg}$ to 15\%. The resolution is somewhat
worse if
$m_0=4m_{1/2}$, but $M_{est}$ distributions for gluino masses differing by
25\% appear to be readily distinguishable.
There is a wide spread of mass values in each plot, due to cascade
decay effects, wrong jet assignments, QCD radiation, {\it etc.}.
However, the overall trend is clear: heavier sparticles should give
harder distributions of $M_{est}$ values (or other measurable distributions),
and so, may provide information on the underlying gluino and squark masses.

\subsection{$\tw_1\tz_2\to 3\ell$ signal}

Can the clean tri-lepton signal from $pp\to\tw_1\tz_2\to 3\ell +\etmiss$
be seen at the LHC as well as at the Tevatron? At first glance, this is
unclear\cite{barb}, since the signal cross section rises by typically
a factor of
$\sim 10$ in going from Tevatron to LHC, while background from $t\bar t$
production increases by a factor of $\sim 160$ (depending on $m_t$).
Furthermore, there are additional sources of trilepton events from
other SUSY reactions at LHC energy.

Detailed simulations of signal and background have been performed in
Ref. \cite{bcpttwo}. In this work, a series of cuts were found that would allow
for a clean extraction of the $\tw_1\tz_2\to 3\ell +\etmiss$ signal.
The procedure is to first establish
the signal by requiring
\begin{itemize}
\item three isolated high $p_T$ leptons.
\end{itemize}
After this requirement, the total event sample is dominated by {\it other} SUSY
sources of trilepton events, which mainly come from gluino and squark
cascade decays. To get rid of these, one requires
\begin{itemize}
\item no jets in the event, plus $\etmiss <100$ GeV,
\end{itemize}
Then one is left with clean trilepton events, with large contamination from
$WZ\to 3\ell+\etmiss$. After requiring a $Z$ mass cut,
\begin{itemize}
\item for OS, same-flavor lepton pairs, $m(\ell\bar{\ell})\ne M_Z\pm 8$ GeV,
\end{itemize}
the $WZ$ contamination is reduces to tiny levels, but a significant
background from $t\bar t$ may remain. Requiring either of the following
conditions,
\begin{itemize}
\item two fastest leptons be SS and flavor of the slow lepton be the same
(but anti-) the flavor of either of the two fast leptons, or
\item two fastest leptons are OS if $p_T(slow\ lepton)>20$ GeV,
\end{itemize}
leaves one with signals on the level of $10-40\ fb$, while SM background
is below the $fb$ level. Thus, at least for $\mu < 0$,
the $3\ell$ signal appears viable as long as
the $\tz_2$ spoiler modes are closed, {\it i.e.} $m_{\tg}\alt 550-650$ GeV.
For positive values of $\mu$, a significant hole still
remains in the $m_0-m_{1/2}$ plane, even where the spoiler modes are not
accessible\cite{BCPTLEP}.
The purity of the remaining event sample allows some detailed mass
information to be extracted. For instance, in $\mu e\bar{e}$ events,
the quantity $m(e\bar{e} )$ is kinematically restricted to be less than
$m_{\tz_2} -m_{\tz_1}$; thus a plot of the upper cutoff of this
distribution can yield precise mass information on sparticles even in the
difficult environment of an LHC detector. Other distributions can
constrain different combinations of the chargino/neutralino
masses\cite{bcpttwo}.

\subsection{Slepton search}

The search for sleptons at the LHC has also been addressed in the
literature\cite{amet,slep}. Detailed simulations of signals and backgrounds
for {\it all} slepton production mechanisms including cascade decays
were performed in Ref. \cite{slep}. There, it was shown that the only
viable channel for observing a slepton pair signal was in the $2\ell +\etmiss$
channel, which usually comes from $\tl_R\bar{\tl_R}\to 2\ell +\etmiss$.
By requiring events with two isolated leptons plus no jets, together with
some additional angular cuts, a signal on the order of $fb$'s could be
seen for $m_{\tl}\alt 250$ GeV, against tiny backgrounds from SM and
other SUSY sources. Extraction of any sort of detailed mass
information from the small sample of remaining signal events appeared to
be difficult.

\section{Supersymmetry at future linear $e^+ e^-$ Colliders}

A future linear $e^+ e^-$ collider will obviously have a higher
discovery reach than LEP II, due to its higher center-of-mass energies,
$\sqrt{s} = 0.5$--1~TeV. Furthermore, experimentation at a linear
collider would be richer compared
to that at LEP-II, due to the following characteristics:
(1) flexible center-of-mass energies,
(2) high beam polarization of 90\% and
beyond. It probably is not well known that the
center-of-mass energy can be flexibly varied, for instance between 200
to 500 GeV within a single collider design. Upgrade to a higher energy
machine (say up to 1~TeV) is possible either by making the accelerator
longer or improving the acceleration gradient of the klystrons.
Therefore, under most of the common designs, one single accelerator will
be able to cover a wide range of center-of-mass energies. One can tune the
center-of-mass energy for many different purposes. For instance, one can
optimize the sensitivity on the slepton mass measurement by choosing the
center-of-mass energy such that $\beta \simeq 0.5$ \cite{TSUK}; or
one can avoid the complication due to simultaneous production of several
sparticles by choosing the center-of-mass energy below one of the thresholds.
The virtues of polarization are two-fold. (a) Use of
polarization can suppress the background substantially even up to two
orders of magnitudes. This allows us to obtain a very pure sample of
signals appropriate for precision studies. (b) One basically doubles the
number of experimental observables using both polarization states. This
enables the efficient measurements of various parameters, which is
proven already in the $A_{LR}$ measurement at the SLD experiment.

In early days,
beamstrahlung processes (emission of high intensity $\gamma$-rays due the
interactions between beams at the collision point), which can smear
the center-of-mass energy and produce large leptonic and hadronic
underlying backgrounds from photon collision, had been a source of concern.
This effect is
negligible at LEP-II or SLC because of smaller energy and much larger
beam size at the collision point. However, with improvements in
accelerator designs and as well as in our understanding of
photon structure function, it
has been shown that beamstrahlung effects
are not harmful for most of the interesting physics
studies. One can achieve \cite{CBP} (1) small beam energy spread
even after including initial state radiation and
beamstrahlung effects,
and (2) a clean environment basically without
underlying events even with photon induced hadronic processes.

Thus, all the virtues of lower energy $e^+ e^-$ colliders
remain \cite{early}, while the high center-of-mass energy and beam
polarization will give us additional tools to study physics. The goals
of the $e^+ e^-$ linear collider experiment will be multiple: (1)
discovery of sparticles, (2) measurement of SUSY parameters, (3)
quantitative verification of supersymmetric invariance of the
interactions.

It is also worth recalling here that an $e^+ e^-$ linear collider is also a
Higgs discovery/study machine especially for supersymmetric models. As
is well-known, the lightest neutral Higgs boson in MSSM is always lighter
than $\alt 130$~GeV after including the radiative corrections due to
the stop loop \cite{MSSM-Higgs}. Even in models with additional singlets
or extra families \cite{dirty}, an upper bound $\alt 160$~GeV
persists under the assumption of perturbativity up to the GUT-scale.
Furthermore, a reasonable size of production cross section is guaranteed
for the lightest (or, sometimes, the second lightest if the lightest one
is dominantly singlet)
neutral Higgs boson even with the mixing to singlet
states \cite{Kamoshita}. If we could further find other Higgs states
via processes as $e^+ e^- \rightarrow h^0 Z^0$, $H^0 Z^0$, $h^0 A^0$,
$H^0 A^0$, or $H^+ H^-$,
it is a definite sign that the Higgs sector is beyond that in
the minimal Standard Model. One can cover up to $m_A \simeq 200$~GeV
with a 500~GeV collider \cite{JANOT}.\footnote{Unfortunately, it is difficult
to see the difference between the minimal Standard model and the MSSM if
$m_A$ is beyond the reach of the discovery at a given center-of-mass
energy \cite{Haber}.}

\subsection{$\tilde{l}$-pair production}

\begin{figure}
\centerline{\psfig{file=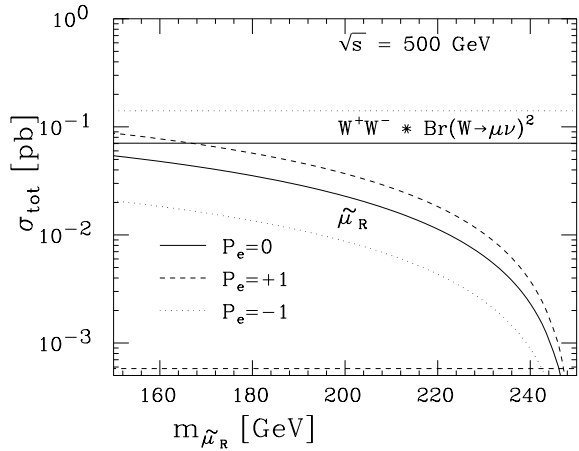,height=6cm,angle=90}}
\caption[smu-JLC]{Total cross sections of $\tilde{\mu}_R$ pair
production at $\sqrt{s} = 500$~GeV, for unpolarized $P_e=0$ and
polarized beams $P_e=\pm 1$. The backgrounds from $W$-pair is also
shown.}
\label{smu-JLC}
\end{figure}

As before, we assume that one of the sleptons is LVSP in this subsection. We
also assume $R$-parity is conserved, and LSP is the lightest neutralino.
Then the only
possible decay mode is $\tilde{l} \rightarrow l \tilde{\chi}_1^0$
assuming lepton family number conservation. The cross sections are shown
in Fig.~\ref{smu-JLC}.

Here we summarize the analysis of Ref.~\cite{Becker}.
One possible set of selection criteria is the following:\footnote{See
the footnote \ref{acopdef} for the definition of the acoplanarity.}
\begin{enumerate}
\item $\theta_{acop} > 65^\circ$.
\item $\not\!p_T > 25$~GeV.
\item The polar angle of one of the leptons should be larger than
46$^\circ$, the other $26^\circ$.
\item $|m_{ll} - m_Z| > 10$~GeV.
\item $E_{l^\pm} < 150$~GeV.
\end{enumerate}
The resulting signal and background cross sections,
 for $\sqrt{s} = 500$~GeV, are listed in
Table~\ref{Becker}. Here a relatively
pessimistic (hard) beamstrahlung spectrum of the ``Palmer-G'' type is
assumed. Conservative detector parameters are chosen: the tracking detector
covers down to 15$^\circ$ with a momentum resolution $\Delta p_T/p_T =
1.5\times 10^{-3} p_T($GeV) and an angular resolution of 1~mrad. The EM
calorimetry covers down to 11.5$^\circ$, with $\Delta E/E =
0.17/\sqrt{E(\mbox{GeV})} + 0.03$.

Assuming a collider energy of $\sqrt{s}=500$ GeV, and
an integrated luminosity of 20~fb$^{-1}$ (again conservative),
a 5$\sigma$ signal can be found up to 225~GeV, as long as the mass
difference between the smuon and LSP is greater than 25~GeV.

\begin{table}
\centerline{
\begin{tabular}{|c|c|c|c|}
\hline
process & $\sigma_{tot}$(fb) & $\epsilon$(\%) & $\sigma_{acc}$(fb)\\
\hline
$\tilde{\mu}_R$(150)-pair & 50 & 24 & 12\\
$\tilde{\mu}_R$(200)-pair & 16 & 31 & 5\\
$\tilde{\mu}_R$(230)-pair & 3 & 36 & 1\\
\hline
$\gamma\gamma \rightarrow \mu \mu$ & 7177. & 0.0 & 0.0\\
$ee \rightarrow \gamma^*/Z^* \rightarrow\mu\mu$ & 608. & 0.0 & 0.0\\
$\gamma\gamma \rightarrow \tau \tau \rightarrow \mu \mu + \nu$'s&
215. & 0.0 & 0.0\\
$ee \rightarrow \gamma^*/Z^* \rightarrow\mu\mu + \nu$'s&
19. & 0.0 & 0.0\\
$ee \rightarrow WW \rightarrow \mu\mu+\nu$'s & 131 & 0.7 & 1.0 \\
$\gamma\gamma \rightarrow WW \rightarrow \mu\mu+\nu$'s & 3.4 & 17.6 &
0.6\\
\hline
total & 8363. & 0.02 & 1.8\\
\hline
\end{tabular}
}
\caption[Becker]{Cross sections and efficiencies for $\tilde{\mu}$-pair
signals and standard model backgrounds \cite{Becker} with an unpolarized beam.
Relatively pessimistic
bremsstrahlung spectrum is assumed.
The LSP mass is $m_{\tilde{\chi}_1^0} =100$~GeV.}
\label{Becker}
\end{table}

For $\tilde{e}$, the production cross section is in general larger because
of the additional diagram with $t$-channel exchange of neutralinos.
The $e \nu_e W$ and $ee Z$ final states are backgrounds to $\tilde{e}$-pair
which are absent for $\tilde{\mu}$-pair. However with an additional cut
$E_e > 15$~GeV, the backgrounds are reduced down to 2.0~fb ($eeZ$) and
1.7~fb ($e\nu_e W$). The discovery reach turns out to be even better than the
$\tilde{\mu}$ case depending on the mass spectrum in the neutralino
sector.

\begin{figure}
\centerline{\psfig{file=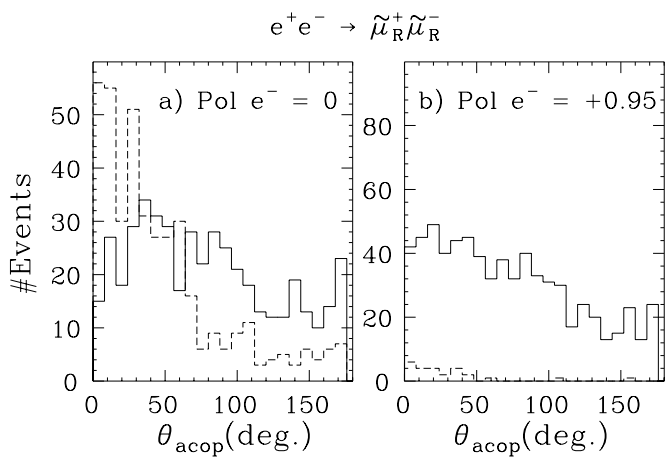,height=6cm,angle=90}}
\caption[acop]{Acoplanarity distribution of lepton pair for $\sqrt{s} =
350~$GeV, $m_{\tilde{l}} = 142$~GeV, $\int {\cal L} dt =
20$~fb$^{-1}$. 
Fig.~b) shows
the dramatic reduction of the backgrounds using the right-handed
electron beam.}
\label{acop}
\end{figure}

\begin{figure}
\centerline{\psfig{file=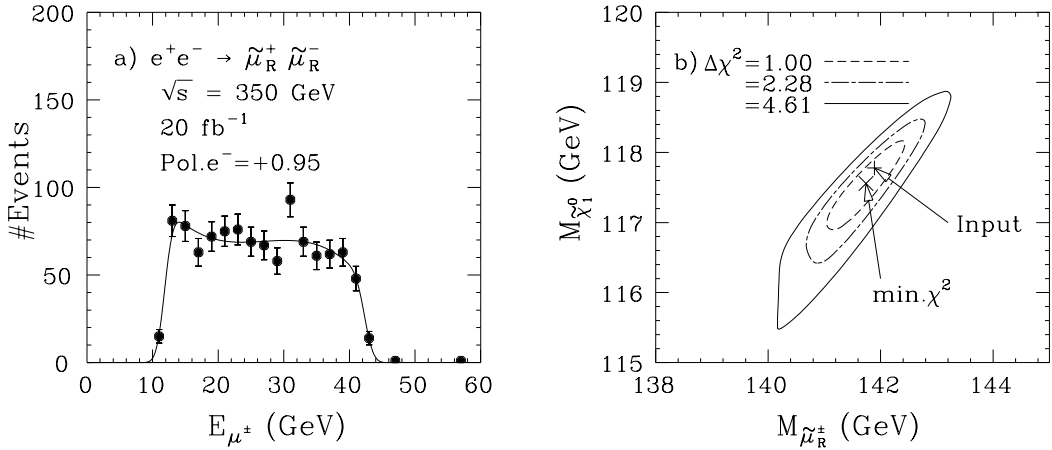,height=6cm,angle=90}}
\caption[El]{(a) Energy distribution of the final muon from
$\tilde{\mu}_R$-pair production, including standard model backgrounds.
(b) Two-parameter fit to the energy distribution on $(m_{\tilde{\mu}},
m_{\tilde{\chi}_1^0})$ plane.}
\label{El}
\end{figure}

Even though the unpolarized beam is efficient enough for the discovery
of sleptons, the background can be further suppressed to a negligible
level using the polarized beam. This is useful for the precise
measurements of the masses and cross sections. Since the right-handed
electron does not couple to $W$, the diagram with $t$-channel
neutrino exchange is absent, and hence the background from $WW$ is
greatly reduced. Then a slightly weaker set of cuts is sufficient
\cite{TSUK}:
\begin{enumerate}
\item 5~GeV $< E_\mu < (\sqrt{s} - 100$~GeV)/2.
\item 20~GeV $< E_{vis} < \sqrt{s} - 100$~GeV.
\item $|m_{ll} - m_Z| > 10$~GeV.
\item $|\cos \theta_{l^\pm}| < 0.9$.
\item $-Q_l \cos \theta_l < 0.75$ where $Q_l$ is the charge of the
lepton and the polar angle is measured from the electron beam direction.
\item $\theta_{acop} > 30^\circ$.
\end{enumerate}
For a sample parameter $m_{\tilde{\mu}_R} = 142$~GeV,
$m_{\tilde{\chi}_1^0} = 118$~GeV, $\sqrt{s} = 350$~GeV, and $\int dt
{\cal L} = 20~\mbox{fb}^{-1}$, $P_e = 95$\%, one could obtain an event
sample with signal purity at 99\% and efficiency of 54.2\%
(Fig.~\ref{acop}).
Given this pure signal samples, one can fit the energy distribution of
the final leptons to measure masses of $\tilde{\mu}$ and
$\tilde{\chi}_1^0$. The resolutions of mass determinations are better
than 1\% (Fig.~\ref{El}).

It is worth noting that one can put an upper bound on the mass of
charginos once one knows the mass of the LSP, assuming the GUT-relation
for gauino masses \cite{TSUK}. Therefore, precision measurement
on the masses will give us a useful clue to the next target center-of-mass
energy. Comparing the cross sections from both right-handed and
left-handed beams, one can test that the gauge quantum numbers of the
observed $\tilde{\mu}_R$ is indeed the same as $\mu_R$, which is a
quantitative support that it is a superpartner of $\mu_R$. Also, the angular
distribution could tell us that the newly discovered particle is a scalar
particle.

Another interesting study can be done once $\tilde{\tau}$-pairs are
found \cite{Nojiri}.  Let us suppose for the moment that $\tilde{\tau}$
decays mainly into $\tau \tilde{\chi}_1^0$ to simplify discussions.  The
cross sections tell us whether the observed $\tilde{\tau}$ is either
left- or right-handed (or their certain mixture).
Also, the polarization of the final state tau leptons can,
in principle, be measured with traditional methods:
the easiest is the measurement of
the energy distribution of the $\pi$ from $\tau \to \pi \nu_\tau$.
We might also use $\tau \to \rho \nu_\tau$ and $a_1 \nu_\tau$ modes.
Therefore, we can see whether the chirality of the final
$\tau$ matches with the type ($L$ or $R$) of the parent stau that was
produced. If it does, the neutralino is dominantly gaugino-like;
otherwise, it is higgsino-like.
Indeed, the energy distribution of the
$\pi$ from $\tau \rightarrow \pi \nu_\tau$ is
sensitive to the higgsino
content of the $\tilde{\chi}_1^0$ \cite{Nojiri}.

\subsection{$\tilde{\chi}_1^\pm$-pair production}

Under the assumption that $\tilde{\chi}_1^\pm$ is the LVSP, it decays
into $\tilde{\chi}_1^\pm \rightarrow \tilde{\chi}_1^0 f \bar{f}'$ where
$f$, $\bar{f}'$ are light fermions in the standard model. The decay
proceeds via (real or virtual) $W$-exchange, slepton- and
squark-exchange. When $W$-exchange dominates, decays to
various final states occur democratically,
resulting in a branching fraction of 70\% for hadronic modes and 10\%
for each of the leptonic modes. This is typical when the mass difference
$\Delta m = m_{\tilde{\chi}_1^\pm} - m_{\tilde{\chi}_1^0}$ is larger
than $m_W$ where the decay into real $W$ dominates. When the mass
difference is smaller, the chargino decays directly into the three-body state.
Branching ratios can vary as a function of slepton and squark masses
just as the case at LEP-II. Both purely
hadronic and mixed hadronic-leptonic mode of chargino pair can be used
for the search.

\begin{figure}
\centerline{\psfig{file=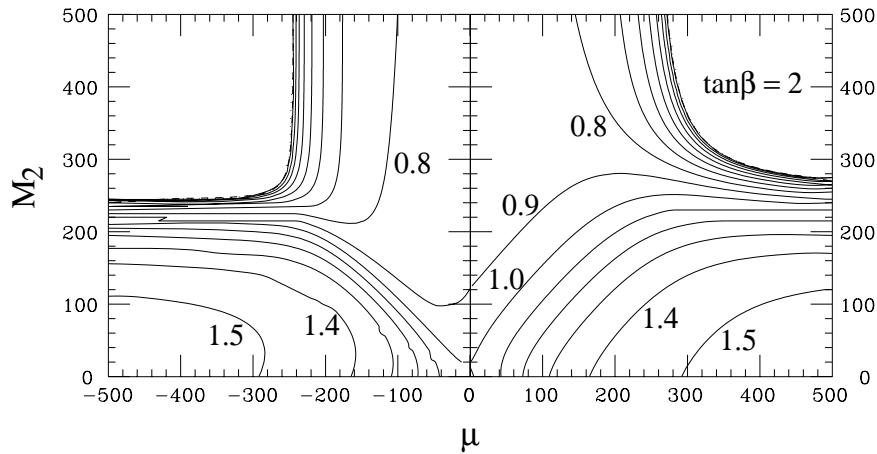,height=6cm,angle=90}}
\caption[chi1000L-JLC]{Contour of the total cross sections in pb of
$\tilde{\chi}_1^\pm$ pair production from the left-handed electron beam
at $\sqrt{s} = 500$~GeV, for $m_{\tilde{\nu}_e} = 1$~TeV. The kinematic
limit is shown in dashed line, while the discovery reach at
20~fb$^{-1}$ is shown in dotted line. They almost overlap with the contour of
0.1~fb.}
\label{chi1000L-JLC}
\end{figure}
\begin{figure}
\centerline{\psfig{file=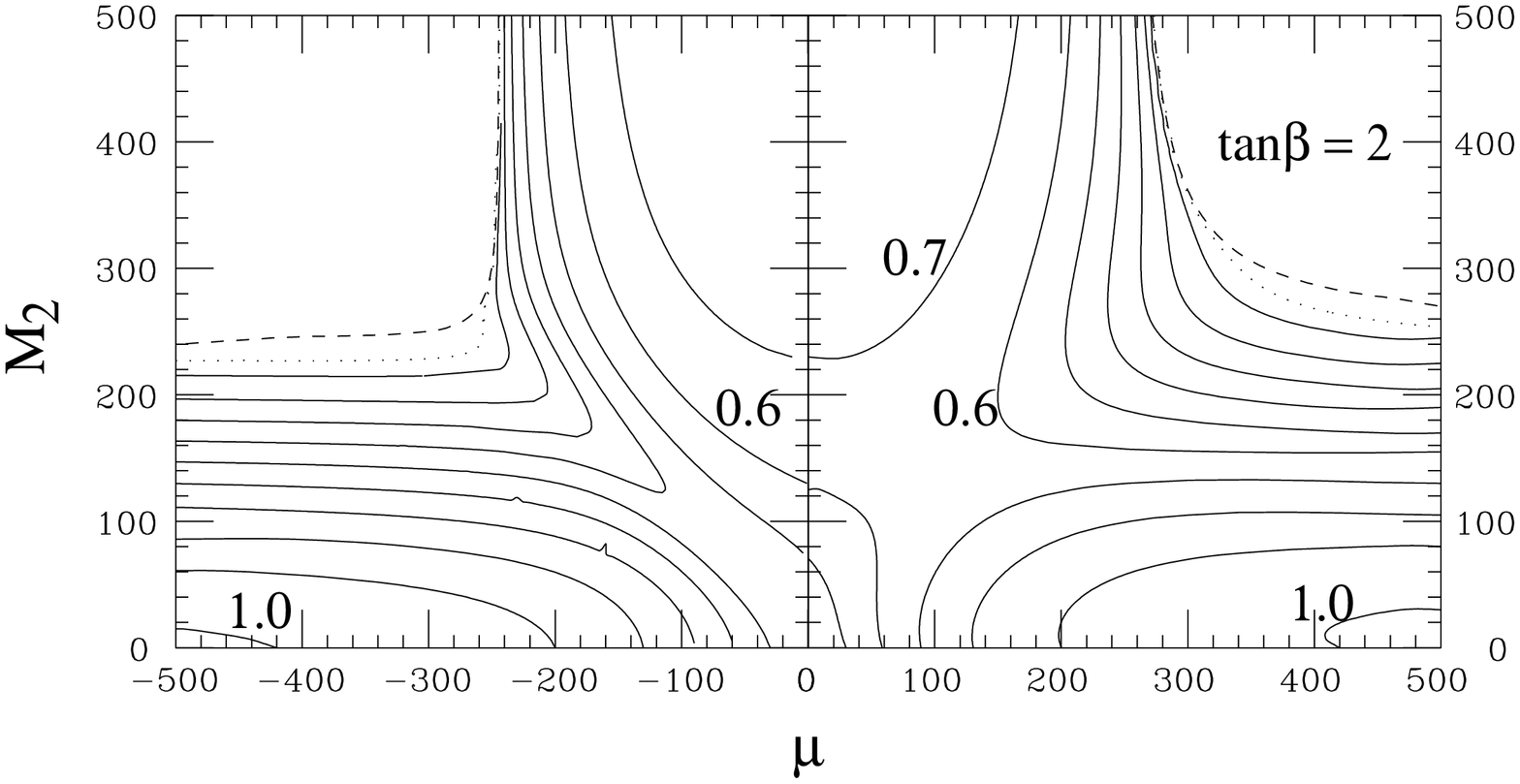,height=6cm,angle=90}}
\caption[chi250L-JLC]{Contour of the total cross sections in pb of
$\tilde{\chi}_1^\pm$ pair production from the left-handed electron beam
at $\sqrt{s} = 500$~GeV, for
$m_{\tilde{\nu}_e} = 250$~GeV. The kinematic limit is shown in dashed
line, while the discovery reach at 20~fb$^{-1}$ is shown in dotted
line. }
\label{chi250L-JLC}
\end{figure}
\begin{figure}
\centerline{\psfig{file=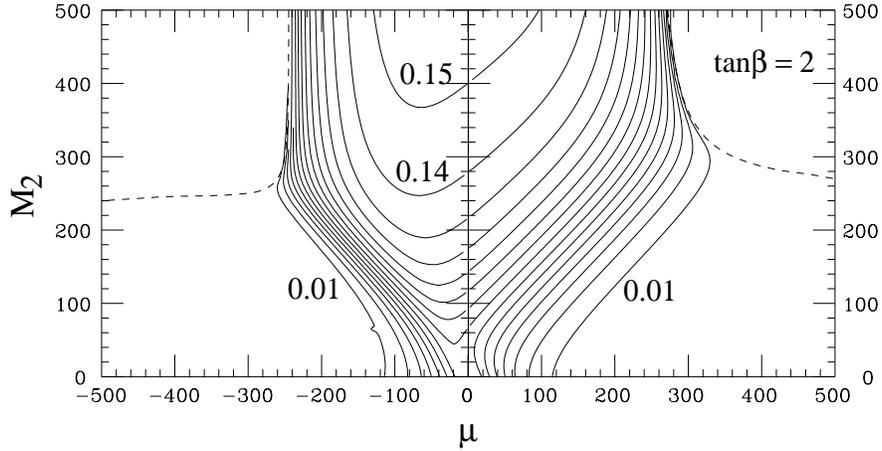,height=6cm,angle=90}}
\caption[chiR-JLC]{Contour of the total cross sections in pb of
$\tilde{\chi}_1^\pm$ pair production from the right-handed electron beam
at $\sqrt{s} = 500$~GeV. The kinematic limit is shown in dashed
line. }
\label{chiR-JLC}
\end{figure}

The cross sections of the chargino pair are shown in
Figs.~\ref{chi1000L-JLC},\ref{chi250L-JLC},\ref{chiR-JLC}. For the
gaugino-dominant
chargino, the cross section from the right-handed beam is very
small. The $t$-channel $\tilde{\nu}_e$-exchange amplitude is destructive
with the $s$-channel $\gamma, Z$-exchange amplitude, so that the cross
section is reduced for lighter $\tilde{\nu}_e$. The higgsino-dominant
chargino has reasonably large cross sections both from left- and
right-handed beams.

The discovery is usually more efficient in purely hadronic mode because
of the higher branching ratios. The $W$-pair background with one $W$
decaying into quarks and the other into $\tau$ is much
less severe at linear collider energies than at LEP-II because two $W$'s
are well
separated in the phase space. One possible set of selection criteria for
the purely hadronic mode at $\sqrt{s} = 500$~GeV is the following
\cite{Grivaz-Finnland}:
\begin{enumerate}
\item number of tracks $>$ 5.
\item The polar angle of the sphericity axis $> 45^\circ$.
\item $\not\!p_T > 35$~GeV.
\item $\theta_{acop} > 30^\circ$.
\item No one single charged particle which carries more than 70\% of
the total energy in the hemisphere.
\item $120~\mbox{GeV} < m_{vis} < 220$~GeV.
\item transverse missing mass $> 200$~GeV.
\end{enumerate}
This set has an advantage that one can look for purely hadronic mode of
chargino pair production when we do not know whether the decay is into a
real $W$ or three-body state. A conservative assumption is made on the
detector capabilities: tracking
resolution of $\Delta p_T/p_T = 1.5\times 10^{-3} p_T/\mbox{GeV}$, ECAL
(HCAL) resolutions of $0.17/\sqrt{E} + 0.03$ ($0.80/\sqrt{E}$) with $E$
in GeV, and coverage of detectors to 18$^\circ$ (10$^\circ$) for
tracking detector (calorimeters).

A sample parameter $M_2 = 200$~GeV,
$\mu = -325$~GeV, $\tan \beta = 4$, gives a signal cross section after cuts
 63~fb (efficiency is 13\%) while the background level is 7.5~fb.
Assuming the integrated luminosity of $\int dt {\cal L} =
20~\mbox{fb}^{-1}$, 5$\sigma$ signal can be obtained up to
$m_{\tilde{\chi}_1^\pm} \simeq 248$~GeV, very close to the kinematic
limit. The mixed hadronic-leptonic mode typically gives an efficiency of
2\% and S/N~$\simeq 1$, and not useful for discovery purpose.

Figs.~\ref{chi1000L-JLC},\ref{chi250L-JLC}
show the discovery reach of the
chargino pair at $\sqrt{s} = 500$~GeV and $\int dt {\cal L} =
20~\mbox{fb}^{-1}$. The only region which is difficult to cover is the
very pure higgsino region $M_2 \gg
\mu$, where the mass of the chargino and LSP become nearly degenerate.
In this region the decay $\tilde{\chi}_1^\pm \rightarrow \tilde{\chi}_1^0 f
\bar{f}'$ has a small $Q$-value, and hence visible energy is small.
There is no quantitative study so far on the discovery reach for small
mass difference. It was estimated that,
with an unpolarized beam, the charginos may evade
discovery if the mass difference is smaller than $\Delta m <
20$~GeV
\cite{Grivaz-Finnland}.\footnote{However, one can use right-handed beam
to reduce the background substantially while the signal cross section
decreases only by a factor of three. Probably one can do a much better job.
More study is necessary on this point. The higgsino LSP does not occur in the
minimal supergravity, but may occur in extended scenarios \cite{POKORSKI}.}

Once one has found charginos, and seen whether the decay is into the
real $W$ or three-body state, one can find much more efficient cuts to
reduce the background for a precision study. When the decay is into the
real $W$ and the LSP, then acoplanar $W$-pair with large missing energy
is the signal. One can identify $W$'s from di-jet invariant masses and
requirement of large missing energy and acoplanarity reduces most of the
backgrounds. Since the decay is two-body, the end points in the energy
distribution of $W$ (sum of two jet energies which form an invariant
mass close to $m_W$) tell us the mass of both the chargino and the LSP.
Therefore, this case is relatively easy.\footnote{An exception is when
the chargino is gaugino-like and $\tilde{\nu}_e$ is light so that the
cross section is low. Then one needs relatively high luminosity of $\int
dt {\cal L} \simeq 50~\mbox{fb}^{-1}$ to achieve the 10\% resolution in
chargino and LSP masses. Fortunately, there are light sleptons in this case,
so that one can measure LSP mass also from slepton study. Using
constraint on the LSP mass, the resolution on the chargino mass can be
still as good as
5\%.} When the decay is directly into the three-body state, the
measurement is more complicated. We will discuss this case in detail
below.

\begin{figure}
\centerline{\psfig{file=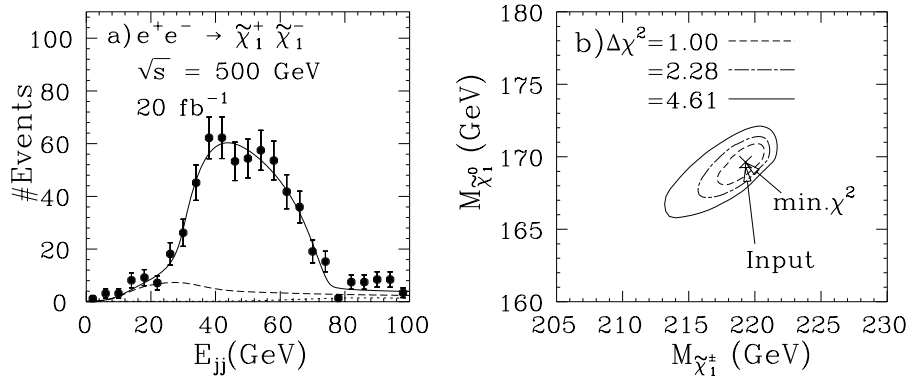,width=12cm,angle=90}}
\caption[chargino]{(a) Di-jet energy distribution from chargino pair
production, (b) contour on the masses of chargino and LSP from a fit of
the di-jet energy distribution.}
\label{chargino}
\end{figure}

We now turn to the measurement of the chargino and LSP masses from
the di-jet energy distribution. We assume three-body decay, and we do
not want to use pure hadronic mode in order to avoid combination
ambiguities of jets as the decay product of one chargino. We employ the
mixed hadronic-leptonic mode below.  For charginos with mixed
hadronic-leptonic mode without real $W$, a possible set of selection
criteria is:
\begin{enumerate}
\item number of tracks $>$ 5, incl. isolated $e$ or $\mu$ ($E_l > 5$~GeV,
energy
deposited within half-angle 30$^\circ$ cone less than 1~GeV).
\item 20~GeV $< E_{vis} < \sqrt{s}-100$~GeV.
\item two jets with $y_{cut} > 5\times 10^{-3}$, $m_{jj} < m_W -
12$~GeV, $E_{jj} < (\sqrt{s} - 100~GeV)/2$.
\item $|\cos \theta_j| < 0.9$, $|\cos \theta_l| < 0.9$, $-Q_l \cos
\theta_l < 0.75$, $Q_l \cos \theta_{jj} < 0.75$.
\item $|m_{l\nu} - m_W| > 10$~GeV for $W$-pair hypothesis.
\item $\theta_{acop} > 30^\circ$ where acoplanarity angle is defined
between the summed momentum of two jets and lepton momentum.
\end{enumerate}
As an example, consider a sample parameter set $M_2 = 400$~GeV, $\mu =
250$~GeV, $\tan \beta = 2$, in the limit of heavy scalar masses. The
chargino and LSP masses are 219~GeV and 169~GeV, respectively, and the
final signal cross section is 234~fb (efficiency 10\%) with the
background from $W$-pair 37~fb and $e\nu W$ 6.6~fb. The efficiency and
S/N ratio are at the comparable level we had for the purely hadronic
mode above. The fit to the di-jet energy spectrum yields the mass
determination at the 1\% level (Fig.~\ref{chargino}).\footnote{The
resolution of the jet energy measurements is crucial here.
We tried to link tracks of the charged particles detected in
the central drift chamber to energy clusters detected in the
electromagnetic calorimeter or hadron calorimeter, and, when linked,
we used the tracking information, since it has better resolution in general.
To be realistic in this linking process, we generated calorimeter
hits with a finite shower size and simulated the cluster overlapping
\cite{Miyamoto,TSUK}.}

\subsection{Systematic Discoveries and Tests on GUT or Supergravity Models}

A nice feature of the study of sparticles at an $e^+ e^-$ collider and
measurement of masses is that one can place an upper bound on the next
sparticle based on modest theoretical assumptions. Furthermore, having
several sparticles at hand allows us to test various predictions of the
GUT or SUGRA models at the several percent level.

\begin{figure}
\centerline{\psfig{file=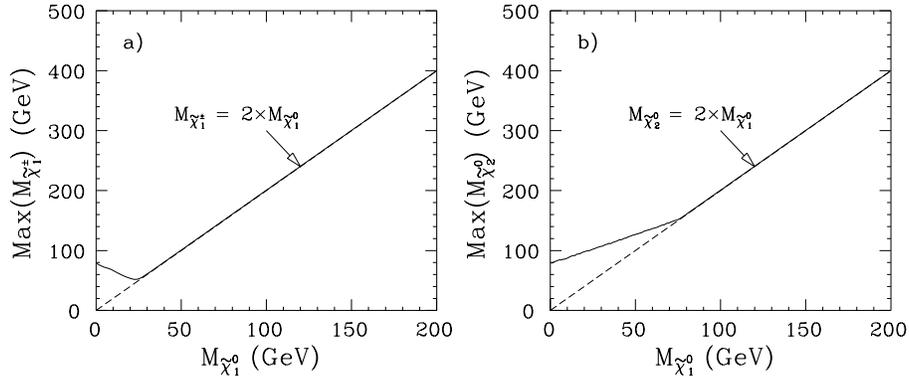,width=12cm,angle=90}}
\caption[upper]{Upper bound on the chargino $\tilde{\chi}_1^\pm$ and
second neutralino $\tilde{\chi}_2^0$ masses as a function of LSP mass
$m_{\tilde{\chi}_1^0}$, assuming the GUT-relation for gaugino masses $M_2
\simeq 2.0 \times M_1$.}
\label{upper}
\end{figure}

Suppose the right-handed sleptons are the LVSPs. Knowing the masses of
$\tilde{\mu}_R$ and $\tilde{e}_R$, one can test the universality of the
scalar masses better than 1\%, which is an assumption of minimal
supergravity framework. The mass of the LSP is also measured better than
1\%. If one assumes the GUT-relation for gaugino masses, one can put
an upper bound on the mass of the chargino $m_{\tilde{\chi}_1^\pm}
\alt 2 m_{\tilde{\chi}_1^0}$ (Fig.~\ref{upper})
and similarly for the second
neutralino. Therefore, we obtain an idea on the next target
center-of-mass energy.
If we will not discover chargino below that mass, at least a
GUT with minimal particle content will be excluded.

If a chargino is the LVSP, we measure cross sections both from
left-handed and right-handed beams, and the masses of the chargino and
LSP. Using $m_{\tilde{\chi}_1^\pm}$, $m_{\tilde{\chi}_1^0}$ and the
cross section from the right-handed beam, one can perform a
three-dimensional fit on $(M_2, \mu, \tan \beta)$ space. Here, the
GUT-relation of the gaugino masses is assumed. Then, we can predict the
masses of other charginos and neutralinos. The obvious limitation of
this fit is we effectively lose information on $\mu$ ($M_2$) when the
chargino is gaugino-rich (higgsino-rich). However in the higgsino-rich
case, we expect another neutralino nearby which should be discovered
not too far above the chargino threshold. In the
gaugino-rich case, one can extract the mass of the $\tilde{\nu}_e$
exchanged in the $t$-channel from the cross section from the left-handed
beam. Since the mass of $\tilde{\nu}_e$ is related to the mass of
$\tilde{e}_L$ by $m^2_{\tilde{e}_L} = m^2_{\tilde{\nu}_e} - m_Z^2 \cos^2
\theta_W \cos 2\beta \leq m^2_{\tilde{\nu}_e} + 0.77 m_Z^2$, one can
place an upper bound on the mass of $\tilde{e}_L$. Therefore, in both
cases one obtains some information on the next sparticle mass.

\begin{figure}
\centerline{\psfig{file=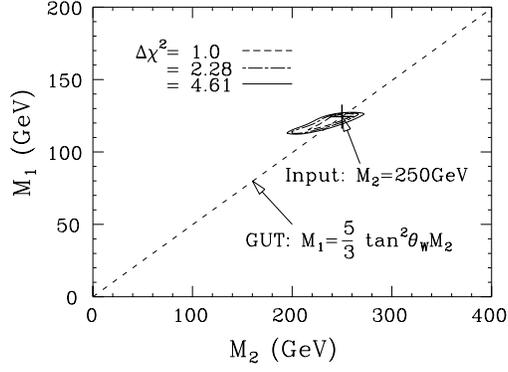,width=7cm,angle=90}}
\caption[GUT]{Test of the GUT-relation of the gaugino masses. The
contours are obtained by a four-dimensional fit on $(M_1, M_2, \mu, \tan
\beta)$.}
\label{GUT}
\end{figure}

\begin{figure}
\centerline{\psfig{file=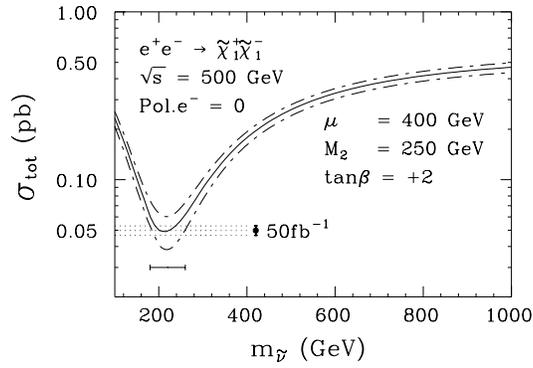,width=7cm,angle=90}}
\caption[snu]{Extracting the mass of $\tilde{\nu}_e$ from the chargino
production cross section, after measuring $(M_1, M_2, \mu, \tan\beta)$ from
the global fit in the previous figure.}
\label{snu}
\end{figure}

If several sparticles are found, stringent tests on theoretical
assumptions can be made. One sample case is the following
\cite{TSUK}. Take the SUGRA parameters $m_0 = 70$~GeV, $M_2 =
250$~GeV, $\mu = 400$~GeV and $\tan \beta = 2$, which is in a
cosmologically interesting region. The low-lying sparticle spectrum is:
\begin{equation}
\begin{array}{rl}
\tilde{\chi}^0_1:& 117.8~{\rm GeV},\\
\tilde{l}_R: & 141.9~{\rm GeV},\\
\tilde{\chi}^\pm_1: & 219.3~{\rm GeV},\\
\tilde{\chi}^0_2:& 221.5~{\rm GeV},\\
\tilde{\nu}_L: & 227.2~{\rm GeV},\\
\tilde{l}_L^\pm: & 235.5~{\rm GeV}.
\end{array}
\end{equation}
Then the right-handed sleptons are found first, and one can infer an
upper bound on the chargino mass as discussed above. Once the chargino
is found, one can use the following four physical observables to
constrain the chargino and neutralino sectors:
\begin{quote}
\begin{tabular}{l}
The mass of LSP.\\
The mass of chargino.\\
The slepton production cross section from the right-handed beam.\\
The chargino production cross section from the right-handed beam.
\end{tabular}
\end{quote}
Since there are four observables, a global fit on the four-dimensional
space
$(M_1$,$M_2$,$\mu$,$\tan\beta)$ is possible without assuming the
GUT-relation $M_1/M_2 = (5/3) \tan^2 \theta_W$. Then the result gives us
an experimental test on the GUT-relation. For the above parameter set,
the test is possible at 3\% level, as seen in Fig.~\ref{GUT}. Once the
chargino and neutralino parameters are known, the chargino production cross
section from the left-handed beam tells us the mass of $\tilde{\nu}_e$
(Fig.~\ref{snu}). Also the comparison between $\tilde{l}_R$ and $\tilde{l}_L$
masses allows us to extract the difference of their masses at the
GUT-scale.

\subsection{Other sparticles and higher order processes}

\begin{figure}
\centerline{
\psfig{file=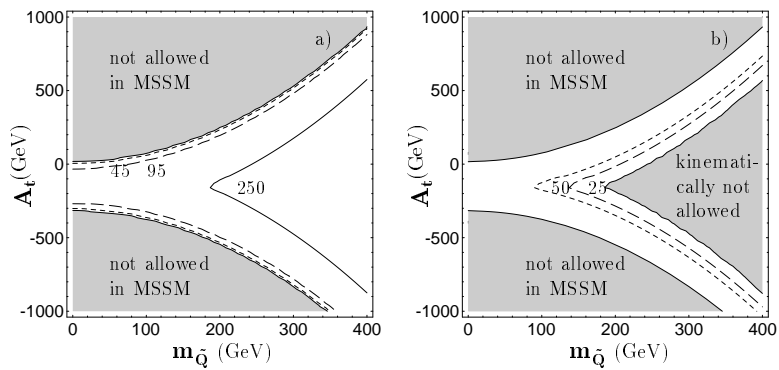,height=5cm}}
\caption[stops]{Contours of a) constant masses of the lightest stop (in GeV),\\
and
b) constant cross sections of $e^+ e^- \rightarrow \tilde t_1 \tilde t_1$
(in fb) as a function of $m_{\tilde Q}$ and $A_t$ for $\tan\beta = 2$ and
$\mu = -300$~GeV in the Minimal Supersymmetric Standard
Model (MSSM). Masses of squark doublet and right-handed stop are assumed
to be the same.
In a), the kinematic production limits of $e^+ e^- \rightarrow \tilde t_1
\tilde t_1$ are given for LEP1, LEP2 (with $\sqrt{s} = 190$~GeV), and the \NLC
(with $\sqrt{s} = 500$~GeV).
In b), the contours are given for 25 and 50 fb, which correspond
approxi\-mately
to the expected experimental sensitivity at $\sqrt s = 500$~GeV
and $\int {\cal L} dt = 10~\mbox{fb}^{-1}$.}
\label{stops}
\end{figure}

\begin{figure}
\centerline{
\psfig{file=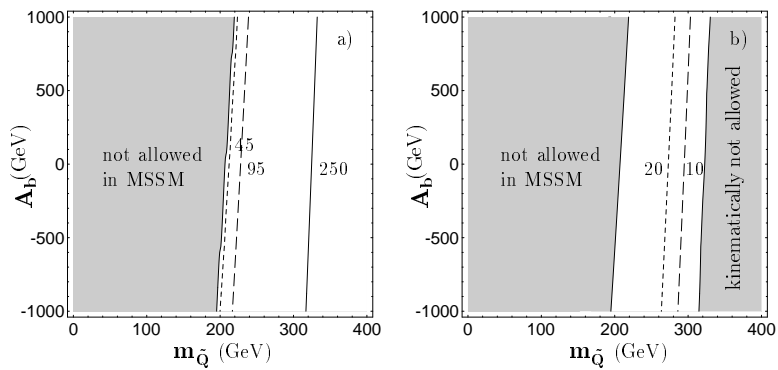,height=5cm}}
\caption[sbottoms]{Contours of a) constant masses of the lightest sbottom
(in GeV),
and
b) constant cross sections of $e^+ e^- \rightarrow \tilde b_1 \tilde b_1$
(in fb) as a function of $m_{\tilde Q}$ and $A_b$ for $\tan\beta = 30$ and
$\mu = -300$~GeV in the Minimal Supersymmetric Standard Model (MSSM).
Masses of squark doublet and right-handed sbottom are assumed
to be the same.
In a), the kinematic production limits of $e^+ e^- \rightarrow \tilde b_1
\tilde b_1$ are given for LEP1, LEP2 (with $\sqrt s = 190$~GeV), and the \NLC
(with $\sqrt s = 500$~GeV).
In b), the contours are given for 10 and 20 fb, which correspond
approxi\-mately
to the expected experimental sensitivity at $\sqrt s = 500$~GeV
and $\int {\cal L} dt = 10~\mbox{fb}^{-1}$.}
\label{sbottoms}
\end{figure}

The third generation squarks $\tilde{t}$, $\tilde{b}$ deserve a special
attention, since they are probably the lightest among the squarks and
hence the most likely candidate for the first signal of squarks at an
$e^+ e^-$ collider. Also they have a mixing between left
right states which is unique to the third generation squarks.
They offer possibilities of measuring $A$-parameters, if $\mu$ and $\tan \beta$
are measured from neutrino, chargino or Higgs sectors.
The dependence of masses and cross sections on $A$-parameters and $m_Q$
are shown in Figs~\ref{stops},\ref{sbottoms} for stop and sbottom production.
For instance, knowing the masses of $\tilde{t}_1$, $\tilde{t}_2$ from
the processes $\tilde{t}_1$-pair and $\tilde{t}_1\tilde{t}_2$
associated production, and their cross sections overdetermine the stop
$2\times 2$ mass matrix.

The search for and study of first- and second-generation squarks was
discussed in \cite{FF}, assuming degenerate masses for left-handed and
right-handed squarks separately.  It was shown that squark mass
measurement at a few GeV level is possible even in the presence of
cascade decays using the kinematic fits.  Here again the beam
polarization plays a crucial role to disentangle left- and right-handed
squarks.

Higher order signal processes (three-body final states)
have not been studied in enough detail for future
linear $e^+ e^-$ colliders.  One can look for final states like
$\tilde{e} e \tilde{\chi}_1^0$, $\tilde{\nu} \tilde{\nu}^* \gamma$,
$\tilde{\chi}_1^\pm \tilde{\nu} e^\mp$ to extend the discovery reach
beyond that using the pair production processes.
The main backgrounds are processes with
$t$-channel exchange of the gauge bosons such as $ee\gamma$, $e\nu_e W$,
$e\nu WZ$ and $eeWW$ final states with or without additional high $p_T$
photons which become increasingly important at higher energies.
More work is necessary here.
(See, however, \cite{Jimbo}).

\subsection{Quantitative verification of supersymmetry}

In this subsection we discuss how one can test whether the new particles
are indeed sparticles whose interactions are restricted by the
supersymmetric Lagrangian. The measurements of parameters discussed
above are based on the {\it assumption}\/ that the newly discovered
particles are sparticles, and we used supersymmetric Lagrangian to
analyze the experimental data. Here we relax this assumption and try to
test the supersymmetric invariance of the Lagrangian experimentally.

We assume that the chargino $\tilde{\chi}_1^\pm$ is the LVSP. Suppose
the chargino is gaugino-rich, {\it i.e.},\/ almost a pure wino
$\tilde{W}$. Then, assuming SUSY, its coupling $g_\chi$ to the electron and
scalar
neutrino is fixed to be the same as the $SU(2)_L$ gauge coupling $g$ even if
SUSY is spontaneously broken. We
study whether this equality can be experimentally tested.

First of all, the chargino production cross section from the
right-handed electron beam should nearly vanish in this limit.
Therefore, we first learn, directly from experiment,
that the hypercharge of the new particle is
zero. Below we assume that it belongs to $SU(2)_L$
triplet as $\tilde{W}$ does. Then the $s$-channel amplitudes of
$\gamma$- and $Z$-exchange are fixed completely by the gauge invariance.
For the production of charginos from the left-handed electron beam,
there is another diagram where $\tilde{\nu}_e$ is exchanged in the
$t$-channel, with a factor
\begin{equation}
\frac{g_\chi^2}{t - m_{\tnu_e}^2}
\end{equation}
in the amplitude. If $m_{\tnu_e}$ is not too large, one can extract both
$g_\chi$ and $m_{\tnu_e}$ from the angular distribution of the charginos.

\begin{figure}
\centerline{
\psfig{file=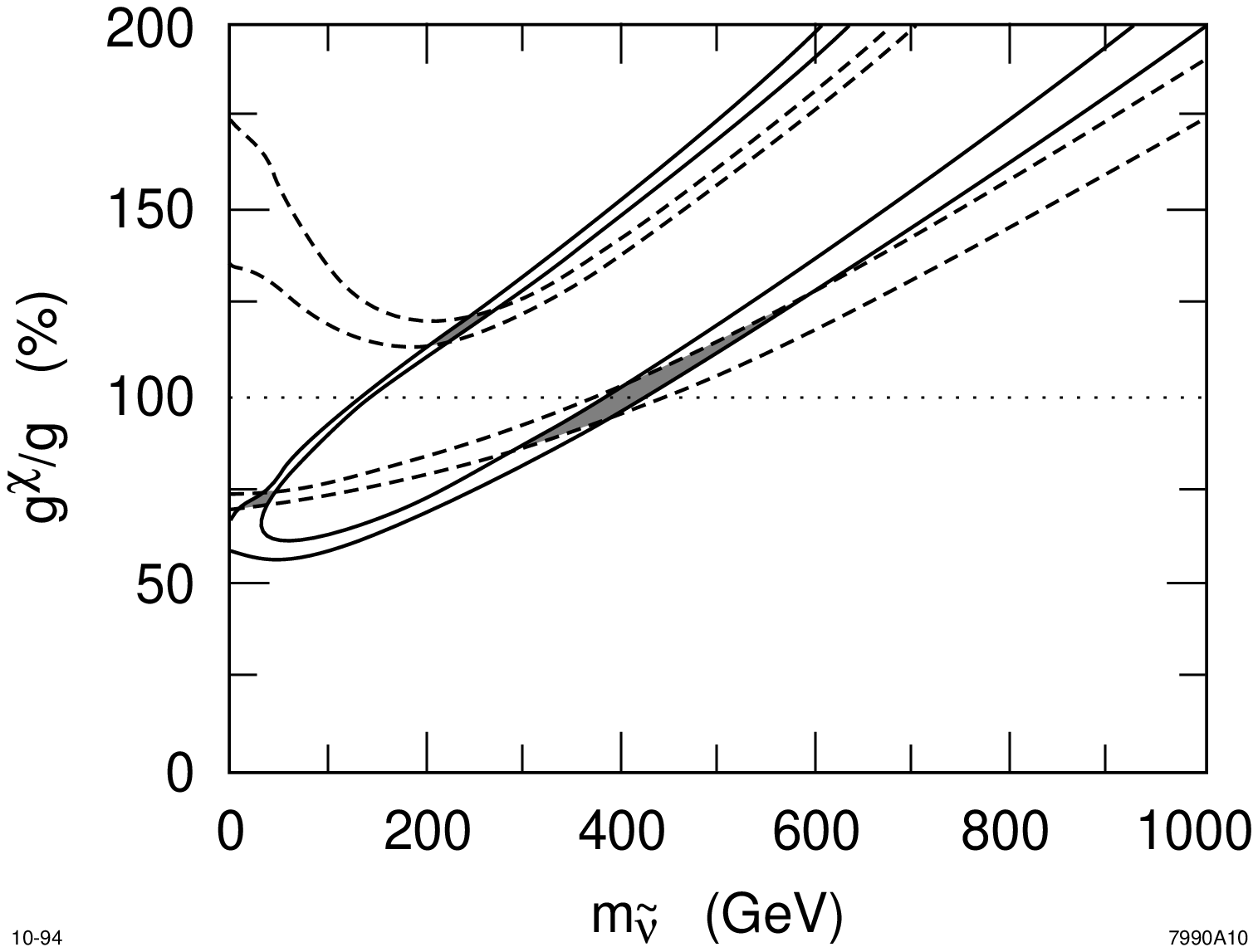,height=5cm}}
\caption[FPMT7b]{Allowed region (shaded) of the $(m_{\tilde{\nu}},
g^\chi)$ plane for the integrated luminosity 100~fb$^{-1}$. Solid
(dashed) curves are contours of constant $\sigma_L$ ($A^\chi$) that
bound the allowed regions. On the dotted lines, the SUSY relation
$g^\chi = g$ is satisfied.}
\label{FPMT7b}
\end{figure}

A sample case was studied in Ref.~\cite{FPMT}, with the following
parameter set
\begin{equation}
(\mu, M_2, \tan\beta, M_1/M_2, m_{\tilde{\nu}_e})
	= (-500, 170, 4, 0.5, 400) .
\end{equation}
In this case the MSSM gives
\begin{quote}
\begin{tabular}{rcl}
$m_{\tilde{\chi}_1^\pm}$ &=& 172~GeV,\\
$m_{\tilde{\chi}_1^0}$ &=& 86~GeV,\\
$\sigma_R$ &=& 0.15~fb,\\
$\sigma_L$ &=& 612~fb,\\
$(\phi_+,\phi_-)$ &=& (1.2$^\circ$, 12.8$^\circ$).
\end{tabular}
\end{quote}
Here $\sigma_R$, $\sigma_L$ are the production cross sections from
right-handed and left-handed electron beams, respectively. The angles
$(\phi_+,\phi_-)$ are the mixing angle for left- and right-handed charginos.
Since the chargino $\tilde{\chi}_1^\pm$ decays
into real $W$ and LSP in this case, the branching ratios are
known. Therefore the total cross section can be determined independently
from the other SUSY parameters. The angular distribution of the chargino
cannot be determined directly due to its decay. For the mixed
hadronic-leptonic mode, one can use the forward-backward asymmetry of
the $W$ instead, where the charge of the $W$ is determined by the lepton
from the other chargino. The cuts are the same as the one used above for
the chargino-pair decaying into the real $W$. Because of the asymmetric
cut in the forward and backward region, the asymmetry is defined by
\begin{equation}
A^{had} = \frac{\sigma_L (0 < \cos\theta < 0.707)
		- \sigma_L(-1 < \cos\theta < 0)}
		{\sigma_L (-1 < \cos\theta < 0.707) } ,
\end{equation}
where the polar angle $\theta$ is that of the reconstructed $W$. This
observable is strongly correlated with the corresponding asymmetry of
the chargino polar angle $A^\chi$. With an integrated luminosity of
100~fb$^{-1}$ with the left-handed electron beam, one can measure the
asymmetry as
\begin{equation}
A^\chi = 0.20 \pm 0.049,
\end{equation}
and the cross section as
\begin{equation}
\frac{\Delta \sigma}{\sigma} = 5.6\%.
\end{equation}
Given four experimental observables, the mass of the chargino
$m_{\tilde{\chi}_1^\pm}$, the LSP mass $m_{\tilde{\chi}_1^0}$, the total
cross section and the asymmetry, one can determine the region on the
space $(m_{\tilde{\nu}_e}, g_\chi/g)$ which reproduce the data, as shown
in Fig.~\ref{FPMT7b}. Since the measurements determine the parameters to
lie in one of the two shaded regions of Fig.~\ref{FPMT7b}, we see that the
equality of $g_{\chi}$ and $g$, and hence SUSY, may be tested to within 25\%.
For further details, and a test for SUSY when the chargino is a mixed
gaugino-Higgsino state, we refer the reader to Ref.\cite{FPMT}.

\subsection{$\gamma\gamma$, $e\gamma$, $e^- e^-$ options}

The $e^+ e^-$ linear colliders have also options to operate as
$\gamma\gamma$, $e\gamma$ or $e^- e^-$ colliders. The discussion of
each mode in this subsection is rather qualitative because
full studies with realistic detectors have not been done. It is only meant to
survey the advantages and disadvantages of each option.

There is a claim that $\gamma\gamma$ or $e\gamma$ colliders can achieve
higher luminosity than $e^+ e^-$ colliders because of the absence of
beam-beam interaction, {\it i.e.}\/, beamstrahlung \cite{Telnov}.
Another advantage of $\gamma\gamma$ mode is that the scalar pair
production occurs in $S$-wave, such that the threshold behavior of the
cross section is $\propto \beta$ rather than $\beta^3$ at an $e^+ e^-$
collider. Also the production cross section is democratic, {\it i.e.}\/,
it only depends on the electric charge of the sparticles. Disadvantages
of $\gamma\gamma$ option is that the $W$-pair cross section is high
$\simeq 100$~pb and stays constant above the threshold, and that the
center-of-mass energy of $\gamma\gamma$ collision has a spread of $>O(10)$\%.
The highest center-of-mass energy is 80\% of the
corresponding $e^+ e^-$ collider.

The search for sleptons has been discussed for $\gamma\gamma$ option,
and is well possible even in the presence of large $W$-pair background
\cite{slepton-gamma-gamma}. For a case $m_{\tilde{l}} -
m_{\tilde{\chi}_1^0} > 50$~GeV, one can even almost completely eliminate
the $W$-pair background by requiring $p_T(l^\pm) > 50$~GeV and
$\theta_{acop} > 90^\circ$, maintaining a reasonable efficiency $\sim 10$\%
for the signal. The
measurement of the slepton mass is worse than $e^+ e^-$ case due to the
energy spread of the backscattered $\gamma$-beam, but still possible
at 5\% level \cite{Kilgore}. The discovery reach is more or less the
same as the $e^+ e^-$ option, since it gains by $S$-wave production, but
loses by the lower center of mass energy.

Chargino search at $\gamma\gamma$ colliders is more difficult
than at $e^+ e^-$ colliders. For instance, when
charginos decay into real $W + \tilde{\chi}_1^0$, acoplanar $W$-pair
with large $\not\!p_T$ is the signature. However, given large $W$-pair
cross section, there is still a long tail in missing $p_T$ distribution
from $WW\gamma$ production where $\gamma$ escapes into the beam pipe
\cite{Eboli}. More thorough study is necessary.

An interesting advantage of $e\gamma$ option is that it has higher reach
than $e^+ e^-$, $\gamma\gamma$ modes on the search for a scalar
electron.\footnote{However, the luminosity in this mode may be limited to
avoid the beam-induced $e^+ e^-$ pair background.}
One can produce $e\gamma \rightarrow \tilde{e}
\tilde{\chi}_1^0$, and the kinematic reach is limited by $m_{\tilde{e}}
+ m_{\tilde{\chi}_1^0}$ rather than $2 m_{\tilde{e}}$ \cite{egamma}. The
signature is a single lepton with a large angle and missing $p_T$. The
most severe background is $e^-\gamma\rightarrow \nu_e W^-$, which can be
suppressed by employing a right-handed electron beam. The kinematic
suppression is only $\beta$, and a search is possible roughly up to the
kinematic limit. Other sparticles are not easy to produce in this
option, mainly only via effective $\gamma\gamma$ collision from
$t$-channel $\gamma$-exchange, and hence the cross section is down by
another factor of $\alpha$.

At an $e^- e^-$ collider, one can produce $\tilde{e}_{L,R}^-,
\tilde{e}_{L,R}^-$ by $t$-channel neutralino exchange \cite{e-e-}. An
advantage is the absence of $W$-pair background, and one can accumulate a
very pure sample of selectron signals both with right- and left-handed
beams. To avoid $W$-pair
backgrounds at an $e^+ e^-$ collider, the use the right-handed
electron beams was preferred for the precision studies.
For the studies of the neutralino sector using selectron
production, this choice of the right-handed electron beam drops the
information on the wino component in the neutralino sector. At $e^-
e^-$ colliders, one can study $\tilde{e}_L^-
\tilde{e}_L^-$ production with both the electron beams left-handed, and
extract information on the wino component in the neutralino sector.
Therefore this option could be used to study sleptons and neutralino
sector further once discovered at the $e^+ e^-$ option \cite{Strassler}.

\section{Overview and complementarity of facilities}

Here, we summarize our results for the SUSY reach of various hadron collider
options (the reach of $e^+e^-$ colliders is essentially the beam energy for
most sparticles). Finally, we emphasize the complementarity
of the hadron and $e^+e^-$ collider options for a complete study of
supersymmetry.

\subsection{Comparison of Hadron Collider options}

Our main results for the SUSY reach of various hadron collider options are
summarized in Table 7. These have generally been obtained within the minimal
supergravity framework, or using MSSM parameter values motivated by
supergravity.

\begin{table}
\small
\caption[] {Estimates of the discovery
reach of various options of future hadron colliders. The signals have
mainly been computed for negative values of $\mu$. We expect that the reach
in especially the $all \to 3\ell$ channel will be sensitive to the sign of
$\mu$.}

\vskip 0.5\baselineskip
\footnotesize
\begin{tabular}{c|ccccc}
&Tevatron II&Main Injector&Tevatron$^*$&DiTevatron&LHC\\
Signal&0.1~fb$^{-1}$&1~fb$^{-1}$&10~fb$^{-1}$&
1~fb$^{-1}$&10~fb$^{-1}$\\
&1.8~TeV&2~TeV&2~TeV&4~TeV&14~TeV\\
\hline
$\etmiss (\tq \gg \tg)$ & $\tg(210)/\tg(185)$ &
$\tg(270)/\tg(200)$ & $\tg(340)/\tg(200)$ & $\tg(450)/\tg(300)$
& $\tg(1300)$ \\

$l^\pm l^\pm (\tq \gg \tg)$ & $\tg(160)$ & $\tg(210)$ &
$\tg(270)$ & $\tg(320)$ & \\

$all \rightarrow 3l$ $(\tq \gg \tg)$ & $\tg(180)$ &
$\tg(260)$ & $\tg(430)$ & $\tg(320)$ & \\

$\etmiss (\tq \sim \tg)$ & $\tg(300)/\tg(245)$ &
$\tg(350)/\tg(265)$ & $\tg(400)/\tg(265)$ & $\tg(580)/\tg(470)$ &
$\tg(2000)$ \\

$l^\pm l^\pm (\tq \sim \tg)$ & $\tg(180-230)$ & $\tg(320-325)$ &
$\tg(385-405)$ & $\tg(460)$ & $\tg(1000)$ \\

$all \rightarrow 3l$ $(\tq \sim \tg)$ & $\tg(240-290)$ &
$\tg(425-440)$ & $\tg(550^*)$ & $\tg(550^*)$ &
$\stackrel{>}{\sim}\tg(1000)$ \\


$\tilde{t}_1 \rightarrow c \tz_1$ & $\tilde{t}_1(80$--$100)$ &
$\tilde{t}_1 (120)$ & \\

$\tilde{t}_1 \rightarrow b \tw_1$ & $\tilde{t}_1(80$--$100)$ &
$\tilde{t}_1 (120)$ & \\

$\Theta(\tilde{t}_1 \tilde{t}_1^*)\rightarrow \gamma\gamma$ &
--- & --- & --- & --- & $\tilde{t}_1 (250)$\\

$\tl \tl^*$ & $\tl(50)$ & $\tl(50)$ & $\tl(50)$ & &
$\tl(250$--$300)$

\end{tabular}
\end{table}

For the $\etmiss$
signal at the Tevatron and its upgrade options, we present two sets of numbers
corresponding to the analysis of two different groups. The higher number
is obtained from the study by Kamon {\it et. al.}\cite{KAMON}, where it is
assumed that the signal is observable using a ``$5\sigma$'' criterion; {\it
i.e.} if the number of signal events ($N_S$)
exceeds $5\sqrt{N_B}$, $N_B$ being the number of background events. To
extend the reach to large values of squark and gluino mass, these authors
use a relatively hard cut $\etmiss \geq 150$~GeV. They argue that the
background dominantly comes from $Z \rightarrow \nu\bar{\nu}+jet$ events
and (conservatively)
take the total background to be 5 times the $Z$ background. Their analysis
includes a detailed simulation of the CDF detector.
The second number for the Tevatron $\etmiss$ reach in Table 7 is
obtained in Ref. \cite{RPV} using ISAJET to simultaneously
generate all sparticles, but
with softer jet and $\etmiss$ ($\etmiss \geq 75$~GeV)
requirements. These authors have estimated backgrounds from $W$, $Z$ and
$t\bar{t}$ production but have used a toy calorimeter for detector simulation.

A major difference from Ref.\cite{KAMON} is the criteria used to obtain
the reach. Since there are systematic uncertainties, both theoretical ({\it
e.g.} from higher order QCD corrections) as well as numerical (from
simplifications in the simulations),
in the computation of the backgrounds, Ref. \cite{RPV}
considers a signal
to be observable if the signal satisfies
$\sigma (signal)\geq 0.25\sigma (background)$ {\it in addition to}
the $5\sigma$ criterion introduced above. Because we are considering very
large integrated luminosities, this difference is important for signals with
large SM backgrounds: for instance, a signal cross section of 200 $fb$ (which
yields 5K events with an integrated luminosity of 25 $fb^{-1}$), which would be
observable over a background of 40 $pb$ with just the $5\sigma$ criterion,
but not with the additional (and somewhat arbitrary) requirement,
$\frac{N_S}{N_B} \geq 0.25$. We should also mention that no attempt has
been made to optimize the cuts in Ref.\cite{RPV}, and that it may be possible
to further enhance the reach by using harder cuts as in
Ref. \cite{KAMON}.

For the reach in the SS dilepton and $all \rightarrow 3\ell$
channels, we have shown
the numbers from the SUGRA analysis (with $\mu <0$)
of Ref.~\cite{RPV,BCKT}. For the high luminosity upgrades of the Tevatron,
the reach in the trilepton channel is dominantly governed by
the relatively clean event sample
from $\tw_1\tz_2$ production since the production of squarks and gluinos is
kinematically suppressed. For positive values of $\mu$ the reach will
be governed by $\eslt$ and SS dilepton events from gluino
production and will, presumably, be somewhat smaller.

The following comments about Table 7 are worth noting:
\begin{itemize}
\item Given $10$pb$^{-1}$ of integrated luminosity (Tevatron run IA), the
highest reach in $m_{\tg}$ is attained via the multi-jet$+\etmiss$ channel.
The rate limited SS and multilepton signals, which will have a significant
reach by the end of Tevatron run IB, yield the maximum reach
at the Main Injector and $TeV^*$ upgrades of the Tevatron. Within
the assumed framework, the reach
in the clean trilepton channel from $\tw_1\tz_2$ production is comparable
to that of multileptons if $\mu <0$. For positive values of $\mu$ the
branching fraction for the leptonic decay of $\tz_2$, and consequently, the
trilepton signal may be strongly suppressed.

\item For the proposed DiTevatron $p\bar p$ collider, the reach in $m_{\tg}$
via the multi-jet$+\etmiss$ channel may be superior to the
reach via multi-lepton channels. We quote DiTevatron reach values
for only one value of integrated luminosity. If substantially higher
luminosities can be achieved, then the reach in many of the channels
can be significantly increased.

\item At the $TeV^*$ and at the DiTevatron,
the hadronically quiet
trilepton events may be observable all the way up to the spoiler, but
only for some ranges of parameters --- in
particular, this is sensitive to the sign
of $\mu$ because the leptonic branching fractions of $\tz_2$ are considerably
larger for $\mu < 0$ as compared to $\mu > 0$. The SS dilepton signal from
gluinos and squarks is also somewhat suppressed for positive values
of $\mu$ because $m_{\tw_1}-m_{\tz_1}$ tends to be smaller
than in the $\mu < 0$ case, reducing the efficiency for passing the
experimental cuts.

\item The analysis of the LHC working group\cite{POLESELLO} has shown
that the LHC can detect gluinos and squarks well beyond 1~TeV in the $\etmiss$
channel, and up to $\sim 2$~TeV if $m_{\tq}=m_{\tg}$. The SS and trilepton
channels also have a reach of about 1~TeV, so that such events,which should
be detected simultaneously with $\etmiss$ events should provide spectacular
evidence for gluino and squark production.

\item At the LHC, the reach of the clean trilepton signal from
$\tw_1\tz_2$ production extends up to where the spoiler decay modes of $\tz_2$
become accessible for negative values of $\mu$. If $\mu > 0$, the signals
are readily observable for rather small and very large values of
the SUGRA parameter
$m_0$; there are, however, significant regions ($m_0 = 400-100$~GeV) where
this signal may be suppressed.
It is also worth emphasizing that because it is possible to obtain a
very clean sample of $\tw_1\tz_2$ events, it should be possible to
reliably reconstruct
$m_{\tz_2}-m_{\tz_1}$ (and, perhaps, also some other combinations of
masses)\cite{bcpttwo} at the LHC. This may also be possible at the $TeV^*$,
but will require the machine and detectors to perform at their limits.

\item The Tevatron and its Main Injector upgrade should be able to search
for $\tst_1$ up to, or just beyond about 100~GeV, regardless of whether
these decay via the tree-level chargino mode or via the loop decay $\tst_1
\rightarrow c\tz_1$\cite{BST}.
Signals for yet heavier stops ($m_{\tst_1} \geq 150$~GeV)
which could have other kinematically allowed decays are under investigation. It
has
also been pointed out, assuming that $\tst_1 \rightarrow c\tz_1$ is the
dominant decay mode of $\tst_1$, that it should be possible\cite{DN}
to search for
it at the LHC via two photon decays of its scalar bound state in much the
same way that Higgs bosons are searched for. With an integrated luminosity of
100 $fb^{-1}$, the reach in this channel has been estimated to be
$m_{\tst_1}\alt 250$~GeV.

\item Finally, it appears that even the $TeV^*$ will not probe sleptons
significantly beyond the reach of LEP. The corresponding reach for the
LHC is about 250~GeV\cite{slep}. The analysis for slepton signals at
the DiTevatron has not been performed, but expectations are pessimistic.
\end{itemize}

\subsection{Complementarity between $e^+e^-$ and hadron colliders}

At the LHC, supersymmetric events will, in general, manifest themselves
via complicated cascades of heavy sparticles, resulting in relatively
spectacular $\etmiss$ and multilepton signatures for SUSY.
A number of
complementary signals ought to be detectable.
While the observation of such
events will unequivocally signal the existence of New Physics, it will
probably be difficult to unravel the complicated cascades in the rather
messy environment of the hadron collider.
Especially, it will not be easy
(except in some cases such as $\tw_1\tz_2$ trilepton signal) to sort
out the sources of various signals or do sparticle spectroscopy.
In contrast, at a 500~GeV $e^+e^-$
collider,
where only relatively light sparticles will be kinematically
accessible, the decay cascades
will likely be less complicated or even absent. Note, however, that if gaugino
masses have a common origin as in a GUT, a reach of about 250~GeV in the
chargino mass is equivalent to a reach of $\sim 700-800$~GeV in $m_{\tg}$.
We have also seen that these machines offer the prospects of precision
measurements of masses, spins and, in some cases, also
couplings of sparticles. Such measurements, which are generally difficult
at hadron colliders, not only serve as the
most direct tests of supersymmetry, but may also
yield information about physics at very high energy scales.

It is also interesting to ask how the information about,
say, chargino couplings and masses learned from $e^+e^-$ experiments can be
used to sort out the cascade decays seen at the LHC. If the turn-on of the
\NLC occurs significantly after the LHC becomes operational, it will be
important to appropriately archive the raw data from LHC experiments for
subsequent reanalysis in light of new knowledge from the \NLC. This interplay
between $e^+e^-$ and $pp$ collider analyses highlights yet another
sense in which these two facilities are complementary.

\section{Conclusions}

We have seen that if low energy supersymmetry is the physics that stabilizes
the electroweak scale, the supersymmetric partners of ordinary particles
will, in many cases,
almost certainly be detectable at the LHC or the future $e^+ e^-$ linear
colliders, but only
if we are lucky, at the Tevatron or at LEP II.
Although this is not the subject of this report, it is worth stressing that
supercolliders are also capable of searching for a variety of other New
Physics that Nature may have chosen to adopt.
While the Tevatron upgrades that we have considered
probe substantial ranges of SUSY parameters, they
do not yield observable signals over what is generally
accepted as the complete range of these parameters. Supercolliders
appear to be essential both for a complete exploration of the parameter space
as well as for the elucidation of any New Physics that might be discovered.
Finally, we cannot overstress the complementary nature of $e^+e^-$ and
hadron supercollider if supersymmetry is indeed present at the weak scale.
Experiments at these facilities will together not only lead to unambiguous
discovery of sparticles, but will allow a comprehensive study of
their properties, which
in turn, may yield information about physics at ultra-high energy scales.

\section{Acknowledgements}

We thank Manuel Drees and Howie Haber for a critical reading of the manuscript,
and for valuable comments.
This work was supported in part by the U.S. Department of Energy under
contract number DE-FG05-87ER40319, DE-AC03-76SF00098 and DE-FG-03-94ER40833.

\bibliographystyle{unsrt}

\end{document}